\documentclass[12pt]{article}
\pdfoutput=1

\usepackage{putex}
\usepackage{graphicx}
\usepackage{caption}
\usepackage{amsmath}
\usepackage{array}
\usepackage{subcaption}
\usepackage{epstopdf}
\usepackage{enumerate}
\usepackage{cite}
\usepackage{youngtab}
\usepackage{tensor}
\usepackage{slashed}
\usepackage[aligntableaux=center]{ytableau}
\usepackage[utf8]{inputenc}
\usepackage[
      colorlinks=true,
      linkcolor=blue,
      urlcolor=blue,
      filecolor=black,
      citecolor=red,
      ]{hyperref}
\usepackage{subcaption}

\newcommand{\grp}[1]{\mathrm{#1}}

\newcommand{\HH}{\mathbb{H}}

\newcommand{\abs}[1]{\left\lvert #1 \right\rvert}

\newcommand {\be} {\begin {equation}}
\newcommand {\ee} {\end {equation}}

\newcommand {\bes} {\begin {equation*}}
\newcommand {\ees} {\end {equation*}}

\newcommand{\es}[2] {\begin{equation} \label{#1} \begin{split} #2 \end{split} \end{equation}}

\newcommand{\CP}{\mathbb{CP}}
\newcommand{\Z}{\mathbb{Z}}

\newcommand{\R}{\mathbb{R}}
\newcommand{\C}{\mathbb{C}}

\newcommand{\cC}{{\mathcal C}}
\newcommand{\cD}{{\mathcal D}}

\newcommand{\cF}{{\mathcal F}}

\newcommand{\cH}{{\mathcal H}}

\newcommand{\cK}{{\mathcal K}}
\newcommand{\cL}{{\mathcal L}}
\newcommand{\cN}{{\mathcal N}}
\newcommand{\cO}{{\mathcal O}}
\newcommand{\cP}{{\mathcal P}}
\newcommand{\cQ}{{\mathcal Q}}

\newcommand{\cS}{{\mathcal S}}
\newcommand{\cW}{{\mathcal W}}
\newcommand{\cM}{{\mathcal M}}
\newcommand{\cU}{{\mathcal U}}

 \newcommand{\f}{\tiny\yng(1)}

\newcommand{\vphi}{\varphi}

\newcommand{\tM}{\widetilde{M}}

\newcommand{\beq}{\begin{equation}}
\newcommand{\eeq}{\end{equation}}

\newcommand{\ov}{\over}

\def\le{\left}
\def\ri{\right}

\newcommand\al{{\alpha}}
\newcommand\ep{\epsilon}
\newcommand\sig{\sigma}
\newcommand\Sig{\Sigma}
\newcommand\lam{\lambda}

\newcommand\om{\omega}

\newcommand\ga{{\ensuremath{{\gamma}}}}
\newcommand\Ga{{\ensuremath{{\Gamma}}}}
\newcommand\de{{\ensuremath{{\delta}}}}

\def\p{\partial}

\def\ie{\begin{equation}\begin{aligned}}
\def\fe{\end{aligned}\end{equation}}

\newcommand{\la}{\langle}
\newcommand{\ra}{\rangle}

\newcommand{\m}{\mu}
\newcommand{\n}{\nu}
\newcommand{\pa}{\nabla}
\newcommand{\A}{{\alpha}}
\newcommand{\B}{{\beta}}
\renewcommand{\C}{{\gamma}}
\newcommand{\D}{{\delta}}

\newcommand{\CC}{{\Gamma}}

\newcommand{\ve}{{\varepsilon}}

\newcommand{\mZ}{{\mathbb Z}}
\newcommand{\mR}{{\mathbb R}}

\newcommand{\Tr}{{\rm Tr\,}}
\newcommand{\mf}{\mathfrak }

\numberwithin{equation}{section}

\def\<{\langle}
\def\>{\rangle}

\begin{document}

\preprint{PUPT-2576}

\institution{Simons}{Simons Center for Geometry and Physics, SUNY, Stony Brook, NY 11794, USA}
\institution{PU}{Joseph Henry Laboratories, Princeton University, Princeton, NJ 08544, USA}

\title{Chern-Simons theory from M5-branes and calibrated M2-branes
}

\authors{M\'ark Mezei,\worksat{\Simons} Silviu S.~Pufu,\worksat{\PU} and Yifan Wang\worksat{\PU}}

\abstract{
We study a sector of the 5d maximally supersymmetric Yang-Mills theory on $S^5$ consisting of $1/8$-BPS Wilson loop operators contained within a great $S^3$ inside $S^5$.  We conjecture that these observables are described by a 3d Chern Simons theory on $S^3$, analytically continued to a pure imaginary Chern-Simons level.  Therefore, the expectation values of these 5d Wilson loops compute knot invariants.  We verify this conjecture in the weakly-coupled regime from explicit Feynman diagram computations. At strong coupling, these Wilson loop operators lift to $1/8$-BPS surface operators in the 6d $(2,0)$ theory on $S^1\times S^5$.  Using AdS/CFT, we show that these surface operators are dual to M2-branes subject to certain calibration conditions required in order to preserve supersymmetry.  We compute the renormalized action of a large class of calibrated M2-branes and obtain a perfect match with the field theory prediction. Finally, we present a derivation of the 3d Chern-Simons theory from 5d super-Yang-Mills theory using supersymmetric localization, modulo a subtle issue that we discuss. }
\date{}

\maketitle

\tableofcontents

\section{Introduction}

In the past decade, supersymmetric localization has provided a plethora of exact results in supersymmetric field theories with various amounts of supersymmetry and in various numbers of dimensions (see \cite{Pestun:2016zxk} for a collection of review articles).  Of particular importance have been the partition functions on round spheres, whose computations in supersymmetric theories motivated the $F$-theorem \cite{Jafferis:2011zi,Klebanov:2011gs} and generalizations thereof \cite{Giombi:2014xxa,Chang:2017cdx,Minahan:2017wkz,Gorantis:2017vzz}, provided new precision tests of the anti-de Sitter / conformal field theory (AdS/CFT) correspondence \cite{Drukker:2010nc,Herzog:2010hf,Jafferis:2011zi,Bobev:2013cja,Jafferis:2012iv,Chang:2017mxc}, as well as various exact results for correlation functions in superconformal field theories (SCFTs) \cite{Giombi:2009ds,Giombi:2009ek,Pestun:2007rz,Pestun:2009nn,Closset:2012vg,Gerchkovitz:2016gxx,Dedushenko:2016jxl, Chang:2017cdx}.  Our work here stems from noticing a potential coincidence in \cite{Kim:2012ava,Kim:2012qf}:  after localization, the partition function of the 5d maximally supersymmetric (${\cal N} = 2$) Yang-Mills (MSYM) theory on a round $S^5$ has an identical form to the partition function of 3d Chern-Simons (CS) theory on a round $S^3$! 
For example, if the gauge group is $U(N)$, then the $S^5$ partition function takes the form of an $N$-dimensional integral \cite{Kim:2012ava,Kim:2012qf} (see also \cite{Hosomichi:2012ek,Kallen:2012va,Qiu:2016dyj})
\es{ZS5}{
	Z_{S^5} \propto \int d^N\lambda \, e^{-\frac{4\pi^3 R}{g_\text{YM}^2} \sum_{i=1}^N \lambda_i^2}
	\prod_{i<j} \left[2 \sinh \left[ \pi (\lambda_i - \lambda_j) \right] \right]^2 \,,
}
where $R$ is the radius of $S^5$, and where we ignored an overall proportionality factor.  For the same choice of $U(N)$ gauge group, the $S^3$ partition function of CS theory at renormalized level $k$ can also be written as an $N$-dimensional integral \cite{Kapustin:2009kz}:\footnote{In the conventions adopted here, the usual CS path integral for compact gauge group $G$ is given by
	\ie
	Z_{\rm CS}=\int D A\, e^{-i{k_b\over 4\pi}\int \Tr\left(AdA-{2i\over 3} A^3\right)} \,,
	\fe
		with hermitian gauge field $A$. This differs from the convention in \cite{Witten:1988hf} by $A_{\rm here}=-i A_{\rm there}$ so that the gauge covariant derivative is $D_{\rm here}=d-iA_{\rm here}$.
		Here $ k_b$ is the bare CS level and the renormalized CS level $k$ is 
	\ie
	k=k_b+ h\, {\rm sgn}( k_b) \,, 
	\fe
	where $h$ is the dual Coxeter number of the gauge group $G$.
}
\es{ZS3}{
	Z_{S^3} \propto \int d^N\lambda \, e^{i \pi k \sum_{i=1}^N \lambda_i^2}
	\prod_{i<j} \left[2 \sinh \left[ \pi (\lambda_i - \lambda_j) \right] \right]^2 \,.
}
The two formulas \eqref{ZS5}--\eqref{ZS3} take the same form provided that we analytically continue the 3d partition function to pure imaginary values of $k$ and identify
\es{kFromgYM}{
	k = i \frac{4 \pi^2 R}{g_\text{YM}^2}  \,.
} 
The appearance of the 3d CS partition function from that of 5d MSYM theory seems rather surprising because the 3d CS theory breaks parity and is topological, while 5d MSYM preserves parity and, although it is IR free, its correlation functions at intermediate energy scales are far from trivial.   

Nevertheless, we will explain the similarity in the form of the expressions in \eqref{ZS5} and \eqref{ZS3} by showing that one can obtain a full 3d CS theory (not just its partition function) on an $S^3$ submanifold of $S^5$. 
In particular, we identify explicitly a sector of renormalized ${1\over 8}$-BPS Wilson loop operators $W^{\rm ren}_R(\cK)$  in the 5d MSYM theory whose shapes are general knots $\cK$ that are restricted to belong to a great $S^3 \subset S^5$. Here, $R$ is a representation of the gauge group $G$.  We propose that these Wilson loops correspond to the usual Wilson loop operators $W^{3d}_R(\cK)$ in the effective 3d CS theory. In particular their 5d expectation values compute knot invariants in $S^3$. For small $g_{\rm YM}^2$, we can of course compute the expectation value of $W^{\rm ren}_R(\cK)$ from 5d Feynman diagrams.  As we show, due to the supersymmetry preserved, the contributions at each order in $g_{\rm YM}^2$ can be put in the form of certain integrals that compute classical knot invariants, such as the Gauss linking integral. This lends direct support for our proposal.

An important point we want to emphasize is that the naively-defined 5d ${1\over 8}$-BPS Wilson loop $W_R(\cK)$ is not topological. However its dependence on the shape of the loop is rather special and can be canceled by a combination of counter-terms associated with the length $L(\cK)$ of the loop  and torsion $T(\cK)$ of the loop in $S^3$. Therefore we can define a topological renormalized Wilson loop $W^{\rm ren}_R(\cK)$  by multiplying $W_R(\cK)$ by a factor depending on $L(\cK)$ and $T(\cK)$.  While the length  $L(\cK)$ is an innocuous 1d local counter-term on the Wilson loop, the torsion  $T(\cK)$ is a 1d Chern-Simons term associated to the $\mf{so}(2)$ normal bundle in $S^3$, and it introduces framing dependence (choice of a trivialization for the normal bundle) in $W^{\rm ren}_R(\cK)$.  This is parallel to what happens in 3d CS theory where a framing of the knots $\cK$ (as well as for the underlying 3-manifold) is required to define the topological Wilson loop observables.\footnote{In other words, there is a framing anomaly for the topological Wilson loops in 3d CS theory.}  In particular, for our proposal to work, the framing dependence of the renormalized 5d supersymmetric Wilson loop must match that of the 3d Wilson loop in CS theory.

Our result has an interesting implication for the 6d superconformal theories with $(2, 0)$ supersymmetry \cite{Witten:1995zh,Strominger:1995ac,Witten:1995em}, which are  labeled by an ADE Lie algebra $\mathfrak{g}$.   
As argued in \cite{Seiberg:1996bd,Seiberg:1997ax,Douglas:2010iu,Lambert:2010iw,Lambert:2012qy}, the 5d ${\cal N} = 2$ MSYM theory with Lie algebra $\mathfrak{g}$ can be obtained from dimensionally reducing the 6d $(2, 0)$ theory on a small circle of radius $R_6 = g_\text{YM}^2 / (4 \pi^2)$, with an appropriate R-symmetry twist along this circle in order to preserve ${\cal N} = 2$ supersymmetry in 5d.  More precisely, the dimensional reduction gives the 5d ${\cal N} = 2$ MSYM theory supplemented by a particular set of higher derivative corrections suppressed in $R_6$.  The existence of a sector of 5d MSYM theory captured by 3d Chern-Simons theory implies the existence of a similar sector of the $(2, 0)$ theory when the latter is placed on $S^5 \times S^1$ in the small $R_6$ limit.  (See also \cite{Dimofte:2011ju,Cecotti:2011iy,Ooguri:1999bv,Terashima:2011qi,Terashima:2011xe,Dimofte:2011py,Galakhov:2012hy,Beem:2012mb,Yagi:2013fda,Gang:2013sqa,Cordova:2013cea} for other ways of relating 3d Chern-Simons theory to $(2, 0)$ theories.)  In this dimensional reduction, the non-trivial observables computed by Chern-Simons theory, namely the $1/8$-BPS Wilson loop operators of 5d MSYM mentioned above, arise as certain $1/8$-BPS two-dimensional surface operators in the $(2, 0)$ theory that wrap the $S^1$ factor.  
We emphasize that although the 5d MSYM is unrenormalizable, an UV complete definition of the 5d Wilson loop observables is given by the surface operators in the $(2,0)$ theory.

When $g_{\rm YM}^2 \sim R_6$ is large, the 6d description by the $(2,0)$ theory becomes more natural.  
 In this case one may contemplate whether it is possible to provide evidence for the existence of a Chern-Simons subsector of the $(2, 0)$ theory using holography.  At large $N$, the $A_N$-type $(2, 0)$ theory is dual to M-theory on $AdS_7 \times S^4$, with $N$ units of four-form flux threading the $S^4$ factor.  As will be explained in Section~\ref{HOLO}, to place the field theory on $S^5 \times S^1$ while preserving SUSY, one has to analytically continue the $AdS_7 \times S^4$ background to Euclidean signature and perform the bulk analog of the R-symmetry twist needed in the field theory.  In this setup, the simplest two-dimensional surface operators are those that in 5d become Wilson loop operators in the fundamental representation of $SU(N)$, and they correspond to M2-branes that  end on the boundary of the bulk geometry.   

A potential difficulty in comparing 11d M-theory to 5d MSYM theory is that the 5d MSYM  description is a reliable approximation only when $R_6$ is small, while the 11d description is reliable in the opposite limit, when $R_6$ is large.  When $R_6$ is small one could work in type IIA string theory, as we do in Section~\ref{IIA}, but this description becomes uncontrolled in the asymptotic region, where the dilaton blows up. Nevertheless, experience in other situations involving supersymmetric localization in 5d suggests that the expressions for supersymmetry-protected quantities computed in 5d MSYM theory hold in fact for all $R_6$, and not just when $R_6$ is small.  For example, it was shown in \cite{Bullimore:2014upa} that one can use 5d MSYM to reproduce the characters of protected chiral algebras that are subsectors of the $(2, 0)$ theories \cite{Beem:2014kka}.  The intuition behind this non-renormalization result is that higher derivative corrections to the 5d MSYM action are likely to be $\cQ$-exact and thus do not affect $\cQ$-invariant observables.\footnote{The supersymmetric higher-derivative corrections to the 5d MSYM theory can be classified into D-terms and F-terms. The former come from descendants of local scalar operators $\cO$ with respect to all of the 16 supercharges of the MSYM.  The F-terms, on the other hand, arise when starting with local operators $\cO$ that preserve a subset of the supercharges, so that one only needs to act with a subset of the supercharges in order to obtain a fully supersymmetric term.  Such terms in the flat space limit are classified in \cite{Bossard:2010pk,Chang:2014kma}.

While the D-terms are obviously $\cQ$-exact with respect to any supercharge $\cQ$, this is not typically the case for the F-terms.  For example, in 4d, if $\cQ$ is a supercharge that does not have a definite chirality, then F-terms constructed as chiral superspace integrals are not $\cQ$-invariant. (For a concrete example, see \cite{Gerchkovitz:2016gxx} where the $S^4$ partition function of 4d $\cN=2$ SYM was shown to depend on F-term deformations.) 

Here, the possible F-term higher derivative correction to the 5d MSYM  consist of: a ${1\over 2}$-BPS F-terms given by the supersymmetric completions of $\Tr F^4$ and $(\Tr F^2)^2$, and a ${1\over 4}$-BPS F-term that involves $D^2(\Tr F^2)^2$, 
 with a particular contraction of the 5d spacetime indices \cite{Bossard:2010pk,Chang:2014kma}.  However it was shown in  \cite{Lin:2015zea} that such terms must be absent in the 5d effective theory from the $S^1$ compactifcation of the $A_1$ $(2,0)$ theory (in this case due to the trace relations it suffices to show it for $\Tr F^4$).  A modification of the argument there would be needed to prove the absence of such F-terms in the $S^1$ compactification of general $(2,0)$ theories.} (Here, $\cQ$ is a supercharge preserved by our protected sector.)   Thus, assuming that the same intuition holds true in the case of interest to us, we conjecture that at any $R_6$, there exists a sector of the $(2, 0)$ $A_N$ theory on $S^5 \times S^1$ that is captured by 3d Chern-Simons theory.  

The holographic duals of some of the operators in this sector (namely those that reduce to Wilson loops in the fundamental representation of $SU(N)$ in 5d) are Euclidean M2-brane that preserve two supercharges, wrap the $S^1$ circle, and at the boundary of the bulk spacetime approach the product between the $S^1$ factor and a knot $\cK$ that lies within an $S^3$ submanifold of $S^5$.  That the dual boundary operators are captured by CS theory implies that their expectation values, identified at leading order in large $N$ as the regularized and appropriately renormalized M2-brane action, are independent of continuous deformations of $\cK$.  While we do not construct explicitly the M2-brane embeddings that extremize the M2-brane action, we nevertheless use supersymmetry to derive the first order equations they obey, and we show, using methods similar to those in \cite{Dymarsky:2006ve,Drukker:2007qr,Giombi:2009ms}, that the regularized on-shell action (supplemented with the same finite counter terms as in the field theory computation) is indeed a topological invariant of $\cK$.  Its value agrees with the corresponding Wilson loop expectation value in the 3d CS theory in the strong coupling regime.  This is thus a test of our conjecture that the $(2, 0)$ theory on $S^5 \times S^1$ contains a 3d Chern-Simons sector.

Given the above nontrivial checks for both the weak and strong coupling limits of our proposal, we proceed to derive the 3d CS sector of the  ${\cal N} = 2$ MSYM on $S^5$
by performing a supersymmetric localization computation  that is different from the one that led to \eqref{ZS5}.
The appearance of 3d Chern-Simons theory from 5d MSYM theory can be already anticipated given various supersymmetric localization results present in the literature. In four dimensions Refs. \cite{Pestun:2009nn, Mezei:2017kmw} showed that, with an appropriate choice of supercharge, the ${\cal N} = 4$ SYM theory placed either on the positively-curved $S^4$ or on its negatively-curved analog $\HH^4$, localizes to a 2d Yang-Mills theory on an $S^2$ submanifold of $S^4$ or an $\HH^2$ submanifold of $\HH^4$, respectively.  In five dimensions, it was shown in \cite{Bonetti:2016nma} that ${\cal N}  =2$ MSYM on $\HH^5$ localizes to Chern-Simons theory on an $\HH^3$ subspace of $\HH^5$.  Given the analogy with the four-dimensional situation, it is then natural to guess that one can also show that the ${\cal N} = 2$ MSYM theory on $S^5$ localizes to Chern-Simons theory on $S^3$.  One of our goals here is to spell out this computation.  More explicitly, using an off-shell formulation of 5d MSYM on $S^5$ that preserves the supercharge $\cQ$, we show that ${\cal N} = 2$ MSYM on $S^5$ localizes to 3d CS theory on the $S^3$ submanifold and supersymmetric Wilson loops that preserve $\cQ$ become ordinary Wilson loops in the CS theory. 
One important subtlety in the localization computation, which we do not fully address in this paper and hope to come back to in the future, is related to the choice of reality condition for the 5d fields and possible complex solutions to the BPS equations.

The rest of this paper is organized as follows.  In Section~\ref{5DSYM} we review the relation between the $(2, 0)$ theory on $S^5 \times S^1$ and ${\cal N} = 2$ MSYM on $S^5$.  
In Section~\ref{sec:WLsector}, we define the ${1\over 8}$-BPS Wilson loops in the 5d MSYM and explain the relation to surface operators in the 6d $(2,0)$ theory. 
Motivated by perturbative results for these Wilson loops which we present in Section~\ref{sec:pert}, we give a proposal for the effective 3d Chern-Simons theory  in Section~\ref{CONJECTURE} along with predictions for the strongly coupled limit of these observables. 
In Section~\ref{HOLO}, we  study the holographic duals of the ${1\over 8}$-BPS surface operators in M-theory and match to the field theory prediction.
Lastly,
in Section~\ref{LOCALIZATION} we describe the localization computation that reduces 5d MSYM on $S^5$ to Chern-Simons theory on $S^3$.   We end with a brief summary and future directions in Section~\ref{CONCLUSION}.


  \section{Review of ${\cal N} = 2$ SYM on $S^5$}
  \label{5DSYM}
  
  In this section, we begin with a review of ${\cal N} = 2$ SYM on $S^5$ and its relation to the 6d $(2, 0)$ theory.

  \subsection{The 5d MSYM action on $S^5$}

In any number of spacetime dimensions, the maximally supersymmetric Yang-Mills theory in flat space can be obtained by dimensional reduction of the 10d super-Yang-Mills theory on a flat torus.  The SYM theory in 5d can be written in terms of an $\cN = 2$ vector multiplet, which consists of a gauge field $A_\mu$, $\m = 1, \ldots, 5$, five scalars $\Phi^I$, $I = 1, \ldots, 5$, as well as fermions $\Psi_A$, $A = 1, \ldots , 4$, whose spinor indices we suppress, all transforming in the adjoint representation of the gauge group $G$\@.  The $\Phi_I$ and the $\Psi_A$ transform, respectively, in the ${\bf 5}$ and the ${\bf 4}$ of the $\mathfrak{so}(5)_R$ R-symmetry algebra.   The 5d flat space Euclidean Lagrangian is
 \es{EucLagrangianSYM}{
  \cL_\text{flat}=&{1\over g_\text{YM}^2} \tr \biggl[
  {1\over 4}F_{\m\n}F^{\m\n}
  +{1\over 2} D_\m \Phi^I D^\m \Phi^I
  +{i\over 2} \bar \Psi^A  \C^\m  D_\m \Psi_A 
  -{1\over 4}[\Phi^I,\Phi^J]^2 
  -{i\over 2} \bar \Psi^A  (\hat \C^I )_A{}^B [\Psi_B,\Phi^I] \biggr] \,,
 }
where  $D_\mu = \partial_\mu - i A_\mu$ is the gauge covariant derivative, $F_{\m\n} = \partial_\m A_\n - \partial_\n A_\m - i[A_\m,A_\n]$ is the gauge field strength,\footnote{The fields $A_\m$ and $\Phi_I$ are taken to be hermitian here.} $\ga^\mu$ and $\hat \C^I$ are spacetime and $\mathfrak{so}(5)_R$ gamma matrices respectively, and
 \es{Conjugate}{
  \bar \Psi^B \equiv (\Psi_A^T C) \hat C^{AB} \,.
 }
For the two antisymmetric charge conjugation matrices $C$ (with spacetime spinor indices) and $\hat C$ (with $\mathfrak{so}(5)_R$ spinor indices) see Appendix~\ref{GAMMACONVENTIONS}.  In Lorentzian signature, the condition \eqref{Conjugate} could be interpreted as a symplectic Majorana condition provided that $\bar \Psi^B$ is identified as the Dirac adjoint of $\Psi_B$.  In Euclidean signature, we take $\bar \Psi^B$ simply to be given by \eqref{Conjugate}.

The action for 5d SYM on $S^5$ can be obtained by covariantizing the expression \eqref{EucLagrangianSYM} and adding curvature corrections.  These curvature corrections are fixed by demanding invariance under the ${\cal N} = 2$ supersymmetry algebra on $S^5$, which is $\mf{su}(4|2)$. The bosonic part of this algebra consists of the $\mf{su}(4) \cong \mf{so}(6)$ rotational symmetry of $S^5$ as well as an $\mf{so}(3)_R \oplus \mf{so}(2)_R \cong \mf{su}(2)_R \oplus \mf{u}(1)_R$ R-symmetry. (We will interchangeably use $\mf{so}(3)_R \oplus \mf{so}(2)_R$ and $\mf{su}(2)_R \oplus \mf{u}(1)_R$ to describe the R-symmetry.)  This R-symmetry algebra is a subalgebra of the $\mf{so}(5)_R \cong \mf{usp}(4)_R$ present in the flat space limit.  The fact that only an  $\mf{so}(3)_R \oplus \mf{so}(2)_R$ R-symmetry is preserved implies that the five scalars $\Phi_I$ split into two groups $\Phi_i$, $i = 1, 2$, and $\Phi_a$, $a = 3, 4, 5$, that may appear asymmetrically in the curvature corrections to \eqref{EucLagrangianSYM}.  Indeed, the Lagrangian of the ${\cal N} = 2$ SYM theory on $S^5$ is 
 \es{nabsym}{
  \cL_{S^5}=& \cL_\text{flat}
   + 
   {1\over g_\text{YM}^2} \tr \bigg[ 
  {4\over 2r^2}(\Phi^a)^2
  +{3\over 2r^2}(\Phi^i)^2
   -{i\over 4r} \bar \Psi^A  (\hat \C^{12})_A{}^B \Psi_B
  -{1\over 3r} \epsilon_{abc}\Phi^a[\Phi^b,\Phi^c] 
  \bigg] \,,
}
where $\epsilon_{abc}$ is a totally anti-symmetric tensor of $\mf{so}(3)_R$ with $\epsilon_{345}=\epsilon^{345}=1$.  
  The bosonic fields are all hermitian but as usual in Euclidean theories, we do not impose reality conditions on the fermions   $\Psi_A$.\footnote{
  	Consequently, all bosonic terms in the 5d Lagrangian are hermitian except for the cubic term
  	\ie
  	\cL_{\rm YM}\supset&{1\over g^2}  \tr \bigg[
  	-{1\over 3r} \epsilon_{abc}\Phi^a[\Phi^b,\Phi^c]
  	\bigg] \,.
  	\fe
  }
  The Lagrangian \eqref{nabsym} is invariant under the SUSY transformations  
  \ie
  \D A_\m=&  -\bar \ve^A  \C_\m \Psi_B 
  \\
  \D \Phi^I=& -i \bar \ve^A   (\hat \C^I)_A{}^B \Psi_B
  \\
  \D \Psi=& 
  -{i\over 2}F_{\m\n}\C^{\m\n} \ve 
  -{1\over 2}[\Phi^I,\Phi^J] \hat \C^{IJ} \ve
  + D_\m \Phi^I \C^\m \hat \C^I \ve
  +{2\over r}\Phi^a \hat \C^{a12}\ve
  +{1\over r}\Phi^i \hat \C^i \hat \C^{12}\ve
  \label{ossvarh}
  \fe
  which are parametrized by a Grassmann even spinor $\ve_A$ (with $\bar \ve$ defined as in \eqref{Conjugate}) obeying the Killing spinor equation
  \ie
  \nabla_\m \ve={1\over 2r}\C_\m \hat \C^{12}\ve \,. 
  \label{kseq}
  \fe
  Under the $\mf{su}(2)_R \oplus {\mf u}(1)_R$ R-symmetry subalgebra of $\mf{so}(5)_R$ that is preserved on $S^5$, the ${\bf 4}$ decomposes as ${\bf 2}_{1/2} + {\bf 2}_{-1/2}$.  Correspondingly, the Killing spinor $\ve$ splits as
  \ie
  \ve=\ve_++\ve_-
  \fe
  according to the eigenvalues of $\ve_\pm$ under $i\hat \C_{12}$, which generates  ${\mf u}(1)_R$ acting on $\mf{so}(5)_R$  spinors.   We identify $\ve_+$ with the supercharges $Q$, and the $\ve_-$ with the supercharges $S$, whose names are motivated by uplifting to 6d which we will describe in more detail in Section~\ref{sec:6d}.  See also Table~\ref{RSYM}.  Note that in order for the SUSY algebra generated by $\ve_A$ to be $\mf{su}(4|2)$, one should additionally impose a reality condition on $\ve_A$---a possible such reality condition is $\bar \ve^A = (\ve_A)^\dagger$.
  
  \begin{table}[htp]
  	\begin{center}
  		\begin{tabular}{c|c|c}
  			& $\mf{su}(2)_R$ irrep & $\mf{u}(1)_R$ charge \\
  			\hline 
  			$A_\mu$ & ${\bf 1}$ & $0$ \\
  			$\Phi_i$ & ${\bf 1}$ & $\pm 1$ \\
  			$\Phi_a$ & ${\bf 3}$ & $0$ \\
  			$\Psi$  & ${\bf 2}$ & $\pm 1/2$ \\
  			\hline
  			$Q, S$ &  ${\bf 2}$ & $\pm 1/2$ 
  		\end{tabular}
  	\end{center}
  	\caption{R-symmetry quantum numbers of the various fields that appear in the Lagrangian as well as of the supercharges.}\label{RSYM}
  \end{table}%

  \subsection{Localized $S^5$ partition function and Chern-Simons matrix model}
  
  In \cite{Kim:2012ava, Kim:2012qf} (see also \cite{Hosomichi:2012ek,Minahan:2015jta,Kallen:2012va}), the $S^5$ partition function was computed using supersymmetric localization using a supercharge that squares to translations along the great circles of $S^5$ that form the orbits of the Hopf fibration.  For simplicity, let us review the result for the $U(N)$ gauge theory---for more general formulas, see \cite{Kim:2012qf}.   For general ${\cal N} \geq 1$ theories on $S^5$, the partition function localizes to self-dual instanton solutions on the $\CP^2$ base of the Hopf fibration.  In the maximally supersymmetric case, it was shown in \cite{Kim:2012ava, Kim:2012qf} that the contributions from the different instanton sectors are proportional to one another, and the partition function factorizes into a product of a perturbative contribution $Z_\text{pert}$ and a contribution from instantons $Z_{\rm inst}$,
  \ie
  Z=Z_{\rm pert} Z_{\rm inst} \,,
  \fe
  where 
  \es{GotZPertInst}{
  	Z_{\text{pert}}= \frac{1}{N!} \int d^N\lambda\, e^{-{4\pi ^3 R\over g_\text{YM}^2}\sum_{i=1}^N \lambda_i^2}  \prod_{i<j}\left(2\sinh {\pi (\lambda_i-\lambda_j)}\right)^2 \,.
  }
  and
  \es{Zinst}{
  	Z_{\rm inst}={1\over\eta(q)^N} \,, \qquad
  	\eta(q) \equiv q^{{1\over 24}}\prod_{n=1}^\infty ( 1- q^n),\qquad q\equiv e^{- \frac{8\pi^3R}{ g_\text{YM}^2}} \,.
  }

  The partition function is not the only quantity one can compute with this technique.  In \cite{Kim:2012qf}, it was shown that the expectation values of Wilson loops operators in representation $R$ of $U(N)$
  \ie
  W_R={1\over \dim R} \tr_R \left[
  \cP{\rm exp}\left(
  \oint ds (i A_\m \dot x^\m+\Phi |\dot x|)
  \right)
  \right]
  \fe
  that are extended along a Hopf fiber (with $\Phi$ a particular scalar that can be taken to be $\Phi_{5}$) are given by
  \ie
  \la  W_R\ra= \frac{1}{N! Z_{\rm pert} \dim R} \int d^N\lambda \,
  e^{- \frac{4\pi ^3 R}{g_\text{YM}^2 }\sum_{i=1}^N \lambda_i^2}  \prod_{i<j}\left(2\sinh {\pi (\lambda_i-\lambda_j)}\right)^2
  \tr_R (e^{2\pi \lambda})
  \fe

  As observed in  \cite{Kim:2012ava, Kim:2012qf}, the above integrals also calculate the partition function and expectation values of $1/2$-BPS Wilson loop operators in 3d ${\cal N} = 2$ Chern-Simons theory on $S^3$ \cite{Kapustin:2009kz} at {\em imaginary} Chern-Simons level
  \es{Gotk}{
  	k = i \frac{4\pi ^2 R}{g_\text{YM}^2}\,.
  }
  As mentioned in the Introduction, we will provide an explanation of these result.

  \subsection{The 6d $(2,0)$ superconformal index}
  \label{sec:6d}
  
  As we will now review, the ${\cal N} = 2$ SYM partition function on $S^5$ is related to the partition function of the 6d $(2, 0)$ theory on $S^1 \times S^5$, with certain boundary conditions along the $S^1$ circle that preserve the $\mf{su}(4|2)$ subalgebra.  (We will explain these boundary conditions shortly.)  This $S^1 \times S^5$ partition function can also be interpreted as an $\mf{su}(4|2)$-preserving superconformal index of the 6d $(2, 0)$ theory up to an overall normalization $e^{-\B E_0}$ determined by the supersymmetric Casimir energy $E_0$.  To understand how this is achieved, let us start from the superconformal algebra of the $(2, 0)$ theory, $\mf{osp}(8^*|4)$, and discuss how $\mf{su}(4|2)$ is embedded in it.  The generators of $\mf{osp}(8^*|4)$ and their scaling dimension $\Delta$ are:  
  \es{Generators}{
  	M_\A{}^\B \qquad\qquad& \text{(rotations / boosts, $\Delta = 0$)} \,, \\
  	P_{\A\B} \qquad\qquad& \text{(translations, $\Delta = 1$)} \,, \\
  	K^{\A\B} \qquad\qquad& \text{(special conformal transformations, $\Delta = -1$)} \,, \\
  	H \qquad\qquad& \text{(dilatation, $\Delta = 0$)} \,, \\
  	R_{AB} \qquad\qquad& \text{($\mf{usp}(4)_R$ R-symmetry, $\Delta =0$)} \,, \\
  	Q_{\A A} \qquad\qquad& \text{(Poincar\'e supersymmetry, $\Delta = 1/2$)} \,, \\
  	S^{\A}{}_{A} \qquad\qquad& \text{(superconformal transformations, $\Delta = -1/2$)} \,,
  }
  where lowercase lower / upper indices $\A$, $\B = 1, \ldots, 4$ correspond to chiral / anti-chiral spinor representations of the Lorentz algebra $\mf{so}(5, 1)$ (or equivalently fundamental / anti-fundamental indices of $\mf{sl}(4)$), and uppercase indices $A$, $B = 1, \ldots, 4$ correspond to the fundamental representation of the $\mf{usp}(4)_R$ R-symmetry.  The various commutation relations obeyed by the generators \eqref{Generators} are collected in Appendix~\ref{ALGEBRAS}.

  The $\mf{su}(4|2)$ algebra can then be obtained as the subalgebra of $\mf{osp}(8^*|4)$ that commutes with, say, the generator $H- \frac{R_{\bf 13}+R_{\bf 24}}{2}$.  This condition selects all generators of $\mf{osp}(8^*|4)$ with the property that
  \es{Condition}{
  	\text{(\# of $\mf{usp}(4)$ indices equal to $1$ or $2$)} - \text{(\# of $\mf{usp}(4)$ indices equal to $3$ or $4$)}  = \Delta \,;
  }
  the generators of $\mf{su}(4|2)$ thus are:
  \es{GenSU42}{
  	\{M_\A{}^\B; \ H-(R_{\bf 13}+R_{\bf 24}); \ R_{\bf 13}-R_{\bf 24}, R_{\bf 12}, R_{\bf 34}; \ Q_{\A {\bf 1} }, Q_{\A \bf{4}}; \ S^\A{}_{\bf{3}}, S^\A{}_{\bf{2}} \}  \,.
  }
(See also Appendix~\ref{SU42DETAILS} for details.)  One can then define the $\mf{su}(4|2)$-preserving superconformal index of the $(2, 0)$ theory as 
  \es{SuperconfIndex}{
  	{\cal I}(\beta)=\tr \left[ (-1)^F  e^{-\beta( H- (R_{\bf 13}+R_{\bf 24})/2} \right] \,,
  }
  where the trace is computed in the Hilbert space of the $(2, 0)$ theory when this theory is placed on a round $S^5$ of radius $R$.  In path integral language, this index is computed by the $S^1 \times S^5$ partition function, with the radius of $S^1$ being related to $\beta$ via $\beta = 2 \pi R_6/R$, with anti-periodic boundary conditions for all fermions along the $S^1$ (implementing the $(-1)^F$ factor), and with a non-trivial holonomy along $S^1$ for a background gauge field that couples to the R-symmetry current $j_{\bf 13}^\mu - j_{\bf 24}^\mu$ (implementing the $e^{\beta(R_{\bf 13}-R_{\bf 24})/2}$ factor).  It is a supersymmetric quantity because it is invariant under the $\mf{su}(4|2)$ algebra.

  It has been conjectured that the index ${\cal I}(\beta)$ (or equivalently the $\mf{su}(4|2)$-preserving $S^1 \times S^5$ partition function) of the $(2, 0)$ theory is equal to the $S^5$ partition function of ${\cal N} = 2$ SYM with a gauge group that is identified with the ADE Lie algebra associated to the 6d parent.\footnote{It is possible to define a more refined superconformal index of the $(2, 0)$ theory by including fugacities for various symmetry generators in $\mf{su}(4|2)$ that commute with a given supercharge and its conjugate \cite{Bhattacharya:2008zy}.  The most general such index involves three fugacities, and it can be computed using the squashed $S^5$ partition function of ${\cal N} = 1$ SYM with a massive adjoint hypermultiplet \cite{Kim:2012qf}.}${}^{,}$\footnote{5d MSYM with a non-simply laced gauge group arises from the $S^1$ compactification of the 6d $(2,0)$ theory with an outer-automorphism twist \cite{Tachikawa:2011ch}. The twist is implemented by a codimension-1 topological symmetry defect longitudinal to the 5d spacetime \cite{Wang:2018gvb}.}  Such a picture is natural in the limit of a small 6d circle, where as $R_6 \to 0$, the effective description becomes ${\cal N} = 2$ SYM on $S^5$ \cite{Douglas:2010iu}. The 5d gauge theory contains instanton particles which are identified with the Kaluza-Klein modes on $S^1$. By matching their masses, one obtains the following relation between the 5d gauge coupling and the 6d circle
  \es{gYMR6Relation}{
  	\frac{4\pi^2}{g_\text{YM}^2}= \frac{1}{R_6}.
  }
  One may worry, however, that this picture may fail to be true away from the small $R_6$ limit, and so one may lose information by focusing on the 5d effective gauge theory description.  It is however believed that, as long as one is interested in supersymmetric observables, the 5d SYM description can be used reliably at all $R_6$ because the higher derivative couplings in the 5d effective theory are expected to be either $\cQ$-exact or vanish on the BPS locus. The localization technique however requires keeping track of the entire tower of KK modes by 5d Nekrasov instanton partition functions, which have the rather simple form \eqref{Zinst} due to maximal SUSY here.  This allows the instanton contributions which is a  series in $e^{-\frac{2 \pi R}{R_6}}$ to be resummed and even to be reexpanded, if we wish, in $e^{-R_6/R}$ for large $R_6$ \cite{Kim:2012qf}.



  \section{Protected sector of 5d MSYM}
  \label{sec:WLsector}

\subsection{$1/8$-BPS Wilson loops in 5d MSYM}
\label{18BPS}

As in any gauge theory,  a nontrivial set of  observables in the 5d MSYM are the Wilson loops.  A Wilson loop  in 5d MSYM is defined on a closed curve $\cK$ in $S^5$ in a certain representation $R$ of the gauge group $G$,
 \es{wl}{
 W_R(\cK)={1\over \dim R} \tr_R  \cP\exp
  \left[ i \oint   \hat A_\m  
 dx^\m  \right] \,, \qquad
  \hat A_\m \equiv A_\m + i\cS_{\m I}(x)\Phi^I
 }
 where $\cS_{\m I}$ parametrizes the position-dependent coupling of the Wilson loop to the 5 scalar fields in the $\cN=2$ vector multiplet.
 
 For special choices of the scalar coupling matrix $\cS$, the loop operator $ W_R(\cK)$ preserves a subset of the supercharges in $\mf{su}(4|2)$. A necessary condition for this to happen is that \eqref{wl} is invariant under $\D^2_\ve$ for some Killing spinor $\ve$. In general, $\D^2_\ve$ takes the form
 \es{deltaepsSQ}{
  \delta_{\ve}^2 = -i v^\mu \partial_\mu - \frac{i}{2} w^{IJ} R_{IJ} \,, \qquad
   v^\mu \equiv \bar \ve \gamma^\mu \ve \,, \qquad
    w^{IJ} \equiv \bar \ve \gamma^{IJ} \hat \gamma^{12} \ve \,,
 }
where $v^\mu$ is a Killing vector and $R_{IJ}$ is an R-symmetry generator.    Assuming $v^\mu$ and $w^{IJ}$ are not identically zero,\footnote{In principle we can consider a nilpotent supercharge in the complexified superalgebra (which does not satisfy the reality condition $\bar \ve^A  = (\ve_A)^\dagger$ of  $\mf{su}(4|2)$).
In this case, the Wilson loop can be anywhere on $S^5$ while preserving this supercharge.  We do not study this case because it cannot give a (tractable) non-trivial topological theory simply due to the fact that loops on $S^5$ cannot be linked.

More generally, instead of the background on $S^5$ preserving $\mf{su}(4|2)$, we can consider the twisted background (valid for any Riemannian five manifold)  for the 5d MSYM by identifying the $\mf{so}(5)$ spacetime symmetry with the $\mf{so}(5)_R$ rotation such that  a unique scalar nilpotent supercharge from $\bf{4}\times \bf{4}$ of $\mf{so}(5)$ is preserved. In this case, we have ${1\over 16}$-BPS supersymmetric Wilson loops that couple to all 5 scalars in the form \eqref{wl}. However the expectation value of such observables in the twisted theory is expected to be unity following the same analysis as in \cite{Zarembo:2002an} for 4d $\cN=4$ SYM.
}
the Wilson loop $W_R(\cK)$ is invariant under $\D^2_\ve$ only if the loop $\cK$ is preserved by the Killing vector $v^\mu$.  
In common scenarios, $\cK$ is in fact a loop generated by the Killing vector $v^\mu$.  A different and more interesting case, which is only possible for theories with a sufficiently large number of supersymmetries, is when $v^\mu$ has a nonzero dimensional fixed-point set $\cM$. If this is the case, it is possible for the loop $\cK$ to lie anywhere within $\cM$ and for $W_R(\cK)$ (with appropriate matter couplings) to preserve the SUSY generated by $\ve$.  This situation is well-studied in the context of Wilson loops in $\cN=4$ SYM where there are ${1\over 16}$-BPS Wilson loops of arbitrary shapes on $S^3\in \mR^4$ \cite{Drukker:2007qr}.  Here we will construct similar loop operators in 5d MSYM on $S^5$. 
 
 The 5d MSYM on $S^5$ is invariant under an $\mf{so}(6)$ isometry, thus the commutant of an $\mf{so}(2)$ subgroup generated by $v^\mu$ can be at most $\mf{so}(2)\times \mf{so}(4)$.  It is then easy to see that the largest submanifold that can be fixed by  $v^\mu$ is a great $S^3$.   For concreteness, let us consider the $S^5$ being parameterized by the embedding coordinates $X_i$ constrained by $\sum_{i=1}^6 X_i^2=R^2$, where $R$ is the radius of $S^5$, and let us take the great $S^3$ fixed by $v^\mu$ to be located at $X_1 = X_2 = 0$.  Up to normalization, the Killing vector $v^\mu$ is then given by
  \es{vmuExplicit}{
   v^\mu  = u_{12}^\mu \,,
  }
where $u_{ij}$ are the Killing vectors corresponding to the $\mf{so}(6)$ symmetry of $S^5$:
 \es{uDef}{
 u_{ij}\equiv X_i {\partial \over \partial X_j }-X_j {\partial \over \partial X_i } \,.
 }
 If we consider the stereographic coordinates $x_{1,2,3,4,5}$ defined by
  \es{stereoS5}{
  X_{1\leq i\leq 5}={ x_{i}\over 1+{x^2\over 4R^2}} \,, \qquad X_{6}=R{ 1-{x^2\over 4R^2}\over 1+{x^2\over 4R^2}} \,,
  }
then the great $S^3$ is parameterized by stereographic coordinates $x_{\rm i} \equiv \{x_3,x_4,x_5\}$. (Note the difference between the fonts used for the indices $i$ and ${\rm i}$.)

More explicitly, we will consider a Killing spinor $\ve$ (to be found shortly) such that 
  \es{dsq}{
   \D_\ve^2  \propto i u_{12} + R_{12} \,,
  }
 where $R_{12}$ generates $\mf{so}(2)_R$ rotation so that $R_{12}(Q)=1$ and $R_{12}(S)=-1$.  Up to a rotation by the $\mf{so}(4)$ isometry of the $S^3$, $\mf{so}(2)$ transverse rotation, as well as $\mf{so}(3)_R\times \mf{so}(2)_R$ transformation, we can fix $\ve$ to correspond to  $\ve(\cQ)$ with  
 \ie
 \cQ={1\over 2}(Q_{1\bf 4}+S^{1\bf 4}-Q_{2\bf 1}-S^{2\bf 1}) \,.
 \label{lsc}
 \fe
 Here the boldfaced indices are raised and lowered by the symplectic form $\Omega^{AB}$ and $\Omega_{AB}$.  
 See Appendix~\ref{SU42DETAILS} for details about the notation.

 If we label the supercharges as $Q^{s_4 s_5}_{s_1 s_2 s_3}$ and $S^{s_4 s_5}_{s_1 s_2 s_3}$ by their spins (eigenvalues)  with respect to the $\mf{so}(6)$ rotation generators $M_{12},M_{34},M_{56}$,  as well as the spins $s_4,s_5$ with respect to the Cartans $R_{12}, R_{34}$ of $\mf{so}(2)_R\times \mf{so}(3)_R$, we can write\footnote{The indices of the $\mf{so}(6)$ chiral spinor in terms of $s_i$ are
 \ie
 (1,2,3,4)=(++-,+-+,---,-++)
 \fe
 The similar relation for the anti-chiral spinor simply comes from flipping each $\pm$. The lower boldfaced indices for $\mf{so}(2)_R\times \mf{so}(3)_R\subset \mf{usp}(4)_R$ in terms $s_4,s_5$ are
 \ie
  (\bf{1},\bf{4})=(++, +-)
 \fe
  See \eqref{indextrans} for details.
}
 \ie
 \cQ={1\over 2}(Q^{+-}_{++-}+S^{-+}_{--+}-Q^{++}_{+-+}-S^{--}_{-+-}) \,.
 \fe

 We are interested in Wilson loops \eqref{wl} on an arbitrary curve $\cK \subset S^3$.\footnote{There are also supersymmetric Wilson loops along the Killing vector field $v^\m$. See Appendix~\ref{app:tauwl}. Here we will focus on the Wilson loops on $S^3$ transverse to $v^\mu$.} The invariance under $\D_\ve$ requires
 \es{SUSYLoop}{
  \dot x^{\rm i} (\D_\ve  A_{\rm i} + i\cS_{{\rm i} J} \D_\ve \Phi^J )\bigr|_\cK=0 \,.
 }
 Using \eqref{ossvarh}, this imposes the condition $\left[ -\bar \ve^B  \C_{\rm i} +\cS_{{\rm i} J} \bar \ve^A   (\hat \C^J)_A{}^B  \right] \Psi_B=0$ for arbitrary $\Psi$.  Taking $\Psi = \hat \C^I \ve$, we can then solve for $\cS$:
  \es{GotcS}{
  \cS_{{\rm i} I}={ \bar \ve   \C_{\rm i}  \hat \C_I \ve \over \bar \ve \ve} \,.
  }
 Explicitly using our supercharge \eqref{lsc}, Eq.~\eqref{GotcS} reads
 \ie
 \cS_{{\rm i} a}=-e^{2\Omega}\left( \left (1-{x^2\over 4R^2}\right) \D_{{\rm i} a}
 +{ x_{\rm i} x_a \over  2R^2}+{\epsilon_{{\rm i} a\C} x_\C  \over R}\right) \,,
 \qquad
 \cS_{{\rm i}1}=  \cS_{{\rm i} 2}=0 \,.
 \label{smat}
 \fe
 The matrix $\cS$ satisfies the relations
 \es{cSRelations}{
   \cS_{{\rm i} a}\cS_{{\rm j} a}=e^{2\Omega}  \D_{{\rm i} {\rm j}} \,,
    \qquad 
   \cS_{{\rm i} a}\cS_{{\rm i} b}=e^{2\Omega}  \D_{ab} \,, 
   \qquad 
    \det \cS= -e^{3\Omega} \,.
  }
Alternatively, we can reexpress $\cS$ in terms of the embedding coordinates as\footnote{We note that exactly the same scalar coupling matrix here defines ${1\over 16}$-BPS Wilson loops in 4d $\cN=4$ SYM \cite{Drukker:2007qr}. Despite the kinematic similarity between the 4d and 5d Wilson loops, as we will see in the later section, the underlying theories and consequently the Wilson loop correlators are very different.  
}
 \es{Smthooft}{
 \cS_{{\rm i} a}={1\over R} \eta_{a}^{ij} X_i \partial_{\rm i} X_j \,,
 }
 where $\eta$ is the anti-self-dual 't Hooft $\eta$ symbol
 \es{etaDef}{
  \eta_3=\begin{pmatrix}
  	0 & 0 & 0 & 1
  	\\
  	0 & 0 & -1 & 0
  	\\
  	0 & 1 & 0 & 0
  	\\
  	-1 & 0 & 0 & 0
  \end{pmatrix} \,, \quad
  \eta_4=\begin{pmatrix}
 	0 & 0 & 1 & 0
 	\\
 	0 & 0 & 0 & 1
 	\\
 	-1 & 0 & 0 & 0
 	\\
 	0 & -1 & 0 & 0
 \end{pmatrix}
 \,, \quad
 \eta_5=\begin{pmatrix}
 	0 & -1 & 0 & 0
 	\\
 	1 & 0 & 0 & 0 
 	\\
 	0 & 0 & 0 & 1
 	\\
 	0 & 0 & -1 & 0
 \end{pmatrix} \,,
 }
 obeying the relations
  \es{thooftid}{
 \eta_{a}^{ij}=-{1\over 2}\epsilon^{ijkl} \eta_a^{kl} \,, \qquad
   \eta_a^{ij}\eta_a^{kl}=\D^{ik}\D^{jl}-\D^{il}\D^{jk}-\epsilon^{ijkl} \,.
  }
  
\begin{figure}[!h]
	\centering
	\includegraphics[scale=0.5]{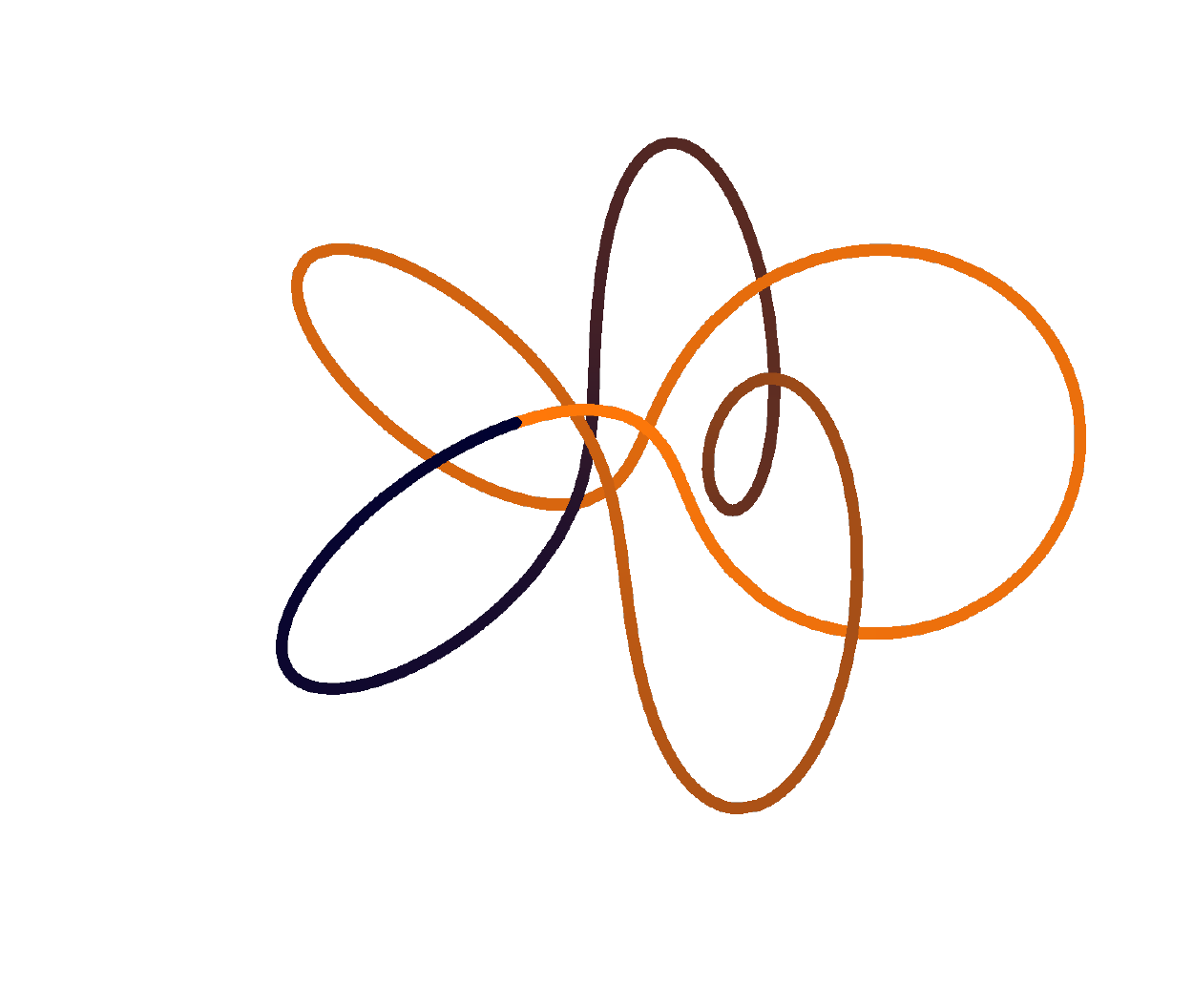}\hspace{2cm}\includegraphics[scale=0.5]{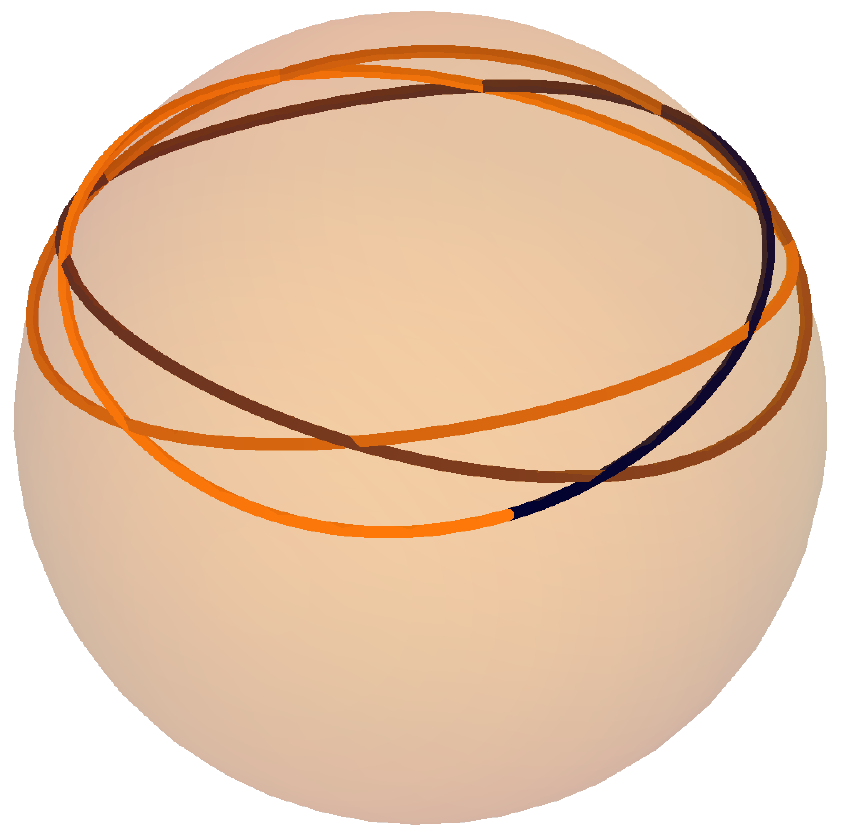}
	\caption{On the left, an example Wilson loop $\cK$ that lies in a great $S^3\subset S^5$, plotted after performing a stereographic projection from $S^3$ to $\R^3$. On the right, we plot the shape of the loop in the internal $S^2\subset S^4$. We interpolate in coloring from light to dark as we go around the  loop in both figures.  The colors are coordinated between the two plots. \label{fig:ExampleKnot}}
	
\end{figure}

Geometrically $\cS$ gives rise to a map from a knot $\cK$ in $S^3$ to a curve on an auxilary $S^2$ parametrized by $\sum_{a=3}^5 \Theta_a^2=1$,
\ie
\cK:x_{\rm i}(t)\to \Theta_a (t)= \dot x_{\rm i} \cS_{ {\rm i} a} (t) \,.
\fe
The image curve on $S^2$ generally have self intersections (see Figure~\ref{fig:ExampleKnot}). As we will see in Section~\ref{HoloArea}, this auxiliary $S^2$ and the image curve $\Theta_a (t)$ will play an important role in the holographic computation of these Wilson loop observables.

From now on, when we refer to the Wilson loop observables \eqref{wl}, we will always take the loop $\cK$ to be contained in the $S^3$ at $x_1 = x_2 = 0$, and we will take the matrix $\cS$ to be given by \eqref{cSRelations} or \eqref{Smthooft}.  The set of all such Wilson loops form a protected sector of the 5d MSYM theory.  Let us understand the symmetries of our protected subsector.  The $S^3$ parameterized by $x_{\rm i}$ has isometry $\mf{so}(3)_l\times \mf{so}(3)_r$, corresponding to the following generators of $\mf{so}(6)$:
 \es{SO3s}{
&\mf{so}(3)_l:\quad M_{34}+M_{56},~M_{35}-M_{46},~M_{45}+M_{36}\,,\\ 
&\mf{so}(3)_r:\quad M_{34}-M_{56},~M_{35}+M_{46},~M_{45}-M_{36}  \,.
  }
 The form of the scalar coupling matrix $\cS$ in \eqref{Smthooft} indicates that $\cS$ is invariant under $\mf{so}(3)_l$. While $\cS$ is not invariant under $\mf{so}(3)_r$, it is invariant under the diagonal subgroup  $\mf{so}(3)_{\rm diag} \subset \mf{so}(3)_r\times \mf{so}(3)_R$  
\ie
 &\mf{so}(3)_{\rm diag}:\quad  M_{34}-M_{56}+2R_{34},~M_{35}+M_{56}+2R_{35},~M_{45}-M_{36}+2R_{45} \,.
 \label{tSO3r}
 \fe
Thus the bosonic symmetry of our Wilson loops on $S^3$ contains $\mf{so}(3)_{\rm diag} \times \mf{so}(3)_l$, which can be thought of as the twisted isometry on $S^3$.  It can be checked that the $\mf{so}(3)_{\rm diag}$ generators can be obtained by anticommuting $\cQ$ with other supercharges in $\mf{su}(4|2)$, so they are $\cQ$-exact.

 Moreover $\mf{so}(2)_R$ acts trivially on $\cS$ but rotates $\cQ$ to 
 \ie
 \tilde\cQ={i\over 2}(Q_{1\bf 4}-S^{1\bf 4}-Q_{2\bf 1}+S^{2\bf 1})
 =
 {i\over 2}(Q^{+-}_{++-}-S^{-+}_{--+}-Q^{++}_{+-+}+S^{--}_{-+-}) \,,
 \fe
 which is also preserved by our Wilson loops. Consequently the Wilson loops actually preserve $2$ out of the $16$ supercharges of $\mf{su}(4|2)$, and thus are ${1\over 8}$-BPS. It is easy to check that
 \ie
 \{ \cQ,\tilde\cQ\}=[\mf{so}(3)_{\rm diag},\cQ]=[\mf{so}(3)_{\rm diag},\tilde\cQ] =[\mf{so}(3)_l,\cQ]=[\mf{so}(3)_l,\tilde\cQ]=0 \,,
 \label{Qsym}
 \fe
 and therefore, the subgroup of $\mf{su}(4|2)$ preserved by our subsector is 
 \ie
  \left[ \mf{su}(1|1)\rtimes \mf{so}(2)_R \right] \times \mf{so}(3)_{\rm diag} \times \mf{so}(3)_l  \,,
 \label{18sym}
 \fe
 where the $\mf{su}(1|1)$ factor is generated by $\cQ,\tilde\cQ$. The $\mf{so}(2)_R$ is as an automorphism of $\mf{su}(1|1)$ and it only acts on the odd generators.

The fact that the generators of $\mf{so}(3)_{\rm diag}$ are $\cQ$-exact means that the $1/8$-BPS Wilson loop observables \eqref{wl} (when restricted to $S^3$ and when the scalar couplings are given by $\cS$) change only by $\cQ$-exact terms if we perform an $\mf{so}(3)_{\rm diag}$ rotation.  Consequently, because these $1/8$-BPS Wilson loop operators are also $\cQ$-invariant, it follows that correlation functions of the form
 \es{CorrFuncGen}{
  \langle W_{R_1}(\cK_1) \cdots W_{R_n}(\cK_n) \rangle
 }
remain unchanged if we act with an $\mf{so}(3)_{\rm diag}$ on any one of the $W_{R_i}(\cK_i)$ operators. This statement should be understood only up to contact terms; the correlation function can change if, as we perform such an $\mf{so}(3)_{\rm diag}$ rotation on a given $\cK_i$, this $\cK_i$ crosses another loop $\cK_j$ with $i \neq j$.  In the next sections, we will provide evidence for a {\it stronger} result, namely that the $1/8$-BPS Wilson loop subsector of 5d MSYM is in fact described by a 3d Chern-Simons theory at a complexified value of the Chern-Simons coupling. Such a result would not only imply the invariance of \eqref{CorrFuncGen} under the action of $\mf{so}(3)_{\rm diag}$ on any of the $W_{R_i}(\cK_i)$ operators, but also much stronger conditions.

  \subsection{The flat space limit and relation to topologically twisting}
  \label{FLAT}
 
  The form of our Wilson  loops \eqref{wl} and the symmetries of the protected subsector \eqref{18sym} are reminiscent of that the topologically twisted theory of \cite{Witten:2011zz}. This is most clear if we take the flat space limit of our setup by sending $R\to \infty$. Then the scalar coupling matrix $\cS$ is simply
  \ie
  \cS_{{\rm i} a}=-  \D_{{\rm i} a}.
  \fe
  Our Wilson loops are now defined by the twisted connection $\hat A_{\rm i}= A_{\rm i}-i \Phi_{\rm i} $   and they lie on the $\mR^3$ parametrized by $x_{\rm i}=(x_3,x_4,x_5)$. 
  
  These are precisely the same observables in the topologically twisted version of 5d MSYM on $\mR^+\times M_4$. Recall that in \cite{Witten:2011zz}, the $\mf{so}(5)_R$ symmetry of the 5d MSYM is broken to $\mf{so}(3)_R\times \mf{so}(2)_R$ by the boundary condition at the end of $\mR^+$. The structure group on the four-manifold $M_4$ is $\mf{so}(4)=\mf{su}(2)_l\times \mf{su}(2)_r$.  The topological twist is implemented by identifying the $\mf{so}(3)_R$ factor with the $\mf{su}(2)_r$ factor. Equivalently, this is achieved by turning on an $\mf{su}(2)_R$ background that coincides with the $\mf{su}(2)_r$ spin connection.  
  
  We emphasize that despite the similarity between our observables and the twisted versions in \cite{Witten:2011zz}, the underlying theories that govern their dynamics, are different. In particular, we have 5d MSYM on $S^5$ with the usual non-topologically twisted background which preserves the maximal amount of supersymmetries.

  \subsection{Wilson loops that preserve more supersymmetries}
  \label{MORESUSY}

As explained in Section~\ref{18BPS}, for each great $S^3$ in $S^5$, we can construct a protected subsector of mutually ${1\over 8}$-BPS Wilson loops.
If we further restrict the curve $\cK$ to lie in particular submanifolds of $S^3$, more supersymmetries can be preserved. 

For instance, if $\cK$ lies on a great $S^2$ inside $S^3$, the corresponding Wilson loop is ${1\over 4}$-BPS. Of course in a given $S^3$ there's a continuous family of great $S^2$'s. For  such Wilson loops to be mutually ${1\over 4}$-BPS, they have to lie on the same $S^2$.  To be concrete, without loss of generality let us take this $S^2$ to be given by $x_1=x_2=x_3=0$ in the $S^5$ stereographic coordinates.  The Wilson loops are defined with twisted connection
\ie
\hat A= A+i \eta^a_{ij} \Phi^{a} X_i dX_j
\fe
restricted to the $S^2$. They
are invariant under the  transverse $\mf{so}(3)$ rotation generated by $M_{ij}$ for $i,j=1,2,3$. They
preserve four supercharges, $Q^{1/4}_\A$ and its Majorana conjugate $\overline Q^{1/4}_\B$ for $\A,\B=1,2$ (which are $\mf{so}(3)$ doublet indices) and
 \es{14Qs}{
  \cQ^{1/4}_1=&{1\over 2}(Q_{1\bf 4}-Q_{2\bf 1})={1\over 2}(Q^{+-}_{++-}-Q^{++}_{+-+}) \,, 
  \\
  \cQ^{1/4}_2=&{1\over 2}(Q_{3\bf 4}-Q_{4\bf 1})={1\over 2}(Q^{+-}_{---}-Q^{++}_{-++}) \,, 
  \\
   \overline Q^{1/4}_1=&{1\over 2}(S^{1\bf 4}-S^{2\bf 1})={1\over 2}(S^{-+}_{-++}-S^{--}_{-+-}) \,, 
  \\
  \overline Q^{1/4}_2=&{1\over 2}(S^{3\bf 4}-S^{4\bf 1})={1\over 2}(S^{-+}_{+++}-S^{--}_{+--}) \,.
  }
Their anticommutators give the $\mf{so}(3)$  as well as an $\mf{so}(2)$ rotation generated by $ M_{56}-R_{34}$. Together these bosonic generators and the supercharges \eqref{14Qs} furnish an $\mf{su}(2|1)$ subalgebra of $\mf{su}(4|2)$. Furthermore, the $\mf{su}(2|1)$ is invariant under 
the twisted isometry on $S^2$ generated by
 \ie
 &\mf{so}(3)'_{\rm diag}:~   M_{45}+R_{45},M_{46}+M_{35},M_{56}-M_{34} \,.
 \fe
Thus the total symmetry of this ${1\over 4}$-BPS sector of loop operators on $S^2$ is
 \ie
 \mf{su}(2|1) \times \mf{so}(3)'_{\rm diag}  
 \label{14alg}
 \fe
 
If $\cK$ is further constrained to be a great circle on $S^3\subset S^5$, we recover the familiar well-studied ${1\over 2}$-BPS Wilson loop. In this case, there are no two non-overlapping Wilson loops that are mutually BPS. We can take this $S^1$ to be given by $x_1=x_2=x_3=x_4=0$ in the stereographic coordinates. The Wilson loop is defined with twisted connection 
\ie
\hat A= A+i {\Phi^5  (X_5dX_6-X_6dX_5) }= A-i { dx_5 \Phi^5\over 1+{x_5^2\over 4R^2}} \,,
\fe
and it is invariant under the transverse $\mf{so}(4)=\mf{su}(2)_l\times \mf{su}(2)_r$ rotation.  This Wilson loop  preserves eight supercharges, namely $Q^{1/2}_\A$ and $Q^{1/2}_{\dot \A}$, and their Majorana conjugates $\overline Q^{1/2}_\B$ and $\overline Q^{1/2}_{\dot \B}$:
\ie
 & \cQ^{1/2}_1\equiv Q_{1\bf 4} =Q^{+-}_{++-}
 ,~
    \cQ^{1/2}_2 \equiv Q_{3\bf 4} =Q^{+-}_{---}
 ,~
 \cQ^{1/2}_{\dot 1}\equiv Q_{2\bf 1}   =Q^{++}_{+-+}
 ,~
  \cQ^{1/2}_{\dot 2} 
  \equiv Q_{4\bf 1}  =Q^{++}_{-++}
  \\
 &  \overline  \cQ^{1/2}_1\equiv S^{1\bf 4} =S^{-+}_{--+}
 ,~
\overline     \cQ^{1/2}_2 \equiv  S^{3\bf 4} =S^{-+}_{+++}
 ,~
\overline  \cQ^{1/2}_{\dot 1}\equiv  S^{2\bf 1}   =S^{--}_{-+-}
 ,~
\overline   \cQ^{1/2}_{\dot 2} 
  \equiv  S^{4\bf 1}  =S^{--}_{+--} \,.
      \fe
Here $\A,\B=1,2$ and $\dot \A,\dot \B=1,2$ are the doublet indices for $\mf{su}(2)_l$  and $\mf{su}(2)_r$ respectively.   The $\mf{su}(2)_l$ in combination with $Q^{1/2}_\A,\overline Q^{1/2}_\B$ generates an $\mf{su}(2|1)_l$ algebra, with the central $\mf{u}(1)$ being generated by $M_{56}+R_{12}+2R_{34}$.
The $\mf{su}(2)_r$ with $Q^{1/2}_{\dot\A},\overline Q^{1/2}_{\dot\B}$ generates an $\mf{su}(2|1)_r$ algebra, with the central $\mf{u}(1)$ is generated by $-M_{56}+R_{12}-2R_{34}$.  The total symmetry of the $1\over 2$-BPS sector is 
\ie
\mf{su}(2|1)_l\times \mf{su}(2|1)_r\times \mf{so}(2) 
 \label{12alg}
\fe   
 where  the $\mf{so}(2)$ is generated by $M_{56}-R_{34}$.

  \subsection{Relation to surface operators in 6d}
  \label{SURFACE} 
  
  Given the embedding of the 5d SUSY algebra on $S^5$ into that of the 6d $(2,0)$ theory compactified on $S^1$, the ${1\over 8}$-BPS Wilson loop operators in 5d MSYM are expected to lift to ${1\over 8}$-BPS surface operators in the 6d theory wrapping the $S^1$. When the 6d theory is free, namely, a single $(2,0)$ tensor multiplet, such a surface operator on $\Sigma=S^1\times \cK$, can be described explicitly as
  \ie
  W_R(\Sigma )=     \exp \bigg[
  i\int_{\Sigma}  \left( B_{\tau\m} + i\cS_{\m I}(x)\Phi^I 
  \right)d\tau  \wedge dx^\m 
  \bigg] \,, 
  \label{ws}
  \fe
  where $B_{\m\n}$ is the self-dual tensor and $\Phi^I$ denotes the 5 scalar fields. The relation to the 5d description is obvious: $B_{\tau \m}$ reduces to $A_\m$ and $\Phi^I$ becomes the 5d scalars. For the interacting case, we do not have such an explicit description of the surface operators in 6d but M-theory and dualities give us crucial guidance.
  
  Recall that the 6d $(2,0)$ (A-type) theory describes the low energy dynamics of a stack of M5 branes. Surface defects in the $(2,0)$ theory can be engineered by M2-brane ending along a codimension-4 locus on the M5 branes. The M2-brane sources the self-dual 3-form field strength and can preserve a fraction of the $(2,0)$ supersymmetry on the M5 branes. If we reduce the M2-M5 configuration along a common direction by circle compactification, we end up with a fundamental string  ending on D4-branes in type IIA string theory. We are familiar with the fact that the endpoint of a fundamental string inserts a Wilson line operator in the fundamental representation in the SYM theory governing the low energy dynamics of the D4 branes. At high energy (strong coupling), the D4 brane worldvolume theory gets completed by that of the M5 brane. We expect observables protected by supersymmetry, such as the BPS Wilson loops, to lift to unique observables in 6d, up to potential counter-terms that are suppressed in the small $S^1$ limit. It is an interesting question to systematically classify such supersymmetric counter-terms but we will not pursue it in this paper (see Appendix~\ref{app:loccount} for a discussion in this direction).

  \section{Knot invariants from 5d perturbation theory}
  \label{sec:pert}
 
  In this section, we use perturbation theory to study the ${1\over 8}$-BPS Wilson loop observables defined in the previous section.  For simplicity, we will focus on the abelian case and comment on the non-abelian extension towards the end of the section.  
  
  \subsection{Abelian theory}
  
  To compute the abelian Wilson loop \eqref{wl} of a given $U(1)$ charge $q$ perturbatively, 
  \ie
  \la W_q(\cK) \ra=\left\la 1+i  q\oint_\cK \hat A-{1\over 2} q^2\oint_\cK \hat A \oint_\cK \hat A +\dots \right\ra
  \label{abpert}
  \fe
  we need to determine the propagators for the gauge fields $A_\m$ and scalars $\Phi^a$ on $S^5$.
  For simplicity, we will perform a change of coordinates sending $x_\m \to 2R x_\m$ and set $R=1$.
  
 To simplify the formulas for the propagators, it is useful to introduce the  chordal distance between two points on $S^5$ with stereographic coordinates $x$ and $y$ and embedding coordinates $X$ and $Y$, respectively, to be
  \es{ChordalDist}{
  s(X, Y)  \equiv \abs{X - Y}  ={2|x-y|\over \sqrt{1+x^2}\sqrt{1+y^2}}\,.
}
The chordal distance is related to the geodesic distance $\theta(X, Y) = \arccos (X \cdot Y)$ by $s(X, Y)=2  \sin \frac{\theta(X, Y)}{2}$.  Recall that the scalars $\Phi_a$ have mass $m^2=4$ on $S^5$.\footnote{The conformal mass on $S^5$ is $m^2={15\over 4}$.  The scalars $\Phi_i$ and $\Phi_a$ are thus not conformally coupled.} The two point function is then given by \cite{Drummond:1977uy}
  \es{PropScal}{
  \la\Phi_a(x)\Phi_b(y) \ra ={g_{\rm YM}^2\over 24\pi ^3} f_1(s) \D_{ab} \,, \qquad
   f_1(s)\equiv  {}_2 F_1(2,2,5/2,1-{s^2/4}) \,,
  }
 where $s = s(X(x), Y(y))$.    The gauge field $A_\m$ on the other hand has two point function \cite{Drummond:1977uy}
  \es{PropVector}{
  \la A_\m(x)A_\n (y) \ra= {g_{\rm YM}^2\over 12\pi^3} \frac{dX}{dx^\mu}  \cdot \frac{dY}{dy^\nu}  f_2(s) \,, \qquad
    f_2(s) \equiv {}_2F_1 (1,3,5/2,1-{s^2/ 4}) \,.
  }
  
  The leading order contribution to \eqref{abpert} comes from the two point function
  \ie
  {\cal I}\equiv &\oint dt_1 \oint dt_2\,   \left\la \dot x^{\rm i} (A_{\rm i} +i S_{{\rm i} a} \Phi^a)(x)
  \dot y^{\rm j} (A_{\rm j} +i S_{{\rm j} b} \Phi^b)(y) 
  \right\ra \,,
  \label{int}
  \fe
which after using $S_{{\rm i} a}(x) = \eta_a^{ij} X_i \frac{dX_j}{dx^{\rm i}}$ and $S_{{\rm j} b}(y) = \eta_b^{kl} Y_k \frac{dY_l}{dx^{\rm j}}$, the Eqs.~\eqref{PropScal}--\eqref{PropVector} for the propagators, as well as the identities \eqref{thooftid} obeyed by the 't Hooft symbols, can be written as
  \es{int2}{
  	{\cal I}&\equiv \oint dt_1 \oint dt_2 ( I_1+I_2)\\
  	I_1&\equiv    
  	{1\over 12\pi^3} \bigg[
  	f_2(s) ( \dot X\cdot  \dot Y) 
  	- 
  	{1\over 2}f_1(s)
  	\left( (  X\cdot   Y) ( \dot X\cdot  \dot Y)
  	-
  	(  X\cdot   \dot Y) ( \dot X\cdot   Y) 
  	\right)
  	\bigg]\\
  	I_2&\equiv{1\over 24\pi^3} f_1(s)
  	\epsilon^{ijkl}
  	X_i \dot X_j Y_k \dot Y_l   \,,
  }
  where we view $X$ and $Y$ as functions of $t_1$ and $t_2$, respectively.  Further using
  \es{MoreIdentities}{
  	 X\cdot   Y &=1-{s^2(X,Y)\ov 2} \,, \qquad
  	 X\cdot   X = Y\cdot   Y =1 \,, \qquad
  	  X\cdot   \dot X = Y\cdot   \dot Y =0\,,
  }
  we can simplify $I_1$ to 
  \es{I1Res}{
  	I_1={1\ov 4\pi^3}\p_{t_1} \p_{t_2} \arccos^2\le(s(X,Y)\ov 2\ri)\,.
  }
  The integral over $t_2$ picks up the discontinuity  ${\rm Disc}_{t_1=t_2}\le[\p_{t_1} \arccos^2\le(s\ov 2\ri)\ri]=-\pi \abs{\dot X}$ and gives
  \es{I1Int}{
  	{\cal I}_1\equiv\int dt_1 dt_2 I_1 =-{1\ov 4\pi^2}\int dt_1  \ \abs{\dot X}=-{1\ov 2\pi}L(\cK)\,,
  }
  where $2\pi L(\cK)$ is the length of the loop $\cK$. In the case when $\cL=\cup_i \cK_i$ is a union multiple loops, $L(\cL)=\sum_i L(\cK_i)$.
  
  We now turn our attention to $I_2$ and show that its integral gives the linking number on $S^3$. Let us first recall the familiar Gauss integral formula for the linking number of two knots $\cK_{1,2}$ in flat $\R^3$ :
  \es{LinkingNum}{
  	{\rm lk}(\cK_1,\cK_2)&={1\over 4\pi}\oint_{\cK_1} dx^{\rm i}  \oint_{\cK_2} {} dy^{\rm j} \epsilon_{{\rm i}{\rm j}{\rm k}} {(x-y)^{\rm k}\over |x-y|^3}
  	={1\over 4\pi}\int dt_1 dt_2 \ \epsilon_{{\rm i}{\rm j}{\rm k}}\,  \dot x^{\rm i} \dot y^{\rm j} \, \nabla_{(y)}^{\rm k} {1\over |x-y|}\,.
  }
  On $S^3$ the generalization of this formula is \cite{deturck2004gauss}:
  \es{LinkingNumS3}{
  	{\rm lk}(\cK_1,\cK_2)&={1\over 4\pi}\int dt_1 dt_2 \ \epsilon_{{\rm i}{\rm j}{\rm k}}  \le(P^{\rm i}_{\,\,\,\,{\rm l}} \,\dot x^{\rm l}\ri) \dot y^{\rm j} \, \nabla_{(y)}^{\rm k} \Phi(\theta) \,, 
  }
  where $P^{\rm i}_{\,\,\,\,\mu} $ is an operator that performs the parallel transport between the tangent spaces at $x$ and $y$,\footnote{An explicit formula is \cite{Osborn:1999az}
    	\es{PProg}{
  		P_{{\rm i}{\rm j}}&=-\frac14\le(2a(\theta)\,\p_{\rm i}\p_{\rm j} \theta^2+b(\theta)\p_{\rm i} \theta^2\p_{\rm j}\theta^2\ri)\,,\\
  		a(\theta)&={\sin\theta\ov \theta}\,, \quad b(\theta)={1-a(\theta)\ov \theta^2}\,.
  	}
  }
  $\theta$ is the geodesic distance (given below \eqref{ChordalDist}), and 
  \es{phiform}{
  	\Phi(x,y)&\equiv{\pi-\theta\ov \sin\theta}\,.
  }
  It is easy to see that the flat space limit of \eqref{LinkingNumS3} is \eqref{LinkingNum}. Both \eqref{LinkingNumS3} and \eqref{LinkingNum} define a topological invariant for a pair of knots. 
  
  In the case of a single knot $\cK_1=\cK_2=\cK$, the same integral  gives 
  \ie
  {\rm sl}_0(\cK)=
  {1\over 4\pi}\oint_\cK dx^{\rm i} \oint_\cK dy^{\rm j} \ \epsilon_{{\rm l}{\rm j}{\rm k}} P^{\rm l}_{\,\,\,\,{\rm i}}  \nabla_{(y)}^{\rm k} \Phi(\theta)
  \fe
  which is   well-defined and finite.\footnote{Despite the apparent pole in the $\Phi$ as $|x-y|\to 0$, the integrand is always finite \cite{Guadagnini:1989am}.} This is sometimes referred to as the writhe or cotorsion of the knot $\cK$ in the literature. Although ${\rm sl}_0(\cK)$ is not topological, there is a natural counter-term, known as the torsion
    \es{TorsionFT}{
  T(\cK)={1\ov 2\pi}\int dt \abs{\dot x}\,\tau \equiv  {1\ov 2\pi}\int dx^{\rm i} \,\epsilon_{{\rm i}{\rm j}{\rm k}} { n^{\rm j} {D\ov Dt} n^{\rm k} \over |\dot x| }
}
  defined using a normal vector field $n^{\rm i}$ of unit norm along $\cK$, where $D\over Dt$ is the covariant derivative along the tangent direction of the knot $\cK_i$.\footnote{The torsion term can be thought of as a 1d CS term along the knot that measures the holonomy (total phase) of the $\mf{so}(2)$ spin connection on the normal bundle of $\cK$ in a Riemannian 3-manifold. We discuss some elements of curve geometry in Appendix~\ref{app:curvegeom}.} It measures $\frac{1}{2 \pi}$ times the phase swept by $n^{\rm i}$ as one goes around the knot once.
  The variation of the torsion under small deformations of $\cK$, cancels that of ${\rm sl}_0(\cK)$, so that
  \ie
  \D( {\rm sl}_0(\cK) +T(\cK) )=0
  \fe
  Indeed if we define  a new knot $\cK_f$, known as the framed contour of $\cK$ by an infinitesimal translation in the $n^{\rm i}$ direction, the ordinary linking number of $\cK$ and $\cK_f$ is nothing but the above combination
  \ie
  {\rm lk}(\cK,\cK_f)={\rm sl}_0(\cK) +T(\cK)\,.
  \fe
  Thus we can define a  topological invariant ${\rm sl}(\cK)$ associated to a framed knot by
  \ie
  {\rm sl}(\cK) \equiv {\rm lk}(\cK,\cK_f)={\rm sl}_0(\cK)+T(\cK)\,.
  \fe
 This quantity is known as the self-linking number of $\cK$. Under a change of framing, the self-linking number shifts by an integer.

  Getting back to our   Wilson loop in 5d MSYM, by plugging in the stereographic coordinates \eqref{stereoS5} into \eqref{LinkingNumS3} and also into the integral ${\cal I}_2\equiv\int dt_1 dt_2\ I_2$, we conclude after some algebra that
  \es{I2Final}{
  	{\cal I}_2=-{1\ov2\pi}\,{\rm sl}_0(\cK)\,.
  }
  If $\cK$ has multiple components $\cK_i$,  
   \es{calI2Final}{
  {\cal I}_2=-{1\ov2\pi} \left( \sum_i {\rm sl}_0(\cK_i)+\sum_{i\neq j}{\rm lk}(\cK_i,\cK_j) \right)
  }
 
  Hence, combing \eqref{I1Int} with \eqref{calI2Final}, we obtain the final expression for ${\cal I}$ in the case where $\cK$ has a single component is:
  \ie
  {\cal I}&=-{1\ov 2\pi}\le({L(\cK) } + {\rm sl}_0(\cK) \ri) \,.
  \fe
  Using Wick contractions, it is easy to see the higher order contributions in \eqref{abpert} complete it into the following formula
  \ie
  \la  W(\{\cK_i,q_i\}) \ra&\equiv \la  \prod_i W_{q_i}(\cK_i) \ra
  =\exp\le[{g_{\rm YM}^2 \over 4 \pi }\le( { \sum_i  q_i^2 L(\cK_i)}+ \sum_i  q_i^2{\rm sl}_0(\cK_i)
  +\sum_{i\neq j} q_i q_j{\rm lk}(\cK_i,\cK_j))
  \ri)\ri]
  \fe
  for the expectation value of a collection of linked ${1\over 8}$-BPS Wilson loops in the 5d abelian theory.  This expression is independent of the choice of framing for the Wilson loops, but it has a mild dependence on the shape of the loop.
  
  We can introduce a renormalized version of our 5d Wilson loop by multiplying it by a counter-term that removes the shape dependence:
 \es{WrenDef}{
  W^{\rm ren}_q(\cK)\equiv W_q(\cK) \exp\le[-{g_{\rm YM}^2 \over 4 \pi }q^2 (L(\cK)- T(\cK))\
  \ri] \,.
  }
  Consequently, their expectation values of the renormalized loop operators are given by
  \ie
  \la  W^{\rm ren} (\{\cK_i,q_i\}) \ra&\equiv \la  \prod_i W^{\rm ren}_{q_i}(\cK_i) \ra
  =\exp\le[{g_{\rm YM}^2 \over 4 \pi }\le( \sum_i  q_i^2{\rm sl}(\cK_i)
  +\sum_{i\neq j} q_i q_j{\rm lk}(\cK_i,\cK_j)
  \ri)\ri] \,.
  \label{abknot}
  \fe
 This expression is topological, but it now transforms under a change of framing because under such a change, we have
  \ie
  {\rm sl}(\cK_i) \to {\rm sl}(\cK_i)+f_i \quad f_i\in \mZ.
  \label{abknotphase}
  \fe

 \subsection{Non-Abelian generalization} 
  
  For the non-abelian generalization of \eqref{abpert} in MSYM with gauge group $G$, the perturbative computation involves the same diagrams at order $g^2_{\rm YM}$. The only difference from the abelian case is that we need to sum over the propagators for each color, giving
  \ie
  \la  W_R (\cK) \ra=1+{g_{\rm YM}^2\over 4\pi} {\dim G\over \dim R}\le({L(\cK) } +{\rm sl}_0(\cK)\ri)+\cO(g_{\rm YM}^4)
  \fe
  for a single knot $\cK$. 
  Similarly, we define the renormalized Wilson loop
  \ie
  W^{\rm ren}_q(\cK)\equiv W_q(\cK) \exp\le[-{g_{\rm YM}^2 \over 4 \pi }  {\dim G\over \dim R} (L(\cK)- T(\cK))
  \ri]
  \fe
  and
  \ie
  \la  W^{\rm ren}_R (\cK) \ra=1+{g_{\rm YM}^2\over 4\pi} {\dim G\over \dim R}{\rm sl}(\cK)+\cO(g_{\rm YM}^4) \,.
  \label{nbknotpert}
  \fe
  At $\cO(g_{\rm YM}^4)$ order, seven more Feynman diagrams contribute, and the number grows further at higher orders \cite{Zarembo:2002an}.

 \section{Proposal for an effective Chern-Simons description} 
 \label{CONJECTURE}

  \subsection{Chern-Simons description of $1/8$-BPS Wilson loops}
  
 The result \eqref{abknot} in the 5d MSYM theory is reminiscent of the formula for the expectation values of Wilson loops in 3d Abelian Chern-Simons theory on $S^3$.  Indeed, a collection of framed Wilson loops $\cK_i$ with charges $q_i$ in the $U(1)_k$ Chern-Simons theory on $S^3$   gives \cite{Witten:1988hf}
  \ie
  \la W_{3d}(\{\cK_i,q_i\})\ra=\exp\le[{i\pi \over k}\le( \sum_i  q_i^2{\rm sl}(\cK_i)
  +\sum_{i\neq j} q_i q_j{\rm lk}(\cK_i,\cK_j)
  \ri)\ri] \,,
  \fe
  with the same framing dependence as in \eqref{abknotphase}.  Therefore we conclude that in the free 5d Abelian theory, the $1/8$-BPS Wilson loop sector is captured by 3d Chern-Simons theory:
  \ie
  \la  W^{\rm ren} (\{\cK_i,q_i\}) \ra=\la W_{3d}(\{\cK_i,q_i\})\ra \,, 
  \fe
 provided that the Chern-Simons level is analytically continued to the pure imaginary value
  \ie
  k=i{4\pi^2\over g_{\rm YM}^2}={2\pi i\over \B } \,.
  \label{kbeta}
  \fe
Recall that the 5d $1/8$-BPS Wilson loops are necessarily contained within a great $S^3$, and it is this $S^3$ that should be identified with the $S^3$ on which the Chern-Simons theory is defined.  The imaginary value of CS level in \eqref{kbeta} requires an analytic continuation of the usual CS path integral as explained in \cite{Witten:2010cx}.  A similar correspondence between 5d $1/8$-BPS Wilson loops and Wilson loops in 3d CS theory holds to order $g_{\text{YM}}^2$ in the non-Abelian theory:  the result \eqref{nbknotpert} matches the corresponding result in the 3d CS theory provided that one again makes the identification in \eqref{kbeta}.

  Instead of computing higher order Feynman diagrams explicitly, motivated by the abelian result and the weak coupling limit of the nonabelian generalization, we conjecture that the final answer is again given by the corresponding 3d CS theory with gauge group $G$ and renormalized level \eqref{kbeta}, such that
  \ie
  \la  W^{\rm ren}_G (\{\cK_i,R_i\}) \ra=\la W^{3d}_G(\{\cK_i,R_i\})\ra \,.
  \label{post}
  \fe
In the next sections, we will provide  evidence for this proposal from holography.  In Section~\ref{LOCALIZATION}, we will then provide the first steps of a direct proof of \eqref{post} that uses supersymmetric localization.

\begin{figure}[!h]
	\centering
	{
		\begin{subfigure}{0.3\textwidth}
			\centering
			\includegraphics[scale=.2 ]{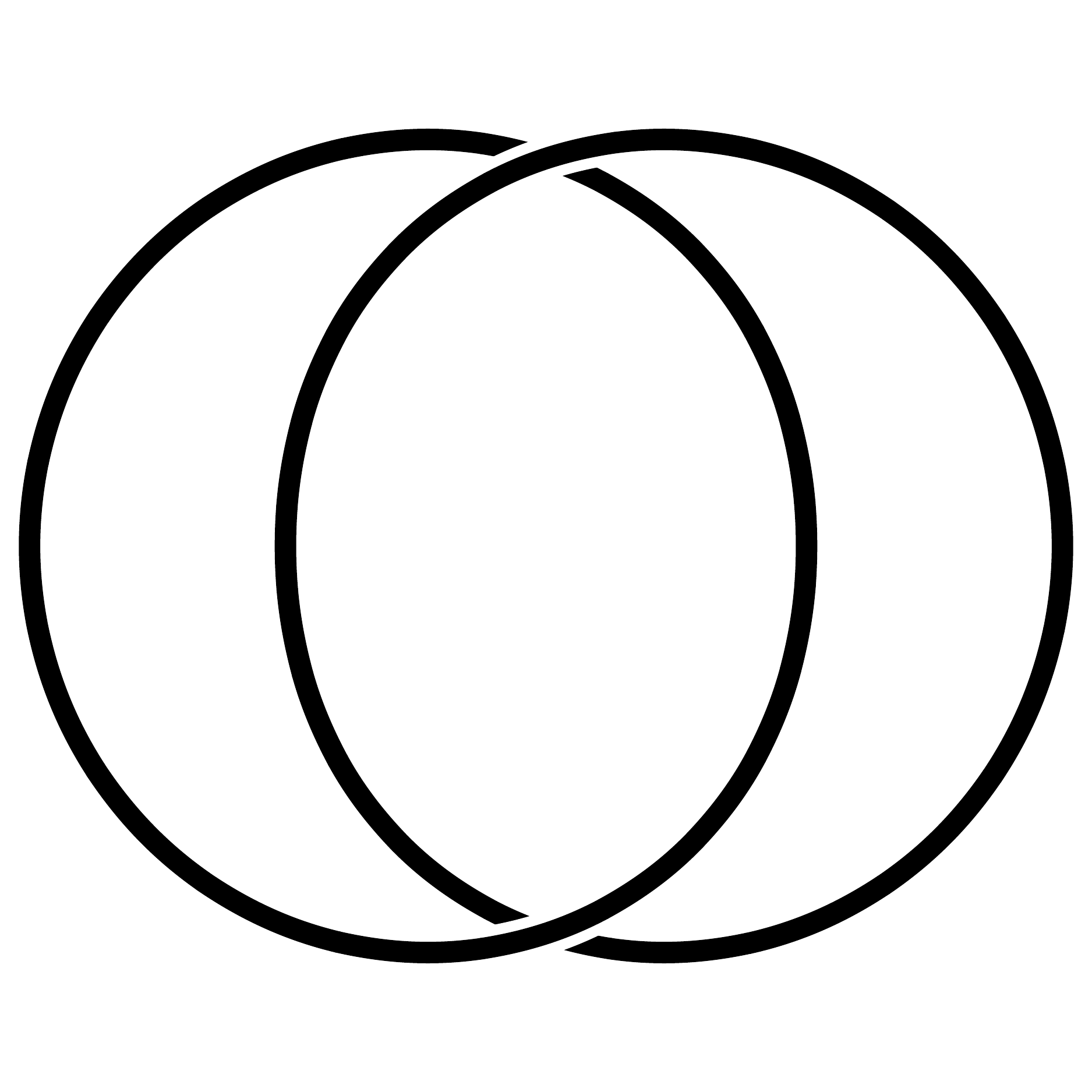}\\
			\caption{$T_{2,-2}$}
		\end{subfigure} 
	}
	\centering
	{
		\begin{subfigure}{0.3\textwidth}
			\centering
			\includegraphics[scale=.18 ]{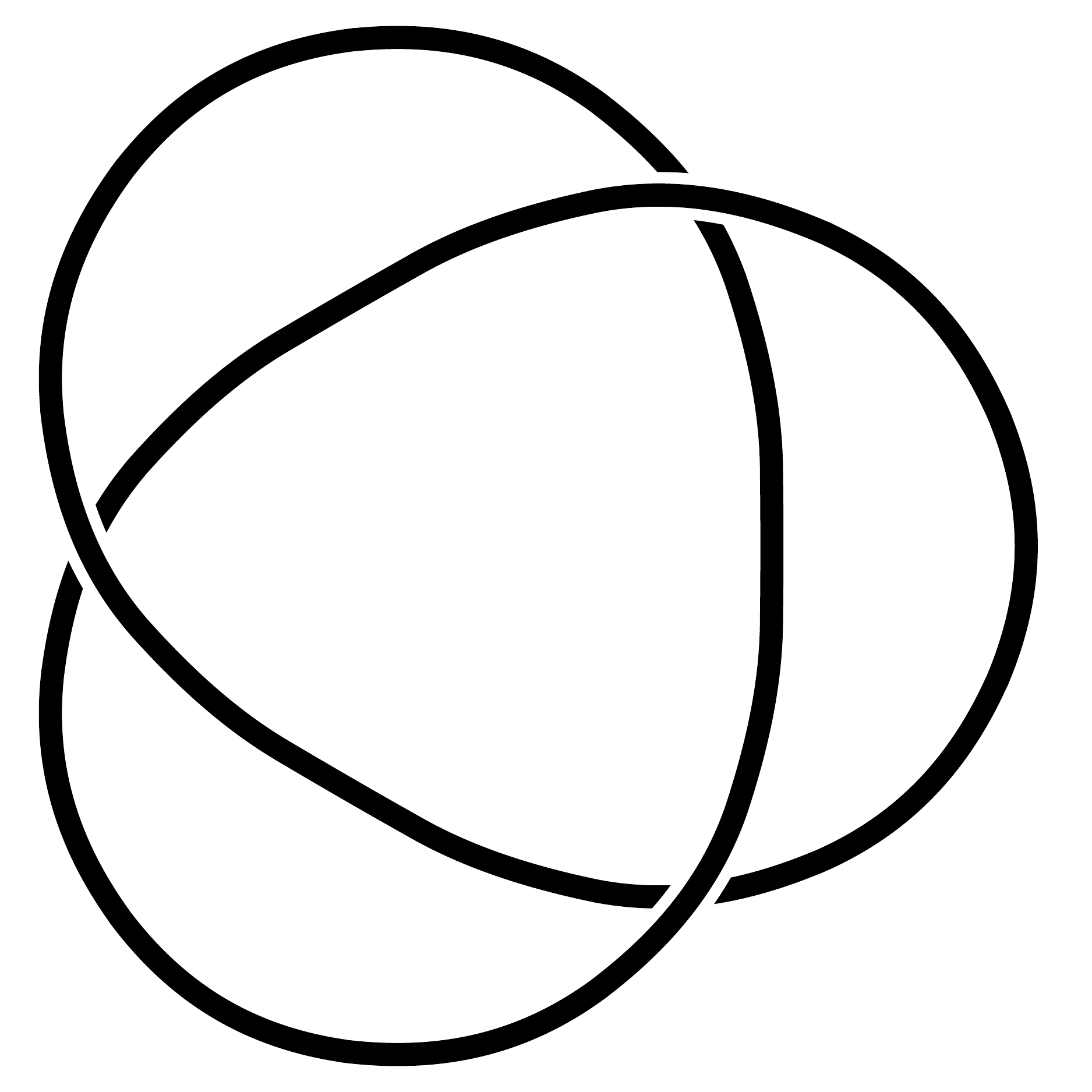}\\
			\caption{$T_{2,-3}$}
		\end{subfigure} 
	}
		\centering
		{
			\begin{subfigure}{0.3\textwidth}
				\centering
				\includegraphics[scale=.18 ]{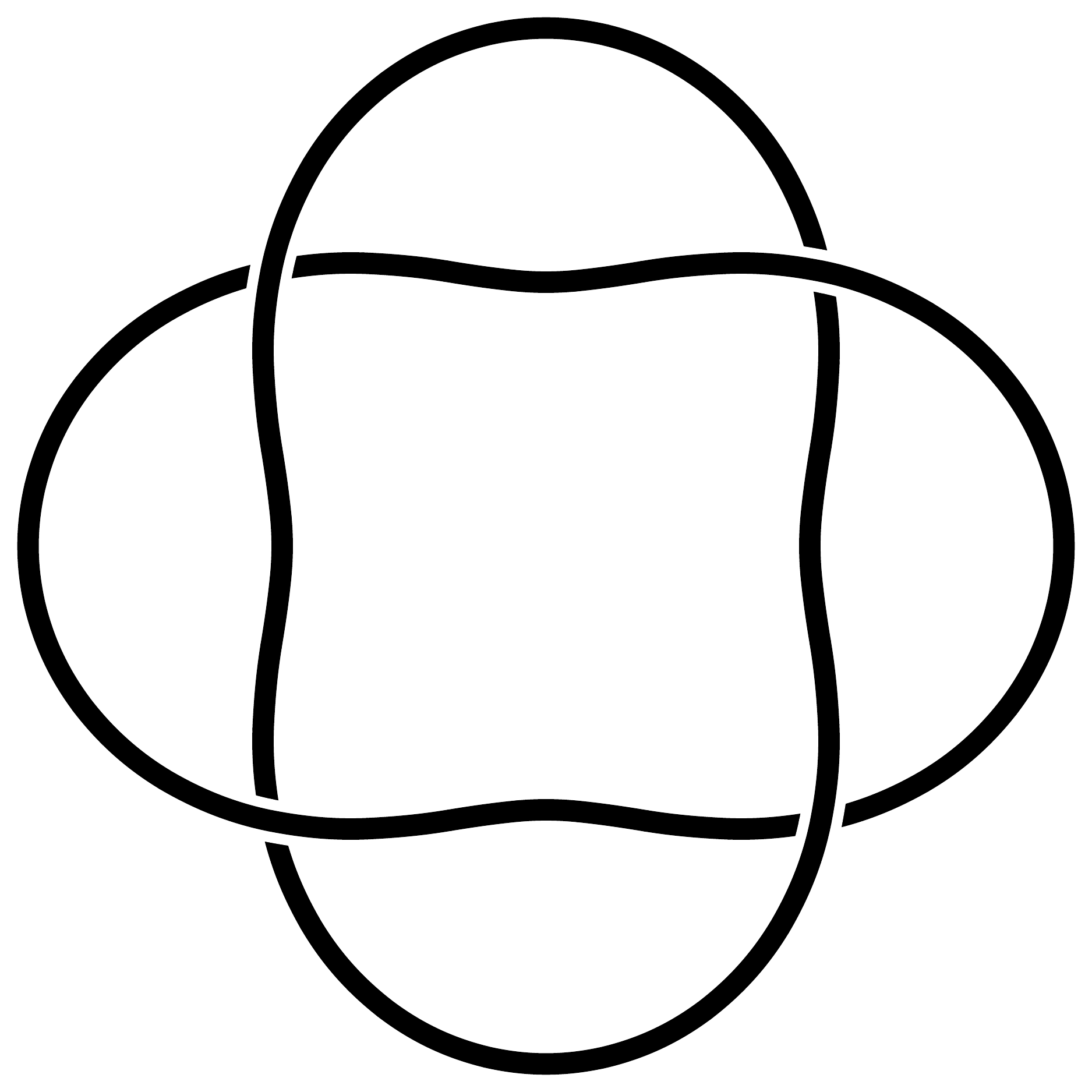}\\
				\caption{$T_{2,-4}$}
			\end{subfigure} 
		}
			\centering
			{
				\begin{subfigure}{0.3\textwidth}
					\centering
					\includegraphics[scale=.18 ]{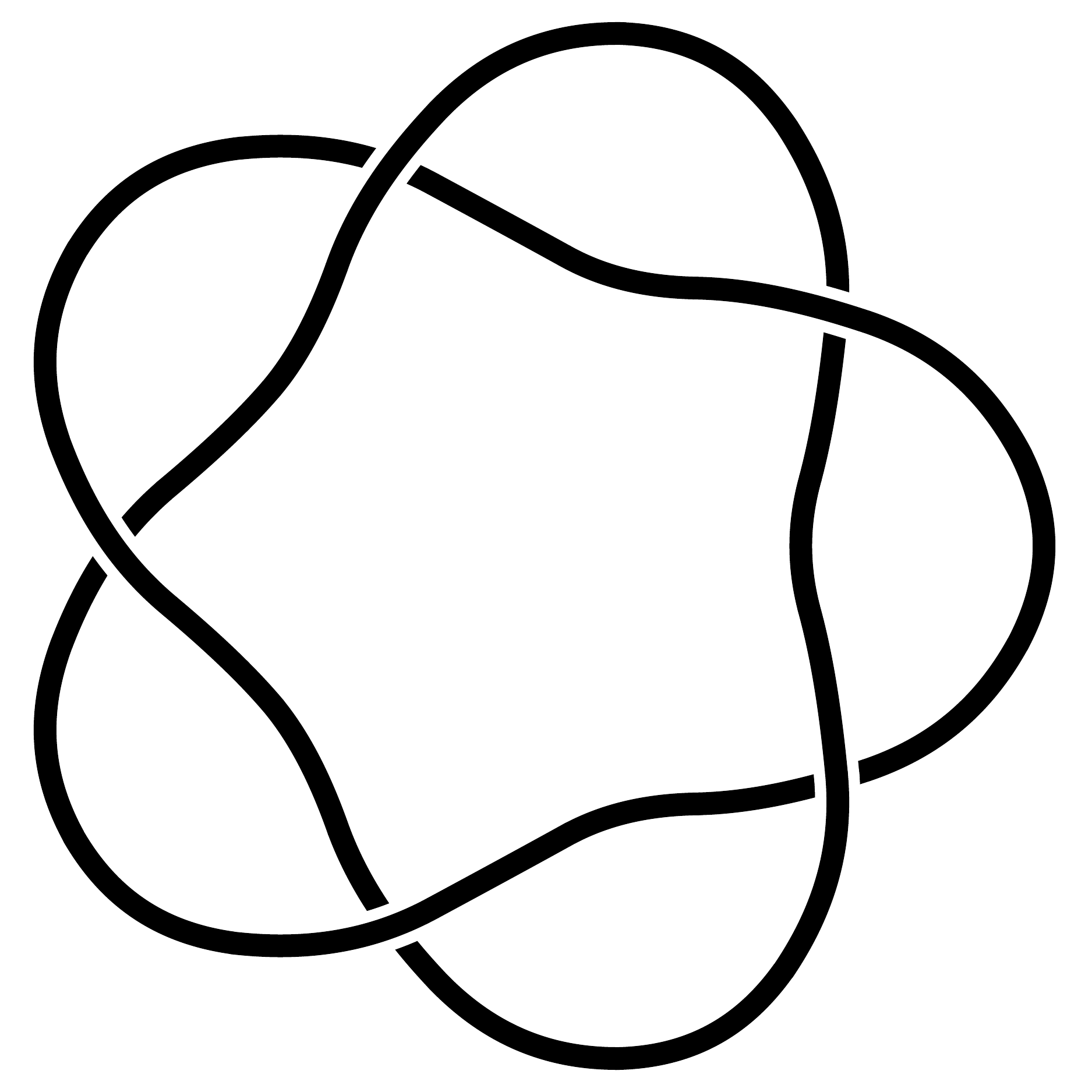}\\
					\caption{$T_{2,-5}$}
				\end{subfigure} 
			}
			\centering
			{
				\begin{subfigure}{0.3\textwidth}
					\centering
					\includegraphics[scale=.18 ]{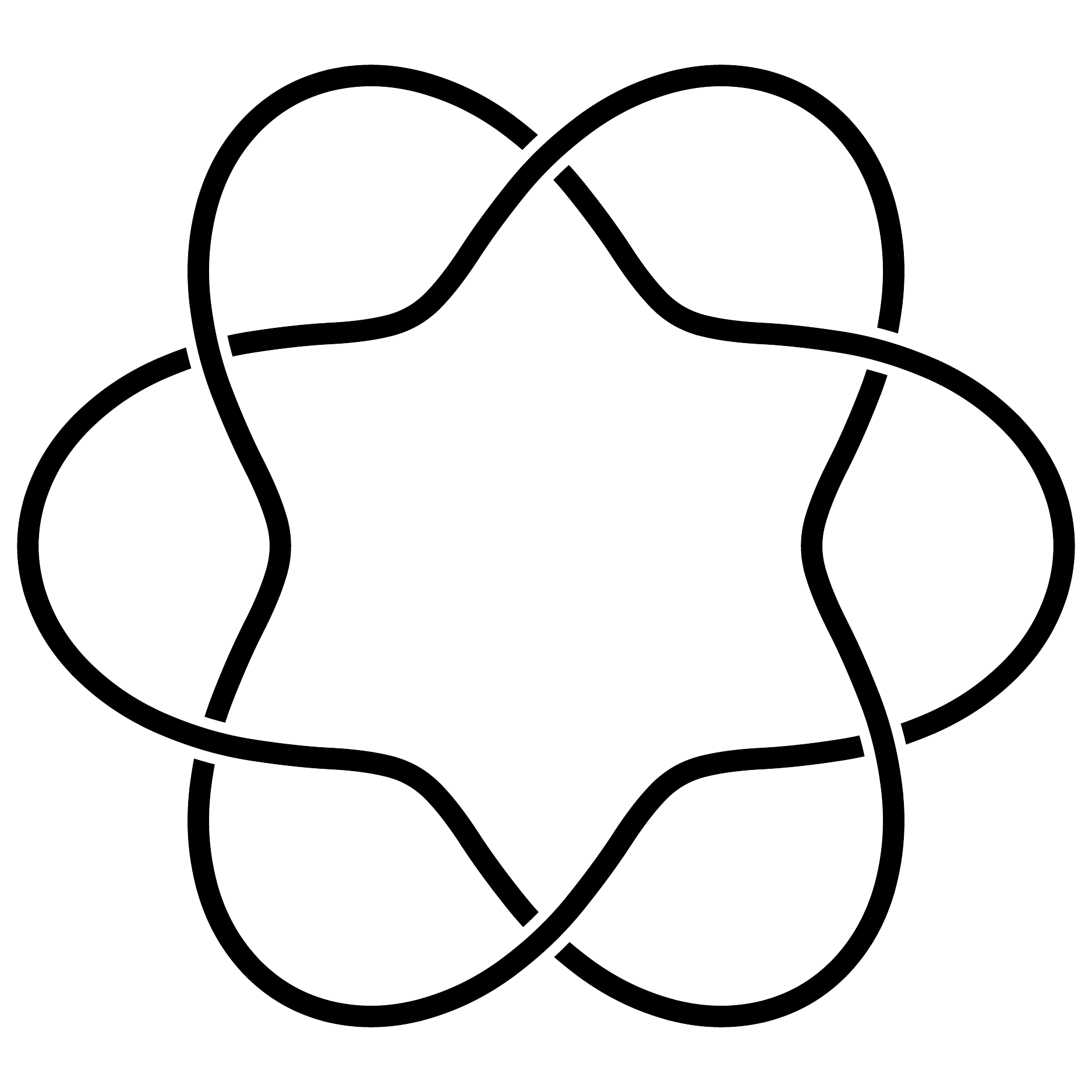}\\
					\caption{$T_{2,-6}$}
				\end{subfigure} 
			}
			\centering
			{
				\begin{subfigure}{0.3\textwidth}
					\centering
					\includegraphics[scale=.18 ]{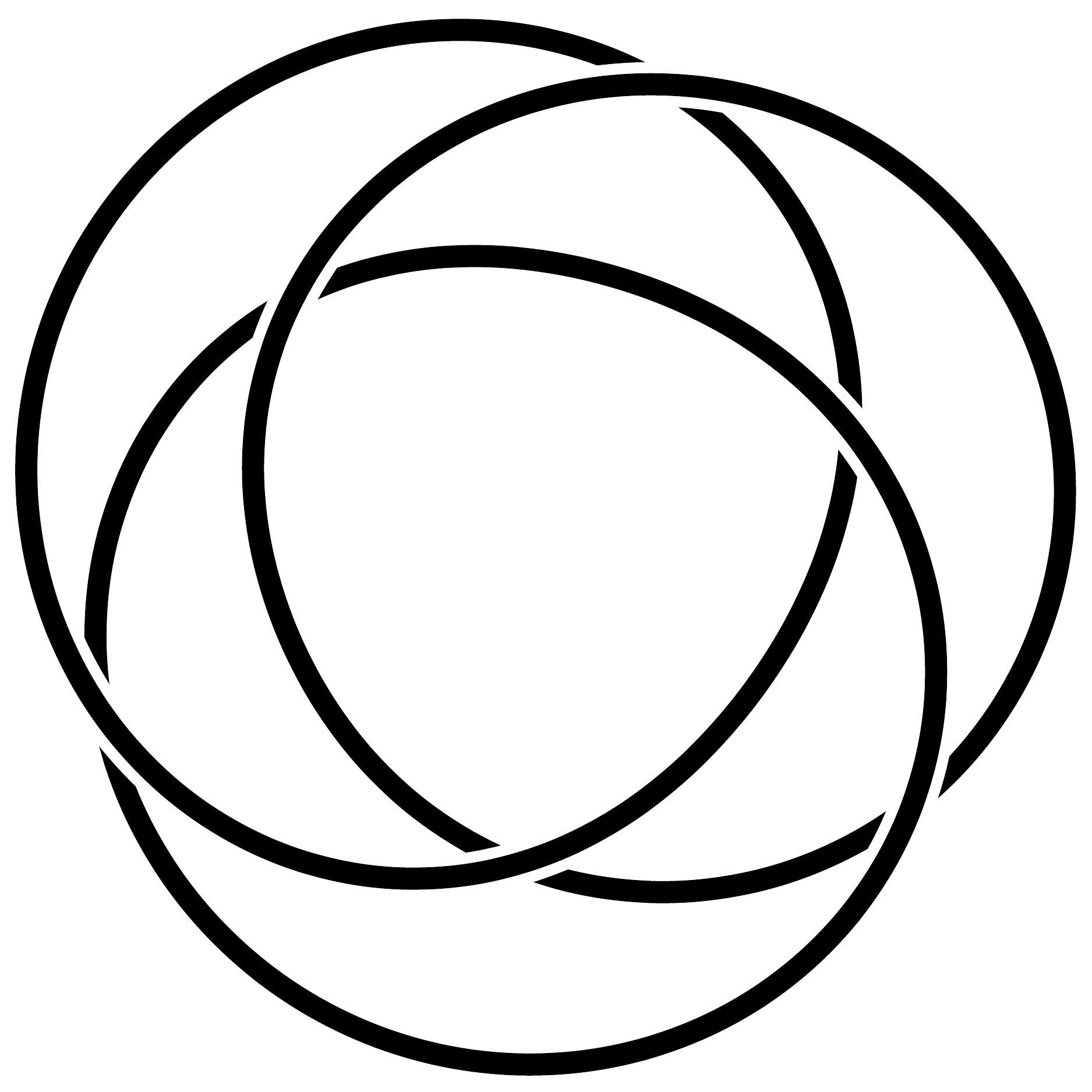}\\
					\caption{$T_{3,-3}$}
				\end{subfigure} 
			}
				\centering
				{
					\begin{subfigure}{0.3\textwidth}
						\centering
						\includegraphics[scale=.18 ]{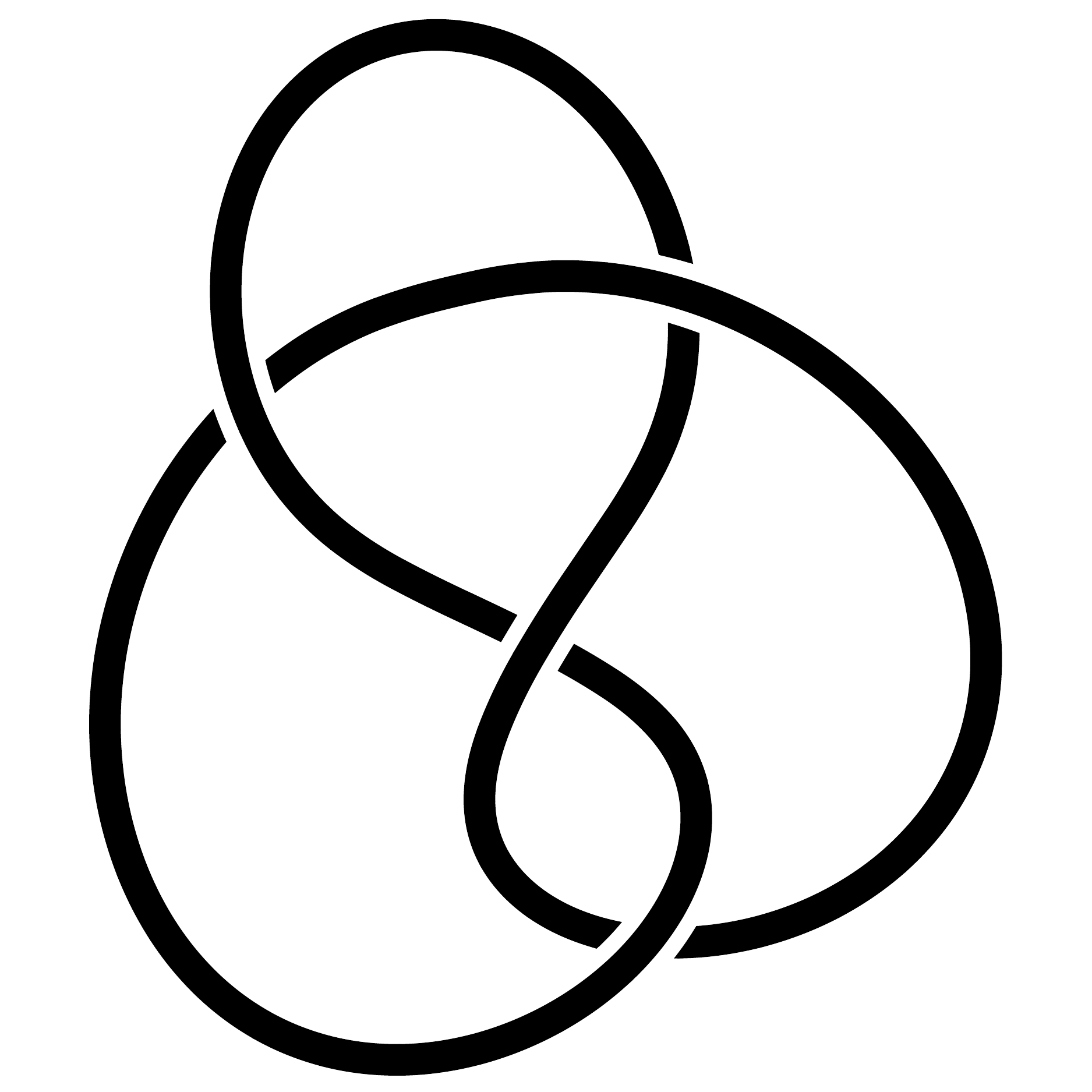}\\
						\caption{$4_1$ hyperbolic knot}
					\end{subfigure} 
				}
				\centering
				{
					\begin{subfigure}{0.3\textwidth}
						\centering
						\includegraphics[scale=.18 ]{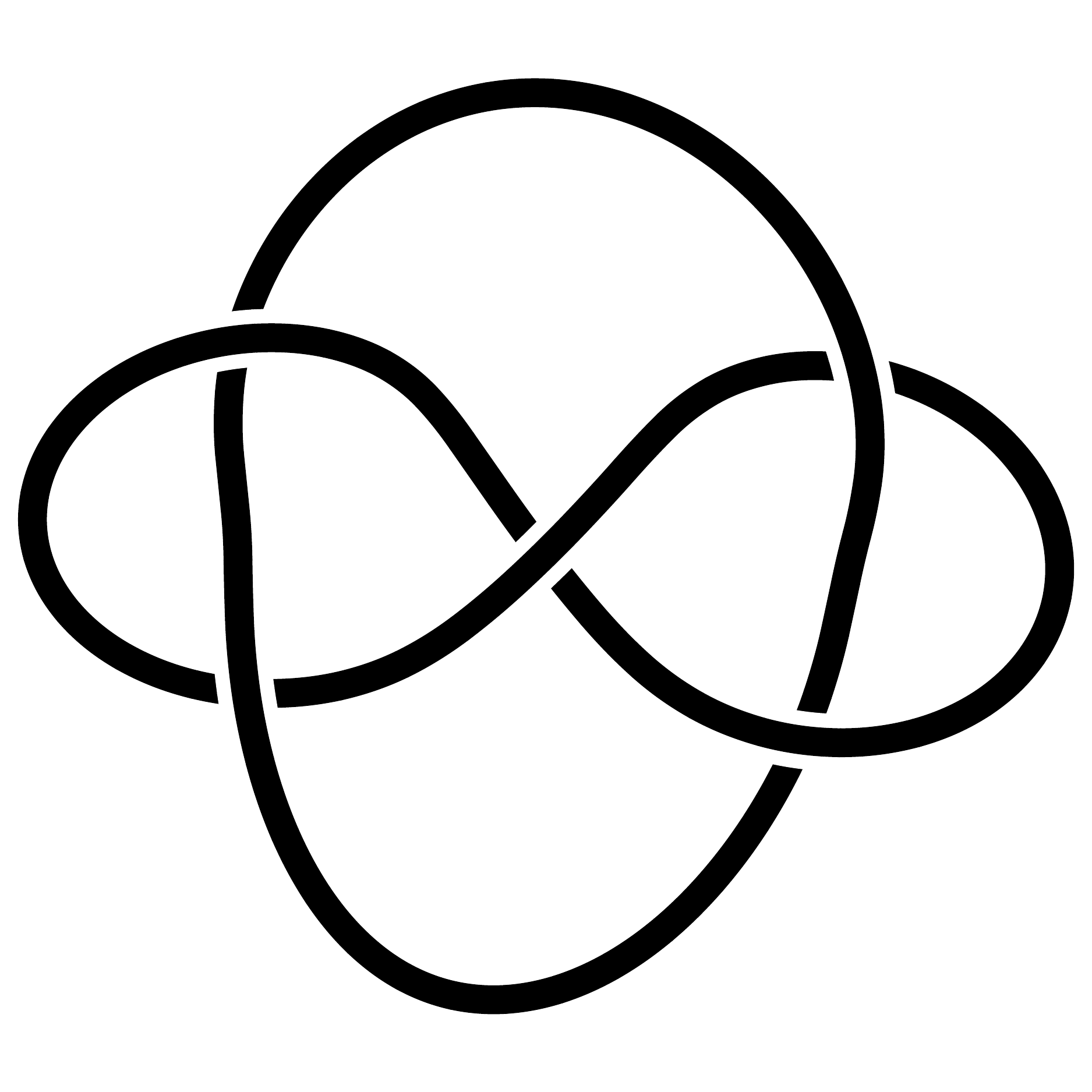}\\
						\caption{Whitehead link}
					\end{subfigure} 
				}
				\centering
				{
					\begin{subfigure}{0.3\textwidth}
						\centering
						\includegraphics[scale=.18 ]{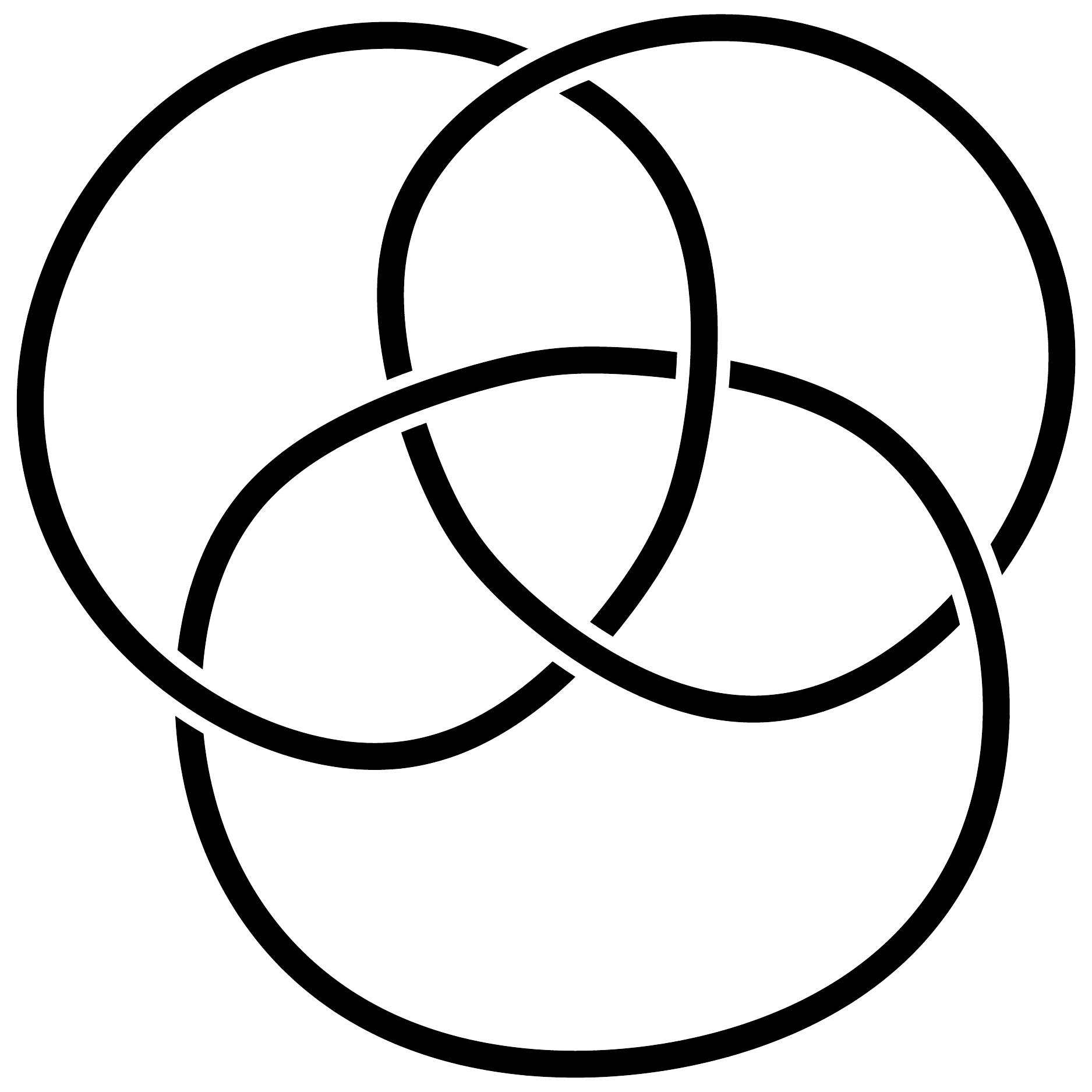}\\
						\caption{Borromean rings}
					\end{subfigure} 
				}
	\caption{A list of basic knots and links. In the first two rows we list the first several torus knots and links. In the third row we give examples of non-torus knot and links. Here the orientation of the links are chosen such that the right most (two for (a), (c), (e) and (h), three for (f) and (i)) vertical strands are oriented upwards. \label{fig:knotlist}}
\end{figure}

    \subsection{Parity properties of the CS sector}
 
The proposal above raises an immediate puzzle:  the fact that the 5d MSYM theory is parity preserving in flat space seems in contradiction with the fact that Chern-Simons theory is parity violating.  So let us briefly comment on the parity properties of this Chern-Simons sector of 5d MSYM as well as of the corresponding 6d $(2,0)$ theory on $S^1 \times S^5$. 

Without loss of generality, we can take the 3d parity operator ${\bf P}_{3d}$ to act by $x_3\to -x_3$ on $S^3$ such that
 \ie
 \hat A_{\rm i} (x_3) \to  P_{\rm i}{}^{\rm j} \hat A_{\rm j}(-x_3) \,,
 \label{p3d}
 \fe
 with $P_{\rm i}{}^{\rm j}={\rm diag}(-1,1,1)$.
 The CS action changes sign under ${\bf P}_{3d}$. Equivalently ${\bf P}_{3d}$ maps the theory with level $k$ to a partner CS theory at level $-k$.

 The 5d MSYM in flat space on the other hand is  invariant under a 5d parity ${\bf P}_{5d}$ that acts not only on the gauge field, but also on the fermions and scalars of the 5d theory. Up to a conjugation by $\mf{so}(5)$ spacetime rotation and $\mf{so}(5)_R$ transformation, ${\bf P}_{5d}$ acts by
 \ie
 &\Psi(x_3) \to i\hat \C_3 \C_3 \Psi(-x_3) \,,
 \\
 &A_{\rm i} (x_3) \to  P_{\rm i}{}^{\rm j}   A_{\rm j}(-x_3) \,, \qquad
 A_{1,2} (x_3) \to A_{1,2}(-x_3) \,,
 \\
 &\Phi_a (x_3) \to  P_a{}^b \Phi_b (-x_3) \,, \qquad \Phi_{1,2} (x_3) \to \Phi_{1,2}(-x_3) \,.
 \label{p5d}
 \fe
 In particular ${\bf P}_{5d}^2=(-1)^F$ where $F$ is the spacetime fermion number. 
 
 On $S^5$, the same parity transformation ${\bf P}_{5d}$ together with the transformation on the scalar coupling matrix $\cS$\footnote{Recall that $\cS$ is defined in terms ratios of bilinears in $\ve$. This transformation of $\cS$ is naturally induced by ${\bf P}_{5d}$ acting on the Killing spinor as $
 	\ve(x_3) \to \ve'(x_3)= i\hat \C_3 \C_3  \ve(-x_3)
 	$. In particular $\ve'$ satisfies  \ie
  \nabla_\m \ve'=-{1\over 2r}\C_\m \hat \C^{12}\ve' \,. 
  \label{kseqflip}
  \fe
 Note the flipped sign relative to \eqref{kseq}.}
  \ie
 &S_{{\rm i} a}(x_3)\to  P_{\rm i} {}^{\rm j}P_a{}^b S_{{\rm j} b}(-x_3)
 \fe
 reduced precisely to ${\bf P}_{3d}$ in the 3d CS sector. However, ${\bf P}_{5d}$ is no longer a symmetry  on $S^5$. Indeed, the curvature couplings
 \ie
 \cL_{\rm YM} \supset \tr \bigg[
 -{i\over 4R}\bar \Psi  \C^{12} \Psi 
 -{1\over 3R} \epsilon_{abc}\Phi^a[\Phi^b,\Phi^c]
 \bigg] 
 \label{pvc}
 \fe
 flip sign under ${\bf P}_{5d}$.\footnote{There is however a parity symmetry ${\bf P'}_{5d}$ of the 5d MSYM on $S^5$ which acts by
 	\ie
 	&\Psi(x_3) \to i\hat \C_3 \C_1 \Psi(-x_3) \,, 
 	\\
 	&  A_{\rm i} (x_3) \to  P_{\rm i}{}^{\rm j}   A_{\rm j}(-x_3) \,, \qquad 
 	A_{1,2} (x_3) \to A_{1,2}(-x_3),
 	\\
 	&\Phi_1 (x_1,x_i) \to -\Phi_1 (-x_1,x_i) \,, \qquad 
 	\Phi_{I\neq 1} (x_1,x_i) \to \Phi_{I} (-x_1,x_i) \,.
 	\fe
 	In the flat space limit, ${\bf P}_{5d}$ and ${\bf P'}_{5d}$ are equivalent by an $\mf{so}(5)_R$ rotation, which is no longer true on $S^5$ due to the curvature couplings \eqref{pvc}.  In the 3d CS sector, ${\bf P'}_{5d}$ has no natural intepretation. Rather it maps one CS sector to another within the 5d MSYM on $S^5$.}  
 
 From the 6d perspective, ${\bf P}_{5d}$ lifts to the ${\bf C P T}$ symmetry of the $(2,0)$ theory. In particular the ${\bf CP}$ is identified with ${\bf P}_{5d}$ while the time reversal $\bf T$ gives rise to the sign flip in \eqref{kseqflip} relative to \eqref{kseq}  \cite{Kim:2012qf}. Consequently, $\bf CPT$ maps one $\mf{su}(4|2)$ subalgebra of $\mf{osp}(8^*|4)$ that commutes with  $H- \frac{R_{\bf 13}+R_{\bf 24}}{2}$ to another that commutes with  $H+\frac{R_{\bf 13}+R_{\bf 24}}{2}$.\footnote{In particular it sends the 6d Poincare supercharges $Q$ to the superconformal supercharges $S$.} These two $\mf{su}(4|2)$ subalgebras are associated with isomorphic yet distinct 5d MSYM sectors of the $(2,0)$ theory that come from two choices of R-symmetry twisting when compactifying the 6d theory on $S^1$.
Consequently, the parity transformation in the 3d CS theory relates two isomorphic 3d CS sectors of the 6d $(2,0)$ theory with opposite CS levels.

  \subsection{Knot invariants in the strong coupling limit}

  In this section, we will use \eqref{post} and known results in CS theory to extract predictions for our ${1\over 8}$-BPS Wilson loops at strong coupling.   In the next section, we will compare these predictions to the analogous quantities computed using holography.  
  
  For simplicity, here we will focus on 5d MSYM with gauge group $G=U(N)$ and Wilson loops in the fundamental representation $R=\f$.\footnote{The $U(N)$ MSYM contains two decoupled  sectors described by $U(1)$ and $SU(N)$ gauge theories respectively. Correspondingly, the 6d uplift is a tensor product of the free $(2,0)$ theory and $A_{N-1}$ interacting theory.
On the holographic side, only the interacting $A_{N-1}$ part is relevant. Here we choose not to separate them as the formula looks simpler for $U(N)$ and  keep in mind that in the strict large $N$ limit we can safely ignore the contribution from the $U(1)$ factor.}   According to our proposal, the protected sector of ${1\over 8}$-BPS Wilson loops is then described by the 3d $U(N)$ CS theory with renormalized level $k$ given by \eqref{kbeta}.  (The bare level is $k_0 = k - N$.)  It is convenient to introduce parameters
  \ie
  q \equiv \exp\le({2\pi i\over k} \ri)= e^\beta \,,  \qquad \lambda \equiv q^{N} = e^{N \B}  \,.
  \label{qlambda}
  \fe

  In terms of $q$ and $\lambda$, the unknot  Wilson loop in the fundamental representation has expectation value 
  \ie
  \la W_{\f} (\cK )\ra 
  ={\lambda^{{\rm sl}(\cK )\ov 2}\ov N}{\lambda^{{1\over 2}}-\lambda^{-{1\over 2}}\over q^{1\over 2}-q^{-{1\over 2}}}  \,.
  \fe
  More generally, for a link $\cL$ made from  knots $\cK_i$,  $i = 1, \ldots, r$, each in the fundamental representation, we have
  \ie
  \la 	W_{\f}(\cL) \ra ={\lambda^{{\rm sl}(\cL)\over 2}\ov N}{\lambda^{{1\over 2}}-\lambda^{-{1\over 2}}\over q^{1\over 2}-q^{-{1\over 2}}} \cH(\cL) \,, \qquad
   \text{with } {\rm sl}(\cL)\equiv \sum_{i }{\rm sl}(\cK_i )+\sum_{i\neq j}{\rm lk}(\cK_i,\cK_j) \,.
  \label{nbknot}
  \fe
  In the literature, these results are often presented with the canonical framing on $S^3$ in which case all self-linking numbers of irreducible knots ${\rm sl}(\cK_i)$ vanish.
  The last factor $\cH(\cL)$ in \eqref{nbknot}
  is the HOMFLY polynomial of $\cL$ which has the following structure
  \ie
  \cH(\cL)=\sum_{i= 0}^\infty p_i(\lambda^{1\over 2}) z^{2i+1-r},\quad z=q^{1\over 2}-q^{-{1\over 2}}
  \fe
  such that $p_i$ are Laurent polynomials in $\lambda^{1\over 2}$ (see for example \cite{Marino:2005sj}).

Graded by both $\lambda$ and $q$ (separately via analytic continuation), the Wilson loop observable \eqref{nbknot} gives a large class of so-called quantum knot invariants for knots on $S^3$. Classical knot invariants such as framed linking numbers can be recovered in the expansion with respect to the effective gauge coupling $1/k$. This is exactly the perturbative expansion we have computed from Feynman diagrams in  Section~\ref{sec:pert}.

As a consistency check, we see that in the Abelian case, we have $N=1$, and consequently $\lambda=q$ and $\cH=1$ \cite{Marino:2005sj}.  Then \eqref{nbknot} matches \eqref{abknot}, as expected.  For the nonabelian case, in the weak coupling limit $\B \ll 1$, we also see \eqref{nbknot} agrees with \eqref{nbknotpert}. 
  Below 
  we would like to consider the opposite limit
  \ie
  N\to \infty,\quad  q~{\rm fixed}\,,
  \label{Mlim}
  \fe
  and consequently $\lambda\to \infty$, because in this limit there is a weakly-coupled dual supergravity description of the theory, as will be discussed in the next section.  In the limit \eqref{Mlim}, the dominant contribution from the HOMFLY polynomial $\cH(\cL)$ of a link $\cL$ comes from the maximal degree in $\lambda^{1\over 2}$, which we denote by $\zeta(\cL)$. Therefore, \eqref{nbknot} becomes
 \es{Prediction1}{
  \la W_{U(N)}(\cL)\ra \approx \lambda^{{1\over 2}  (1+ {\rm sl}(\cL) +  \zeta(\cL))  } \,, \qquad
   \text{as $N \to \infty$.}
}
We will compare the holographic results to this formula.


\section{Holographic dual of the topological surface operators}
\label{HOLO}

As explained in the previous sections, when the gauge group is $SU(N)$, the 5d MSYM theory can be obtained as a twisted reduction of the $A_{N-1}$ $(2, 0)$ theory on $S^5 \times S^1$, which at large $N$ has a weakly coupled supergravity description.  The Wilson lines in the Yang-Mills theory become surface operators in the $(2, 0)$ theory completely wrapping $S^1$.  We would like to check using the holographic prescription that their expectation values are shape-independent and agree with our conjectured result in Eq.~\eqref{Prediction1}.

\subsection{The 11d supergravity background}

The supergravity background corresponding to the $A_{N-1}$ $(2, 0)$ theory on $S^5 \times S^1$ is a background of 11d SUGRA that has an $S^5 \times S^1$ slicing and that preserves $16$ supercharges.  It is obtained by a small modification of the $AdS_7 \times S^4$ background that describes the superconformal $(2, 0)$ theory on conformally flat spaces which preserves 32 supercharges.  So let us first describe the $AdS_7 \times S^4$ and relevant properties, and then the modification required.

\subsubsection{$AdS_7 \times S^4$ Lorentzian background}

To establish conventions, let us start by introducing the 11d supergravity action in mostly plus Lorentzian signature\footnote{We use conventions in which $\gamma^M \gamma^N + \gamma^N \gamma^M = 2 \eta^{MN}$.  For a Majorana spinor $\chi$, the conjugate $\bar \chi$ is defined as $\bar \chi = \chi^T C$, where $C$ is a charge conjugation matrix (a unitary matrix obeying $C^T = -C$ and $(\gamma^\mu)^T = - C \gamma^\mu C^{-1}$.  We also have $\bar \chi = \chi^\dagger i \gamma^0$, which implies $\chi^\dagger  = \chi^T C i \gamma^0$.} (see for example \cite{Freedman:2012zz}):
 \es{11dAction}{
  S_\text{bos} &= \frac{1}{2 \kappa_{11}^2} 
   \int d^{11} x\, \sqrt{-G} \left( R - \frac 1{48} F_{MNPQ}F^{MNPQ} \right)
    - \frac{1}{12 \kappa_{11}^2} \int A_3 \wedge F_4 \wedge F_4 \,, \\
   &{}-\frac{1}{2 \kappa_{11}^2} \int d^{11} x\, \sqrt{-G} \left[  \bar \psi_M \gamma^{MNP} D_N \psi_P +\frac{1}{96} \bar \psi_M \left( \gamma^{PQRS M N} F_{PQRS} + 12 \gamma^{PQ} F_{PQ}{}^{M N} \right) \psi_N + \cdots \right] 
 }
where $G_{MN}$ is the metric,\footnote{We use $x^M$ for the 11d coordinates and upper case indices from the middle of the alphabet ($M$, $N$, $P$, etc.) for all tangent space indices in 11d.} $A_3$ is the 3-form gauge potential with field strength $F_4 = dA_3$, $\psi$ is the gravitino, and $\kappa_{11}$ is the 11d gravitational constant related to the Planck length $\ell_p$ through
 \es{kappa11}{
  2 \kappa_{11}^2 = (2 \pi)^8 \ell_p^9 \,.
 }
The ellipses in \eqref{11dAction} denote higher order terms in the gravitino field.  
This action is invariant under local SUSY transformations\footnote{The linearized gravitino variation can equivalently be  written as 
 $\delta\psi_M = D_M \epsilon + \frac{1}{288} \left(F_{PQRS} \gamma_M \gamma^{PQRS} - 12 F_{MPQR} \gamma^{PQR} \right)$.}
 \es{LinGravitino}{
  \delta e_M^a &= \frac 12 \bar \epsilon \gamma^a \psi_M \,, \\
  \delta \psi_M &= D_M \epsilon + \frac{1}{288} \left(\gamma^{PQRS}{}_M 
   - 8 \gamma^{QRS} \delta_M^P \right) F_{PQRS} \epsilon \,, \\
  \delta A_{MNP} &= -\frac 32 \bar \epsilon \gamma_{[MN} \psi_{P]} \,.
 }

A solution of the equations of motion following from \eqref{11dAction} is $AdS_7 \times S^4$ with 4-form flux threading $S^4$:
 \es{MetricAdS7S4}{
  ds^2 = L^2 ds_7^2 + \frac 14 L^2 ds_{S^4}^2 \,, \qquad
   F_4 = \frac{3 L^3}{8} \vol_{S^4} \,,
 }
where $ds_7^2$ and $ds_{S^4}^2$ are the line elements on unit curvature radius $AdS_7$ and $S^4$, respectively, $\vol_{S^4}$ is the volume form on $S^4$, and $L$ is a constant related to the field theory quantity $N$ via 
 \es{LNRelation}{
  \frac{L^3}{\ell_p^3} = 8 \pi N \,.  
 }
For our purposes, it is convenient to write the $AdS_7$ metric as
 \es{AdS7Metric}{
  ds_7^2 =  -(\cosh \rho)^2 dt^2 + d\rho^2 +  (\sinh \rho)^2 ds_{S^5}^2  
 }
using a radial coordinate $\rho$, a (non-compact, Lorentzian) time coordinate $t$, and a round five-sphere.  This metric is convenient for describing the $(2, 0)$ theory on $\R \times S^5$, as the conformal boundary obtained as $\rho \to \infty$ in \eqref{AdS7Metric} is precisely $\R \times S^5$.  

It will be convenient to parameterize the five-sphere in \eqref{AdS7Metric} by coordinates $X_i$, $i = 1, \ldots, 6$ obeying $\sum_{i=1}^6 X_i^ 2= 1$ and the internal four-sphere in \eqref{MetricAdS7S4} by coordinates $\Theta_a$, $a = 1, \ldots, 5$,  obeying $\sum_{a=1}^5 \Theta_a^2 = 1$.  The line elements are then
 \es{LineElemSphere}{
  ds_{S^5}^2 = \sum_{i=1}^6 dX_i^2 \,, \qquad ds_{S^4}^2 = \sum_{a=1}^5 d\Theta_a^2 \,.
 }
Alternatively, we can view $S^4$ as a circle parameterized by an angle $\vphi = \arg(\Theta_1 + i \Theta_2)$ fibered over a unit three-ball, with the circle shrinking at the boundary of the ball.  The metric and volume form are
 \es{MetricVolumeForm}{
  ds_{S^4}^2 = d\theta^2 + \sin^2 \theta ds_{S^2}^2 + \cos^2 \theta d\vphi^2 \,, \qquad
   \vol_{S^4} = \sin^2\theta\, \cos \theta \, d\theta \wedge \vol_{S^2} \wedge d \vphi  \,,
 }
where $ds_{S^2}^2$ and $\vol_{S^2}$ are the metric and volume forms of a unit radius two-sphere, and $\theta \in [0, \pi/2]$.

In order to perform our desired modification of the background \eqref{MetricAdS7S4}, we should develop a more thorough understanding of its symmetries.  The bosonic symmetries are $\mf{so}(6, 2) \times \mf{so}(5)$ isometries represented by $28 + 10 = 38$ Killing vectors $v = v^M \partial_M$.  Of particular importance will be the Killing vector $\partial_t$ generating translations in $t$ as well as the generators $u_{ij}$ and $w_{ab}$ of $\mf{so}(6)$ and $\mf{so}(5)$, respectively,
 \es{CartansSO6}{
  u_{ij} = X_i \frac{\partial}{\partial X_j} - X_j \frac{\partial}{\partial X_i}  \,, \qquad
   w_{ab} = \Theta_a \frac{\partial}{\partial \Theta_b} - \Theta_b \frac{\partial}{\partial \Theta_a} \,.
 }

The fermionic symmetries of the background \eqref{MetricAdS7S4}, which complete the $\mf{so}(6, 2) \times \mf{so}(5)$ bosonic symmetries into the supergroup $\mf{osp}(8^*|4)$, correspond to the solutions of $\delta \psi_M = 0$.  These equations have 32 linearly independent solutions for the Killing spinors $\epsilon$.  They can be written as
 \es{eps}{
  \epsilon(x) = N(x) \eta \,,
 }
where $\eta$ is an arbitrary 32-component constant spinor and $N(x)$ is a specific position-dependent matrix.  The spinors $\epsilon$ transform under $\mf{so}(6, 2) \times \mf{so}(6)$ as $({\bf 8},{\bf 4})$, and thus a convenient basis in this 32-dimensional space is given by simultaneous eigenspinors under the Cartan generators of $\mf{so}(6, 2) \times \mf{so}(5)$.  In particular, we label the spinors by the eigenvalues under 
 \es{CartanChoice}{
  \text{Cartans of $\mf{so}(6, 2)$}&: \quad \partial_t,\ u_{12},\ u_{34},\ u_{56} \,, \\
  \text{Cartans of $\mf{so}(5)$}&: \quad w_{12},\ w_{34}  \,.
 }

Note that when $\delta \psi_M = 0$, Eq.~\eqref{LinGravitino} can be used to write the Lie derivative of $\epsilon$ with respect to a Killing vector $v^M$ as
 \es{LieDer}{
  {\cal L}_v \epsilon 
   &= \left[ v^M D_M  + \frac 14 \partial_M v_N \gamma^{MN}  \right] \epsilon \\
   &= \left[ -\frac{v^M}{288} \left(\gamma^{PQRS}{}_M 
   - 8 \gamma^{QRS} \delta_M^P  \right) F_{PQRS}  + \frac 14 \partial_M v_N \gamma^{MN}  \right] \epsilon
    \equiv -M_v(x) \epsilon \,,
 } 
where $M_v(x)$ is a position-dependent matrix that depends on the Killing vector $v^M$.  Because the Killing spinors have the form \eqref{eps}, it follows that $\tM_v \equiv N^{-1}(x) M_v(x) N(x)$ is position-independent, and the eigenvalue equation ${\cal L}_v \epsilon(x) = \lambda_v \epsilon(x)$ becomes $\tM_v \eta = \lambda_v \eta$.  Here, $\lambda_v$ denotes the eigenvalue with respect to the Killing vector $v$.

The Killing vectors in \eqref{CartanChoice} are normalized such that the possible eigenvalues of the spinors are $\pm \frac{i}{2}$.  Using the matrices $\tM_v$ derived as above, an explicit computation shows that the eigenvalue $\lambda_{\partial_t}$ is given by
 \es{lamProd}{
   \lambda_{\partial_t} = -4\lambda_{u_{12}} \lambda_{u_{34}} \lambda_{u_{56}} \,.
 } 
Thus, we have a total of 32 possible choice for the remaining eigenvalues, as we can arbitrarily specify $(\lambda_{u_{12}}, \lambda_{u_{34}}, \lambda_{u_{56}}, \lambda_{w_{12}}, \lambda_{w_{34}} )$ and then determine $\lambda_{\partial_t}$ from \eqref{lamProd}.  We can thus label the Killing spinor with eigenvalues $(\lambda_{u_{12}}, \lambda_{u_{34}}, \lambda_{u_{56}}, \lambda_{w_{34}}, \lambda_{w_{12}} ) = \frac{i}{2} (s_1, s_2, s_3, s_4, s_5)$ as $\epsilon^{s_4 s_5}_{s_1 s_2 s_3}$, where the $s_i = \pm$.  When $s_1 s_2 s_3 = -1$, then $\lambda_{\partial_t} = -i/2$ and these spinors correspond to $Q$ generators, and when $s_1 s_2 s_3 = 1$, then $\lambda_{\partial_t} = i/2$ and these spinors correspond to $S$ generators.

\subsubsection{Euclidean background}
\label{EUCLIDEAN}
 
We would now like to modify the Lorentzian background of the previous section in such a way that 1) the $t$ direction becomes a periodic circle parameterized by $\tau$, and 2) the background preserves half the supercharges.  These two requirements are achieved if after the Euclidean continuation, we also perform a twist that makes half the Killing spinors $\tau$-independent.  This condition is ensured by the relation:
 \es{tTotau}{
  \partial_\tau = -i (\partial_t  + w_{12}) \,.
 }
Equivalently, if we write $w_{12} = \partial_\vphi$ for some angular coordinate $\vphi$, then we can replace $t$ and $\vphi$ in \eqref{MetricAdS7S4} everywhere with $-i\tau$ and $\vphi - i \tau$, respectively.  This change of variables does not affect the symmetries of the background, but it does make certain Killing spinors independent of $\tau$:  in particular, the $Q$-type generators with $s_5 = +$ are independent of $\tau$, and so are the $S$-type generators with $s_5 = -$.  In terms of $\tau$, the metric and four-form can be obtained from \eqref{MetricAdS7S4}:
 \es{AdS7Metric2}{
  ds^2 &=  L^2 \left[ (\cosh \rho)^2 d\tau^2 + d\rho^2 +  (\sinh \rho)^2 \sum_{i=1}^6 dX_i^2  \right] 
   + \frac {L^2}{4} \left[ d\theta^2 + \sin^2 \theta ds_{S^2}^2 + \cos^2 \theta (d\vphi - i d\tau)^2 \right] \\
 F_4 &= \frac{3L^3}{8} \sin^2\theta\, \cos \theta \, d\theta \wedge \vol_{S^2} \wedge (d \vphi - i d\tau) \,,
 }
where we used the parameterization \eqref{MetricVolumeForm} of the internal four-sphere.  For the gauge potential $A_3$, we can choose a gauge in which we define it separately on two different patches as
 \es{A3}{
  A_3 = \begin{cases}
    \frac{L^3}{8} \sin^3 \theta\, \vol_{S^2} \wedge (d\vphi - i d\tau) \,, &\text{on patch excluding $\theta = \pi/2$} \,, \\
     \frac{L^3}{8} \sin^3 \theta\, \vol_{S^2} \wedge (d\vphi - i d\tau) - \frac{L^3}{8} \vol_{S^2} \wedge d\vphi \,, &\text{on patch excluding $\theta =0$} \,.
    \end{cases}
 }

We now make $\tau$ compact by imposing the identification $\tau \sim \tau + 2 \pi R_6$.\footnote{Here, we work in units in which the radius $R$ of the five-sphere is set to $R=1$.}  This identification breaks half of the supersymmetries, and it preserves the other half.  In particular, it preserves the supersymmetries generated by the $\tau$-independent  $\epsilon$'s mentioned above:  the $Q$-type generators with $s_5 = +$ and the $S$-type generators with $s_5 = -$:
 \es{epsList}{
  \text{$Q$'s}:\quad &\epsilon^{++}_{++-}, \epsilon^{++}_{+-+}, \epsilon^{++}_{-++}, \epsilon^{++}_{---}, \epsilon^{-+}_{++-}, \epsilon^{-+}_{+-+}, \epsilon^{-+}_{-++}, \epsilon^{-+}_{---} \,, \\
  \text{$S$'s}:\quad &\epsilon^{--}_{--+}, \epsilon^{--}_{-+-}, \epsilon^{--}_{+--}, \epsilon^{--}_{+++}, \epsilon^{+-}_{--+}, \epsilon^{+-}_{-+-}, \epsilon^{+-}_{+--}, \epsilon^{+-}_{+++} \,.
 }
These are the fermionic generators of $\mathfrak{su}(4|2)$.

\subsubsection{Killing spinor}

In Section~\ref{18BPS}, we defined the supercharge \eqref{lsc}.   This supercharge corresponds to the Killing spinor 
 \es{epsChoice}{
 \epsilon = \epsilon_{+-+}^{-+} + \epsilon_{-+-}^{+-} + \epsilon_{-++}^{++} + \epsilon_{+--}^{--}  \,.
 }
While the Killing spinors $\epsilon^{s_4 s_5}_{s_1 s_2 s_3}$ are uniquely determined by the corresponding eigenvalue equations up to overall normalization factors, the relative factors in \eqref{epsChoice} are determined by the condition that the spinor $\epsilon$ obeys the Majorana condition in 11d Lorentzian signature, $\epsilon^\dagger = \epsilon^T i C \gamma^0$, and that it is invariant under the isometries generated by
 \es{Inv}{
  &u_{12} + w_{34} \,, \qquad u_{13} + w_{35} \,, \qquad u_{23} + w_{45} \,, \\
  &u_{12} + u_{34} \,, \qquad u_{13} - u_{24} \,, \qquad u_{23} + u_{14} \,,
 }
in accordance with the commutation relations \eqref{Qsym} satisfied by the field theory supercharge~$\cQ$.

\subsection{Reduction to type IIA}
\label{IIA}

In the limit of small $R_6$, it may be useful to also consider the type IIA reduction of the 11d background presented in the previous section.   If we take the 10d gravitational constant to be related to the 11d one via $\kappa_{11}^2/\kappa_{10}^2 =2 \pi \ell_s g_s =  2 \pi L R_6$ (i.e.~we compactify on a circle of circumference $2 \pi \ell_s g_s = 2 \pi L R_6$), where $\ell_s = \sqrt{\alpha'}$ is the string length and $g_s$ is the string coupling, then the type IIA string frame metric is
 \es{metricIIA}{
   ds^2 &= L^2 \sqrt{\cosh^2 \rho - \frac{\cos^2 \theta}{4}} \Biggl[  d\rho^2 +  \sinh^2 \rho\, ds_{S^5}^2  
   + \frac 14 \left(  d\theta^2 + \sin^2 \theta ds_{S^2}^2 \right) + \frac{1}{4} \frac{\cosh^2 \rho \cos^2 \theta}{\cosh^2 \rho - \frac{\cos^2 \theta}{4}} d\vphi^2  
   \Biggr] \,.
 }
The type IIA dilaton $\phi$ as well as the R-R one-form gauge potential $A_1$, the R-R three-form gauge potential $A_3$, and the NS-NS two-form gauge potentials are given by
 \es{OtherFields}{
  e^{\phi} &= \left( \cosh^2 \rho - \frac{\cos^2 \theta}{4} \right)^{3/4}  \,, \\
  A_1  &= -i    \frac{L \cos^2 \theta}{ 4 \cosh^2 \rho -\cos^2 \theta}d\vphi \,, \\
  A_3 &= \frac{L^3}{8} \sin^3 \theta\, \vol_{S^2} \wedge d\vphi  \,, \\
  B_2 &= -i \frac{L^2}{8} \sin^3 \theta\, \vol_{S^2} \,.
 }

As we can see, this background does not contain an AdS factor, so it does not describe a conformal field theory;  it describes ${\cal N} = 2$ SYM on $S^5$.  The background is smooth everywhere, but it becomes strongly coupled in the UV, at large $\rho$, so in the type IIA duality frame we cannot reliably describe the small $R_6$ behavior.  Note also that the isometry of the internal part of this background is only $\mf{su}(2) \times \mf{u}(1)$, matching the R-symmetry of ${\cal N} =2$ SYM on $S^5$.

\subsection{Minimal surfaces and calibration}

\subsubsection{General Setup}
 \label{GENSETUP}

In Lorentzian signature, for an M2-brane with worldvolume $\cM$ parameterized by coordinates $\sigma^m$, $m=0, 1,2$, the action is 
 \es{M2braneLor}{
  S^\text{Lor} = \tau_{\text{M2}} \left[ -\int_\cM d^3\sigma\, \sqrt{-g} + \int_\cM A_3 \right]  \,,
 }
where $g$ is the determinant of the induced metric on the worldvolume of the brane
 \es{gFromG}{
  g_{mn} = \partial_m x^M \partial_n x^N G_{MN} \,,
 }
and $\tau_\text{M2} = \frac{1}{(2 \pi)^2 \ell_p^3}$ is the M2-brane tension. (The dimensionless combination $\tau_\text{M2} L^3=\frac{2N}{\pi}$ in field theory variables---see~\eqref{LNRelation}.)  Using \eqref{LinGravitino}, it can be checked that the action \eqref{M2braneLor} is invariant under the supersymmetries generated by Killing spinors $\epsilon$ obeying
 \es{SUSYCondition}{
  -\frac{1}{6 \sqrt{-g}} \epsilon^{mnp} \partial_m x^M \partial_n x^N \partial_p x^P \gamma_{MNP} \epsilon = \epsilon \,, 
 }
with $\epsilon^{012} = 1$.

For a Euclidean M2-brane embedding, one has to continue \eqref{M2brane} to Euclidean signature.  If the coordinates parameterizing the brane worldvolume are  $\sigma^m$, with $m = 1, 2, 3$ 
 \es{M2brane}{
  S = \tau_{\text{M2}} \left[  \int_\cM d^3\sigma\, \sqrt{g} - i \int_\cM A_3 \right] \,.
 }
This action is invariant under
 \es{SUSYConditionEuc}{
  -\frac{i}{6 \sqrt{g}} \epsilon^{mnp} \partial_m x^M \partial_n x^N \partial_p x^P \gamma_{MNP} \epsilon =  \epsilon  \,,
 }
with $\epsilon^{123} = 1$.   From now on we work with Euclidean embeddings.

To explore the consequences of supersymmetry, let us multiply Eq.~\eqref{SUSYConditionEuc} by $\epsilon^\dagger$ on the left: 
\es{SUSYCond}{
  -\frac{i}{6 \sqrt{g}} \epsilon^{mnp} \partial_m x^M \partial_n x^N \partial_p x^P  \epsilon^\dagger \gamma_{MNP} \epsilon =  \epsilon^\dagger \epsilon \,.
 }
Dividing this relation by the $\epsilon^\dagger \epsilon$ and multiplying it by $\sqrt{g}$, we obtain
  \es{SUSYCond2}{
   \frac{1}{6} \epsilon^{mnp} \partial_m x^M \partial_n x^N \partial_p x^P  J_{MNP} =  \sqrt{g} \,, \qquad
  J_{MNP} = -i \frac{\epsilon^\dagger \gamma_{MNP} \epsilon}{ \epsilon^\dagger \epsilon}   
   \,.
 } 
The relation \eqref{SUSYCond2} then implies that we can compute the volume of the manifold $\cM$ by simply integrating $J$:
 \es{Volume}{
  \int_{\cM} \sqrt{g} = \int_{\cM} J \,.
 } 
Therefore, the Euclidean action for a supersymmetric M2-brane is
 \es{EucActionFinal}{
  S = \tau_{\text{M2}}  \int_\cM  (J -i  A_3)  \,.
 }

In fact, one can show that the integral of $J$ over the manifold $\cM$ always provides a bound on its volume.  Indeed, because the matrix $\gamma_\cM \equiv -\frac{1}{6 \sqrt{-g}} \epsilon^{mnp} \partial_m x^M \partial_n x^N \partial_p x^P \gamma_{MNP}$ that multiplies $\epsilon$ on the LHS of \eqref{SUSYConditionEuc} squares to the identity matrix, as can be easily checked, we must have that for any surface $\cM$, 
 \es{gammaSigmaIneq}{
 1 \geq \abs{\frac{\bar \epsilon \gamma_\cM \epsilon}{\bar \epsilon \epsilon}}  
  = \abs{\frac {\epsilon^{mnp}  \partial_m x^M \partial_n x^N \partial_p x^P  J_{MNP} }{6\sqrt{g}} } \,,
 }
which implies that
 \es{Ineq}{
  \Vol(\cM) = \int_\cM d^3 x\, \sqrt{g} \geq \abs{\int_\cM J} \,.
 }
The inequality \eqref{Ineq} is thus saturated when $\cM$ is a BPS (or anti-BPS) surface.   If we can then also show that $J$ is a closed 3-form (as will be the case for us), then $J$ is a calibration.

\subsubsection{Explicit formulas}
\label{EXPLICIT}

We are interested in surfaces $\cM$ that are located at $  X_5 = X_6 =\Theta_1 = \Theta_2= 0$ (and hence $\theta = \pi/2$) and wrap the $\tau$ direction.  Such surfaces would be described by
 \es{ParamSurfaces}{
  \tau = \sigma^3\,,\qquad 
  X_i(\sigma^1, \sigma^2)\,, \qquad \Theta_a(\sigma^1,\sigma^2) \,, \qquad \rho(\sigma^1, \sigma^2) \,,
 }
with $i = 1, \ldots, 4$ and $a = 3, \ldots, 5$ with the constraints $\sum_{i=1}^4 X_i^2 = \sum_{a=3}^5 \Theta_a^2 = 1$.  In other words, these surfaces lie within a product between an $S^3 \subset S^5$ in spacetime and $S^2 \subset S^4$ in the internal space.  Thus, in computing the form $J$ introduced in \eqref{SUSYCond2} we can restrict ourselves to the space $  X_5 = X_6 =\Theta_1 = \Theta_2= 0$.

Within this space, plugging in the Killing spinor $\epsilon$ given in \eqref{epsChoice} into the definition of $J$ in \eqref{SUSYCond2}, we find
 \es{omega}{
   J =  L^3 d\tau \wedge \left[ \frac 12 d\left( \sinh^2 \rho\, \eta \right) + \frac {1}4 \vol_{S^2} \right]  \,, 
 }
where the one-form $\eta$ and the two-form $\vol_{S^2}$ are
 \es{Goteta}{
  \eta = \Theta^a X^i dX^j (\eta_a)_{ij} \,, \qquad
   \vol_{S^2} = \Theta_3 d\Theta_4 \wedge d\Theta_5 + \Theta_4 d\Theta_5 \wedge d\Theta_3 + \Theta_5 d\Theta_3 \wedge d\Theta_4\,. 
 }
One can check that the form \eqref{omega} is closed, so it is a calibration (on the space $X_5 = X_6 =\Theta_1 = \Theta_2= 0$).

For supersymmetric M2-brane embeddings we have
 \es{M2braneFinal}{
  S = S^\text{I} +S^\text{II}  \,, \qquad
    S^\text{I} =  \tau_\text{M2}  \int_\cM J \,, \qquad
     S^\text{II} = - \tau_\text{M2} \frac{L^3}{8} \int_\cM \vol_{S^2} \wedge d\tau\,,
 }
 where $\Sig$ is the M2-brane world volume anchored on the cutoff surface at constant $\rho=\rho_c$.\footnote{We think of this cutoff surface as a probe M5-brane as in the setup of Ref.~\cite{Maldacena:1998im}. Because the M2-brane has a nontrivial profile in the internal space, the M5-brane has to also have a nontrivial shape in the internal directions so that the M2-brane can end on it. We leave for future work an analysis of whether this probe M5-brane can preserve supersymmetry.\label{probeM5foot}} 
The quantity $S^\text{II}$ is simply equal to  the area of the projection of the bulk surface onto the internal two-sphere. In the absence of self-intersections of the M2-brane, this area is equal to the area (with signs) of the region on $S^2$ that is enclosed by the boundary curve $\Theta^\text{(bdy)}_{a} (\sigma_1) \equiv \Theta_a(\sigma_1, \sigma_2(\rho_c,\sig^1))$, which is measured by a Wess-Zumino action $S_\text{WZ}[\Theta^\text{(bdy)}] $.  Thus, we write
 \es{S2}{
  S^\text{II} =  2 \pi \tau_\text{M2} R_6 \frac{L^3}{8} S_\text{WZ} [\Theta^\text{(bdy)}_{a}]  \,.
 }

In order to find the precise shape of the supersymmetric surface, one needs to solve a set of first order equations that can be shown to imply the second order equations obtained by varying \eqref{M2braneFinal}.   Following a similar derivation of the first order equations as the one presented in \cite{Dymarsky:2006ve}, first notice that \eqref{omega} implies 
 \es{GJJRelation}{
  G^{\tau\tau} J_{\tau M}{}^N J_{\tau N}{}^P = - \delta_M^P \,.
  }
Then, for the embedding~\eqref{ParamSurfaces}, let us work in conformal gauge where $ \sqrt{g} g^{mn} = \sqrt{g_{33}} \delta^{mn}$, with $m, n = 1,2 $ and $\sqrt{g} = g_{11} \sqrt{g_{33}} = g_{22} \sqrt{g_{33}} $ and $g_{12} = 0$.  Let us define
 \es{PositiveQuantity}{
  {\cal P} &= \frac{2 \pi L R_6}{4}  \int d^2 \sigma\, \sqrt{G_{\tau\tau}} G_{MN} \left( a^M a^N + b^M b^N \right) \,, \\
    a^M &\equiv \partial_1 x^M - \sqrt{G^{\tau\tau}} J_\tau{}^M{}_P \partial_2 x^P   \,, \qquad
    b^M \equiv  \partial_2 x^M +\sqrt{G^{\tau\tau}} J_\tau{}^M{}_P \partial_1 x^P  \,.
 }
The quantity ${\cal P}$ obeys ${\cal P} \geq 0$ by virtue of the M2-brane embedding being a Riemannian manifold.  Expanding out the expression for ${\cal P}$ and using \eqref{GJJRelation}, we see that ${\cal P} \geq 0$ is equivalent to 
  \es{PAgain}{
   {\cal P} = \frac{2 \pi L R_6}{2}
    \int d^2 \sigma \,  \sqrt{G_{\tau\tau}} G_{MN} 
     \left[ \partial_1 x^M \partial_1 x^N + \partial_2 x^M \partial_2 x^N \right] - \int_\cM J \geq 0 \,.
  } 
The first term in \eqref{PAgain} is nothing but $\Vol(\cM)$ in conformal gauge, so \eqref{PAgain} is equivalent to $\Vol(\cM) \geq \int_\cM J$.  As we showed in Section~\eqref{GENSETUP},  BPS M2-brane embeddings saturate this inequality.  But from the definition of ${\cal P}$ in \eqref{PositiveQuantity} we see that the inequality is saturated if and only if $a^M = b^M = 0$ pointwise.  These conditions can be equivalently rewritten as $G_{MN}a^N = G_{MN} b^N = 0$, or more explicitly,
  \es{FirstOrder}{
  J_{\tau MN} \frac{\partial x^N}{\partial \sigma^1} &= - \sqrt{G_{\tau\tau}} G_{MN} \frac{\partial x^N}{\partial \sigma^2} \,, \\
  J_{\tau MN} \frac{\partial x^N}{\partial \sigma^2}  &=   \sqrt{G_{\tau\tau}} G_{MN} \frac{\partial x^N}{\partial \sigma^1} \,.
 }
These are the first order equations obeyed by the BPS M2-brane embeddings.

The equations \eqref{FirstOrder} would in general have to be solved numerically.  It will be useful, however, to also have an expansion near the boundary of AdS.  Assuming that $\sigma^2  = 0$ is the boundary, we can write
 \es{ansatzBoundary}{
  X_i(\sigma^1, \sigma^2) &= X_i^{(0)}(\sigma^1) + (\sigma^2)^2 X_i^{(2)}(\sigma^1) + \cdots \,, \\
  \Theta_a(\sigma^1, \sigma^2) &= \Theta_a^{(0)}(\sigma^1) + (\sigma^2)^2 \log \sigma^2\, \Theta_a^{(2L)}(\sigma^1) + (\sigma^2)^2 \Theta_a^{(2)}(\sigma^1) + \cdots \,, \\
  e^{\rho(\sigma^1, \sigma^2)} &= \frac{2 \rho^{(0)}}{\sigma^2} + \rho^{(2)} \sigma^2 + \cdots \,.
 }
Solving \eqref{FirstOrder}, we obtain
 \es{Solns}{
  \Theta_a^{(0)} &= \eta_a{}^{ij} \frac{  X_i^{(0)} \dot X_j^{(0)}}{\abs{\dot X^{(0)}}} \,, \\
  \rho^{(0)} &= \frac{1}{\abs{\dot X^{(0)}}} \,, \\
  X_i^{(2)} &= \frac{\abs{\dot X^{(0)}}^2}{4} X_i^{(0)} - \frac{3 \dot X^{(0)} \cdot \ddot X^{(0)}}{4 \abs{\dot X^{(0)}}^2}  \dot X_i^{(0)}
   + \frac 14 \ddot X^{(0)} \,, \\
  \rho^{(2)} &= \frac{ \dot X^{(0)} \cdot \dddot X^{(0)} }{3  \abs{\dot X^{(0)}}^3} 
   + \frac{\ddot X^{(0)} \cdot \ddot X^{(0)} }{2  \abs{\dot X^{(0)}}^3}  - \frac{\left(  \dot X^{(0)} \cdot \ddot X^{(0)} \right)^2 }{3  \abs{\dot X^{(0)}}^5}  \,,
 }
etc.
Note the appearance of 't Hooft symbols $\eta_a^{ij}$ defined in \eqref{etaDef} in the the first equality above. To leading order at large $e^\rho$, the RHS of the first equation in \eqref{Solns} is just the supersymmetric scalar coupling matrix $\cS$ in \eqref{Smthooft} that defines our ${1\over 8}$-BPS Wilson loops in 5d MSYM, contracted with the tangent vector to the loop.

Finally, the cutoff surface  should  be regarded as a probe M5-brane, and the M2-brane ending on it determines the shape of the Wilson loop. Correspondingly, we are holding $X^\text{(bdy)}_{i}(\sig^1)=X_{i}(\sig^1,\sig^2(\rho_c,\sig^1))$ fixed and not  $X_i^{(0)}(\sig^1)$. We can express $X_i^{(0)}$ in a series form (the expansion parameter being $e^{-2\rho_c}$) by first determining $\sig^2(\rho,\sig^1)$ from the last equation of \eqref{ansatzBoundary}, and then solving the equation $X^\text{(bdy)}_{i}(\sig^1)=X_{i}(\sig^1,\sig^2(\rho_c,\sig^1))$ perturbatively using the first equation of  \eqref{ansatzBoundary}.  We found it convenient to first express everything in term of $X_i^{(0)}$, and then re-expand the results to get everything in term of $X^\text{(bdy)}_{i}$.

\subsection{Type IIA perspective}

Instead of considering the M2-brane embeddings wrapping the $\tau$ circle, one can equivalently consider a fundamental string worldsheet $\cM'$ in the type IIA background presented in Section~\ref{IIA}.   In Euclidean signature, the fundamental string action is 
 \es{ActionString}{
  S = \frac{1}{2 \pi \alpha'} 
   \left[ \int_{\cM'} d^2 \sigma \, \sqrt{g} 
    - i \int_{\cM'} B_2 \right] \,.
 } 
This action is identical to \eqref{M2brane} for M2-branes wrapping the $\tau$ circle, as can be checked from the relations $\ell_s g_s = L R_6$ and $\ell_p = \ell_s g_s^{1/3}$, which imply 
 \es{tauM2Rel}{
  2 \pi L R_6 \tau_{M2} = \frac{1}{2 \pi \alpha'} \,.
 } 

Further specializing to worldsheets located at $X_5 = X_6 = 0$ and $\theta = \pi/2$, we find from \eqref{metricIIA}--\eqref{OtherFields} that the string worldsheet moves in a 6d ambient space with metric
 \es{Metric6d}{
   ds_{6d}^2 &= L^2 \cosh \rho  \left[  d\rho^2 +  \sinh^2 \rho\, ds_{S^3}^2  
   + \frac 14 \sin^2 \theta ds_{S^2}^2  \right]
 }
and $B$-field
 \es{BField6d}{
  B_2 = -i \frac{L^2}{8} \vol_{S^2} \,.
 }

The reduction of the calibration three-form \eqref{omega} to this space is a calibration two-form
 \es{CalibReduction}{
  J^{(2)} =  L^2 \left[ \frac{1}{2} d (\sinh^2 \rho \, \eta) + \frac 14 \vol_{S^2} \right] \,,
 }
and $J^{(2)}_M{}^N$ is equal to an almost complex structure on the 6d space \eqref{Metric6d}:  indeed, we have $J^{(2)}_{M}{}^{N} J^{(2)}_{N}{}^{P} = - \delta_M^P$.  The action of a calibrated string worldsheet can be written as
 \es{stringAction}{
  S = S^{\text{I}} + S^\text{II} \,, \qquad 
  S^\text{I} =  \frac{1}{ 2\pi \alpha'} \int_{\cM'} J^{(2)} \,, \qquad
     S^\text{II} = - \frac{1}{2 \pi \alpha'} \frac{L^2}{8} \int_{\cM'} \vol_{S^2} \,,
 }
which precisely equals \eqref{M2braneFinal} upon using \eqref{tauM2Rel}.

The first order equations obeyed by the calibrated string worldsheet are just an equivalent way of writing~\eqref{FirstOrder}:
 \es{FirstOrderIIA}{
  J^{(2)}_{MN} \frac{\partial x^N}{\partial \sigma^1} &= -G^\text{6d}_{MN} \frac{\partial x^N}{\partial \sigma^2} \,, \\
  J^{(2)}_{MN} \frac{\partial x^N}{\partial \sigma^2}  &=   G^\text{6d}_{MN} \frac{\partial x^N}{\partial \sigma^1} \,,
 }
 where now $G^\text{6d}_{MN}$ is the 6d metric in \eqref{Metric6d}.  These are the equations of pseudo-holomorphic curves in the complex structure $J^{(2)}_M{}^N$.

\subsection{Area from calibration} \label{HoloArea}

In this section we compute the renormalized area of the 1/8-BPS M2-brane, which matches the proposal \eqref{Prediction1}.  First, we evaluate the regularized bulk action with the M2-brane anchored on the cutoff surface at $\rho=\rho_c$. To get a finite result, we have to use holographic renormalization, and we find that a minimal scheme gives a precise match with \eqref{Prediction1}. Finally, we comment on the absence of the Graham-Witten anomaly \cite{Graham:1999pm} for our BPS M2-branes. 

\subsubsection{The regularized bulk action}

In \eqref{M2braneFinal} we decomposed the regularized M2-brane action $S^\text{(reg)} $ into two terms: $S^\text{(reg)} = S^\text{I} + S^\text{II} $, where $S^\text{I}$ comes from the calibration 3-form integrated over the M2-brane,  $S^\text{II}$ is the contribution of  the $\int_\Sig A_3$ term. After performing the trivial integral of $d\tau$ over the M-theory circle, $S^\text{I}$ involves two terms, one coming from the form $d \tau \wedge d\le(\sinh^2\rho \, \eta\ri)$ while the other from $d \tau \wedge \vol_{S^2}$---see~\eqref{M2braneFinal} and \eqref{omega}. The second term equals $-2 S^{\text{II}}$, while the first term has to be evaluated by explicit computation.\footnote{Since the term $d \tau \wedge d\le(\sinh^2\rho \, \eta\ri)$ in $S^{\rm I}$ is exact, it reduces to an integral on the cutoff surface at $\rho=\rho_c$. Though the divergent part of \eqref{SM2} as we send $\rho_c\to \infty$ is obvious from the integral of $\eta$ on the boundary curve, to extract the subleading finite part in \eqref{SM2} requires the near-boundary asymptotic expansion \eqref{ansatzBoundary}--\eqref{Solns} of the solution to \eqref{FirstOrder}.} To perform the computation, we follow the strategy outlined in the last paragraph of Section~\ref{EXPLICIT}. While the intermediate results look extremely complicated, there are amazing simplifications occuring when converting from  $X_i^{(0)}$ to $X^\text{(bdy)}_{i}$. The final result is:\footnote{In terms of quantities on the asymptotic boundary, the result is more complicated, namely
\es{SM2B}{
S^\text{(reg)}_\text{M2}={\pi R_6 \tau_\text{M2} L^3\ov 4}\le[\int d\sig^1 \ \le[\abs{\dot{X}^{(0)}}
 \le(e^{2\rho_c}-\le(X^{(0)}{}''\ri)^2-1\ri)-2{d\ov d\sig^1}\le(\dot{X}^{(0)}\cdot \ddot {X}^{(0)} \ov  \abs{\dot{X}^{(0)}}^3\ri)\ri]+S_\text{WZ}[\Theta^{(0)}]\ri]\,,
}
where for any quantity $f(\sigma^1)$, we used the notation $f'\equiv {\dot{f}\ov  \abs{\dot{X}}}$ for the reparameterization invariant derivative. (For example, $X''={1\ov  \abs{\dot{X}}} {d\ov d \sigma^1} {\dot{X}\ov  \abs{\dot{X}}}$.)
Note that the total derivative term is not  invariant under the reparameterizations of the boundary curve. }  
\es{SM2}{
S^\text{(reg)}_\text{M2}={\pi R_6 \tau_\text{M2} L^3\ov 4}\le[\int d\sig^1 \ \abs{\dot{X}^\text{(bdy)}} \le(e^{2\rho_c}-2\ri)+S_\text{WZ}[\Theta^\text{(bdy)}]\ri]\,.
}
We would like to show that this quantity equals a constant independent of the shape\footnote{By shape dependence, we mean  potential changes of the Wilson loop observables under continuous deformations of the underlying link $\cL$ of knots $\cK_i$ such that no crossing happens. In knot theory literature, this is the notion of ambient isotopy.} of the boundary plus local counterterms that can be removed by the regularization procedure.

%

\subsubsection{The counter term action}

The next step in computing physical quantities in QFT is to add to the regularized quantity $S^\text{(reg)} $ the contribution of local counter terms, $S_\text{counter}$. Holographic renormalization proceeds by constructing $S_\text{counter}$ on the cutoff surface $\rho=\rho_c$. This is usually done in the lower dimensional $AdS$ supergravity theory after reducing on the internal manifold ($S^4$ in our case), because the higher dimensional perspective on holographic renormalization is somewhat underdeveloped (see, however, \cite{Taylor:2001pp,Taylor:2001fe}). In our case it is easy to see that $-{\tau_\text{M2}\ov 2} \text{Area}(\p \Sig)$ cancels the divergent contribution to the M2-brane action, where following the philosophy explained above, by $\text{Area}(\p \Sig)$ we only mean the area projected onto the $AdS$ directions and neglect the motion of the M2-brane on $S^4$:\footnote{If we took into account the motion on $S^4$ as well the answer in \eqref{SCT1} would instead take the form
\es{SCT1B}{
-{\pi R_6 \tau_\text{M2}L^3\ov 4}\int d\sig^1 \ \abs{\dot{X}^\text{(bdy)}} \le(e^{2\rho_c}+\frac12 \le(\Theta^\text{(bdy)}{}'\ri)^2\ri)\,,
}
where the extra $\frac12 \le(\Theta^\text{(bdy)}{}'\ri)^2$ term is a finite local counter term, and can be cancelled without any difficulty as we explain in Appendix~\ref{app:loccount}.
}${}^,$\footnote{We note that imitating the Legendre transformation prescription for $AdS_5\times S^5$ of Ref.~\cite{Drukker:1999zq}, we obtain the same counter term as in \eqref{SCT1}.  We write the background \eqref{AdS7Metric2} in coordinates
\es{DGOCoords}{
ds^2&=\cosh^2 \rho d \tau^2 + \sinh^2 \rho d\Omega_5^2 + {DY^a DY^a \ov 4 Y^2}\,,\\
Y^a&\equiv e^{-2\rho} \Theta^a\, \qquad DY^a=dY^a-A^{ab}Y^b\,, \qquad A^{12}=-A^{21}=-id\tau\,,
}
and write the counter term $\int d\sig^1 P_a Y^a$, which evaluates to the same result as \eqref{SCT1}. (Choosing different coordinates would lead to a result that differs by finite counter terms.)
Unlike Ref.~\cite{Drukker:1999zq}, we do not have a brane construction that reproduces our background, and we do not have a string-duality-based derivation of this prescription either.  \label{foot:Legendre}  } 
\es{SCT1}{
S_\text{counter}&=-{\tau_\text{M2}L^3\ov 2} \text{Area}(\p \Sig)\\
&=-{\pi R_6 \tau_\text{M2}L^3\ov 4}\int d\sig^1 \ \abs{\dot{X}^\text{(bdy)}} e^{2\rho_c}\,.
}
In addition to the divergent counterterm \eqref{SCT1}, we have the freedom of adding some finite local counter terms; this freedom is  analyzed thoroughly in Appendix~\ref{COUNTER}. 


\subsubsection{The renormalized action}

Adding together $S_\text{M2}$ and $S_\text{counter}$, we obtain:
\es{SM3}{
S_\text{M2}^\text{(ren,min)}&\equiv S_\text{M2}+S_\text{counter}= {\pi R_6 \tau_\text{M2} L^3\ov 4}\le(-2\int d\sig^1 \ \abs{\dot{X}^\text{(bdy)}} +S_\text{WZ}[\Theta^\text{(bdy)}]\ri)\\
&={ \beta N}\le(-L({\cal K})+{S_\text{WZ}[\Theta^\text{(bdy)}]\ov 4\pi} \ri)\,,
}
where in the second line we used $ \tau_\text{M2} L^3={2N\ov \pi}$ and $\beta=2\pi R_6$. To ease the notation, we drop the superscript (bdy) in the rest of the section.

The Wess-Zumino term is the area of the projection of the M2-brane to the internal $S^2$. One way to relate it to various geometric quantities related to the Wilson loop is to take its variation under geometric deformations of the loop. A well-known property of the Wess-Zumino action is that while it cannot be written as a local covariant expression, its variation gives  a local term. The Wess-Zumino term can be written as:
\es{WZExpl}{
S_\text{WZ}[\Theta]&=\int d^2\sig\ \frac12 \ep^{ijk}\ep^{IJ} \Theta_i \p_I \Theta_j \p_J \Theta_k\,.
}
Its variation can be written as:
\es{dWZExpl}{
\de S_\text{WZ}[\Theta]&=\int d^2\sig\  \ep^{ijk}\ep^{IJ} \Theta_i \p_I \de \Theta_j \p_J \Theta_k
=\int d\sig^1 \ \ep^{ijk} \Theta_i  \de \Theta_j \p_1 \Theta_k\,,
}
where we repeatedly used the fact that  $\de x$ lies in the plane of  $(\p_1  \Theta,\, \p_2  \Theta)$ and hence \\$\ep_{ijk}\ep^{IJ} \de \Theta_i \p_I  \Theta_j \p_J \Theta_k=0$.  Using that $\Theta_a = \eta_a{}^{ij} \frac{  X_i \dot X_j}{\abs{\dot X}}$ (see \eqref{Solns}), \eqref{dWZExpl} can be written in a reparameterization-invariant form as\footnote{We note that writing everything in terms of reparameterization invariant derivatives requires care, as for example $\de(X')\neq (\de X)'$.  Instead, $\de (X') = \de \frac{\dot X}{\abs{\dot X}}
 = \frac{\de \dot X}{\abs{\dot X}} - \dot X \frac{(\de \dot X) \cdot \dot X}{\abs{\dot X}^3}
  = (\delta X)' - X' \left[  (\delta X)'\cdot X' \right]$.}
\es{WZVariation}{
\de S_\text{WZ}[\Theta]&=-\int ds \ \abs{\dot{X}} \ep_{ijk}\de \Theta_i \Theta_j \Theta'_k
=\int d\sig^1 \ \abs{\dot{X}} \le(\de X' \cdot X'+ \epsilon_{ijkl} \de X'_i X_j X'_k X''_l\ri)\,.
}
This quantity can be written in terms of the variation of the length $L(\cK)$ and torsion $T_\text{FS}(\cK)$ of the loop $\cK$, where the index FS on the torsion means that it is computed in the Frenet-Serret frame reviewed in Appendix~\ref{app:curvegeom}.\footnote{We use the torsion $T_\text{FS}(\cK)$ in  the Frenet-Serret frame for its explicit integral form. Torsion $T(\cK)$ in a general frame differs from $T_\text{FS}(\cK)$ by an integer.}   The variations of these two quantities are 
\es{GeomVariation}{
2\pi \de  L(\cK)&=  \int d\sig^1\ \de \abs{\dot{X}}=\int d\sig^1 \ \abs{\dot{X}}  \,\le(\de X'\cdot X'\ri) \,, \\
2\pi \de  T_\text{FS}(\cK)&= \int d\sig^1\  \de\le(\abs{\dot{X}} \tau_\text{FS}\ri)= \int d\sig^1 \ \le(\abs{\dot{X}} \, \epsilon_{ijkl} \de X'_i X_j X'_k X''_l+\text{(tot. der.)}\ri)\,.
}
Combining \eqref{GeomVariation} with \eqref{WZVariation}, we learn that
\es{WZRelation}{
\de \left[ {S_\text{WZ}[\Theta]\ov 4\pi} - \frac{L({\cK})+T_\text{FS}({\cK})}2\right] = 0\,.
}

Eq.~\eqref{WZRelation} means that the quantity in the square brackets remains unchanged under continuous deformations of the link:\footnote{The condition is that the curvature $\kappa$ does not vanish, so that the Frenet-Serret frame is well-defined.} 
\es{WZRelationB}{
{S_\text{WZ}[\Theta]\ov 4\pi}-\frac12\le( L(\cK)+ T_\text{FS}(\cK)\ri)&=-{p_\text{FS}\ov 2}\,,
}
where $p_\text{FS}$ is a constant associated to the family of links that can be deformed into each other. Below we will argue that $p_\text{FS}$ is in fact an integer.
 
Using the formulas \eqref{Expr} and \eqref{Relations1}, we can rewrite 
the 2nd term on the LHS of \eqref{WZRelationB}
in a more suggestive way that only involves $\Theta$
 \es{LTKrel}{
 L(\cK)+ T_\text{FS}(\cK)&={1\ov 2\pi}\int d\sig^1 \ \abs{\dot{X}}\le(1+\tau_\text{FS}\ri)
 ={1\ov 2\pi}\int d\sig^1 \ \abs{\dot{X}}\le(1-{\epsilon_{ijkl} {X_i} X_j' X_k'' X_l'' \ov(X'')^2 -1}\ri)\\
 &\hspace{-0.6cm}\stackrel{\text{(BPS loop)}}{=}-{1\ov 2\pi}\int d\sig^1 \ \abs{\dot{X}}\,{{\epsilon_{abc} {\Theta_a} \dot \Theta_b \ddot \Theta_c \ov \abs{\dot{\Theta}}^3}}
 =-{1\ov 2\pi}\int ds \ \abs{\dot{\Theta}} \kappa_g(\Theta)
 \equiv-K[\Theta]\,,
 }
 where we used the standard formula for the geodesic curvature of a curve on a surface $\kappa_g(\Theta)={{\epsilon_{abc} {\Theta_a} \dot \Theta_b \ddot \Theta_c \ov \abs{\dot{\Theta}}^3}}$. Then \eqref{WZRelationB} becomes
 \es{WZRelationC}{
 {S_\text{WZ}[\Theta]\ov 4\pi}+\frac12K[\Theta]=-{p_\text{FS}\ov 2}\,,
 \qquad (p_\text{FS}\in\Z)\,,
 }
 which can be thought of an extension of the Gauss-Bonnet theorem\footnote{It is a generalization of the usual Gauss-Bonnet theorem in the sense that the bounding curve is an immersed (as opposed to embedded) closed curve on the target $S^2$. Unlike an embedded curve, an immersed curve can have self-intersections.}  that associates a topological invariant to the map $\Theta$
 and the integrality of $p_{\rm FS}$ follows from \cite{arnold,rogen1998gauss}. 
 
 Recall that the map $\Theta(\sigma_1,\sigma_2)$ captures how the M2-brane moves in the internal $S^2\subset S^4$.
The value of the constant $p_\text{FS}$ depends on the topology of the map $\Theta$ that extends the boundary values $\Theta^\text{(bdy)}(\sigma_1)$ at $\sigma_2=0$. 
As usual, given a fixed $\Theta^\text{(bdy)}$, ${S_\text{WZ}[\Theta]\ov 4\pi}$ has an integer ambiguity under different choices of extensions, 
which will shift $p_\text{FS}$ by an even integer in \eqref{WZRelationC}
($p_\text{FS}{\,\rm mod}\,2$ is independent of the extension).
 We will not attempt to determine $p_\text{FS}$ here given a general boundary knot $\cK$. Instead we will compute $p_\text{FS}$ for  some examples in the next section.  
 
 Plugging  \eqref{WZRelationC} back into \eqref{SM3} we obtain for a general link
\es{WHolo1}{
 \la W^\text{(min)}(\cL)\ra =\lambda^{\frac12(L({\cal L})- T({\cal L}))}\,\lambda^{{p\over 2} } \,.
}
We emphasize that $p$ by construction is a topological invariant of the knot $\cL$ that also depends on a choice of framing. 
In the FS frame, we have $p=p_{\rm FS}$ and  $T(\cL)=T_{\rm FS}(\cL)$.  In writing \eqref{WHolo1}, we have used the fact that $T(\cL)-p$ is  framing independent in order to write the answer in a frame-independent way.\footnote{This follows from \eqref{WZRelationB} since $S_{\rm WZ}[\Theta]$ does not depend on the choice of framing. The latter is in turn a consequence of the invariance of the $\cS$ matrix under twisted $\mf{so}(4)$ rotation $ \mf{so}(3)_l \times \mf{so}(3)_{\rm diag}$ in \eqref{SO3s}  and \eqref{tSO3r} on $S^3$.}

\subsection{Comparison with Chern-Simons theory}

\subsubsection{General remarks}

In the holographic result \eqref{WHolo1}, we see that the first factor includes all the shape dependence in terms of local geometric quantities, which can be absorbed in the definition of the Wilson loop observable, $W^\text{ren}(\cL)\equiv \lambda^{-\frac12(L({\cal L})- T({\cal L}))} W^\text{(min)}(\cL)$. Remarkably, the minimal holographic renormalization scheme agrees with the field theory scheme used in \eqref{WrenDef}. 

 In Appendix~\ref{app:loccount} we analyze the local counterterms allowed by the six-dimensional origin of the theory, and conclude that the torsion term cannot be uplifted to a local counterterm in 6d.  Consequently, the coefficient of the torsion term in the expression for the non-renormalized loop is universal.   The fact that the coefficient obtained in our holographic computation matches that in Chern-Simons theory gives a strong check of our proposal that the $1/8$-BPS Wilson loops are described by Chern-Simons theory.   While the coefficient of the length term is not fixed by these considerations, the likely
 explanation for its match with field theory is that the minimal holographic renormalization scheme is the only one preserving supersymmetry.

For the renormalized Wilson loop operator $W^\text{ren}(\cL)$, our holographic result \eqref{WHolo1} implies
\es{WHolo2}{
 \la W^\text{ren}(\cL)\ra =\lambda^{{p\over 2} }\qquad (p\in\Z)\,,
}
which is a framed topological invariant of $\cL$.  Comparing to the field theory prediction \eqref{Prediction1}, we see that we indeed get $\lambda^{{1\over 2} }$ raised to an integer power, and we see that both \eqref{WHolo2} and \eqref{Prediction1} have the same framing dependence (because both $p(\cL)$ and ${\rm sl}(\cL)$ have the same framing dependence as the torsion $T(\cL)$). We do not know a general formula for $p$ given an arbitrary loop $\cL$ on $S^3$,
 but based on the subsequent special cases analyzed, it is reasonable to conjecture that 
\es{pValue}{
p=1+ {\rm sl}(\cL) +  \zeta(\cL) \,.
}
Below we gather evidence for this equality in various examples.  In the examples below, all formulas are given in the  Frenet-Serret frame. Converting to other frames is straightforward.

\subsubsection{Latitude loop and match with the literature}

From now on we will parametrize the loops with $t$ instead of $\sig^1$. Let us take a latitude loop given in embedding coordinates by:
\ie
  X = \begin{pmatrix}  a  \cos  t &  a \sin t & \sqrt{1-a^2} &  0 \end{pmatrix}  \,,   \label{LatLopp}
  \fe 
  for some $0<a\leq1$. Its image in the internal $S^2$ is
  \es{LatImage}{
 \Theta =\begin{pmatrix} 
  \sqrt{1-a^2}\cos t & \sqrt{1-a^2}\sin t & -a \end{pmatrix} \,.
 }
  This latitude loop is an unknot $\cU$.  The HOMFLY polynomial for any unknot is $\cH({\cU})=1$, and its self linking number in the FS frame vanishes. Thus, our CS prediction for its expectation value is 
 \es{UnknotExp}{
  \la W^{\rm ren}_{U(\infty)}({\cU})\ra =\lambda^{{1\over 2}  } \qquad \Longleftrightarrow \qquad p_\text{FS}=1\,.
}
  
In the holographic approach, we can argue that $p_\text{FS} = 1$ in two ways.  The first way is as follows.  First, notice that for a great circle loop, when $a=1$, the M2-brane sits at the South Pole of the internal $S^2$ and wraps the equator of the spatial $S^3$ for all values of the radial coordinate $\rho$.  For $0 < a<1$, the M2-brane approaches the curve \eqref{LatImage} at the boundary, which is a circle in the Southern hemisphere of $S^3$.  For the M2-brane to minimize its area, its shape in the internal $S^2$ is given by a curve \eqref{LatImage} at any fixed $\rho$ with the parameter $a = a(\rho)$ now a function of $\rho$.  At the deepest point in the bulk (i.e.~smallest value of $\rho$), we should have $a = 1$, so that the tip of the M2-brane is at the South Pole of $S^2$.   This leads us immediately to the conclusion $p_\text{FS}=1$ from the curve counting explanation below \eqref{WZRelationC}.
  
  Another way to argue that $p_\text{FS} = 1$ is to find the shape of the M2-brane numerically by solving the first order equations \eqref{FirstOrder} in this case.  We performed this exercise and compute the regularized on-shell action, obtaining a very precise numerical match with the formula
\es{SLat}{
S^\text{(reg)}_\text{M2}&={\beta N\ov 2}\le[ a e^{2\rho_c}-(a+1)\ri]\,,\\
 \la W^\text{(min)}(\cK)\ra& =e^{-S^\text{(ren,min)}_\text{M2}}=\lambda^{\frac12(a+1)}\,,
}
from which we can read off $p_\text{FS}=1$.
Ref.~\cite{Minahan:2013jwa} also computed the latter result for $a=1$ (in which case the Wilson loop is $1/2$-BPS), and found agreement with the localization answer $\la W^\text{(min)}(\cK)\ra=\lam$ from \cite{Kim:2012qf}.

\subsubsection{Hopf link and large-$N$ factorization}\label{HOPF}

\begin{figure}[!h]
	\centering
	\includegraphics[scale=0.5]{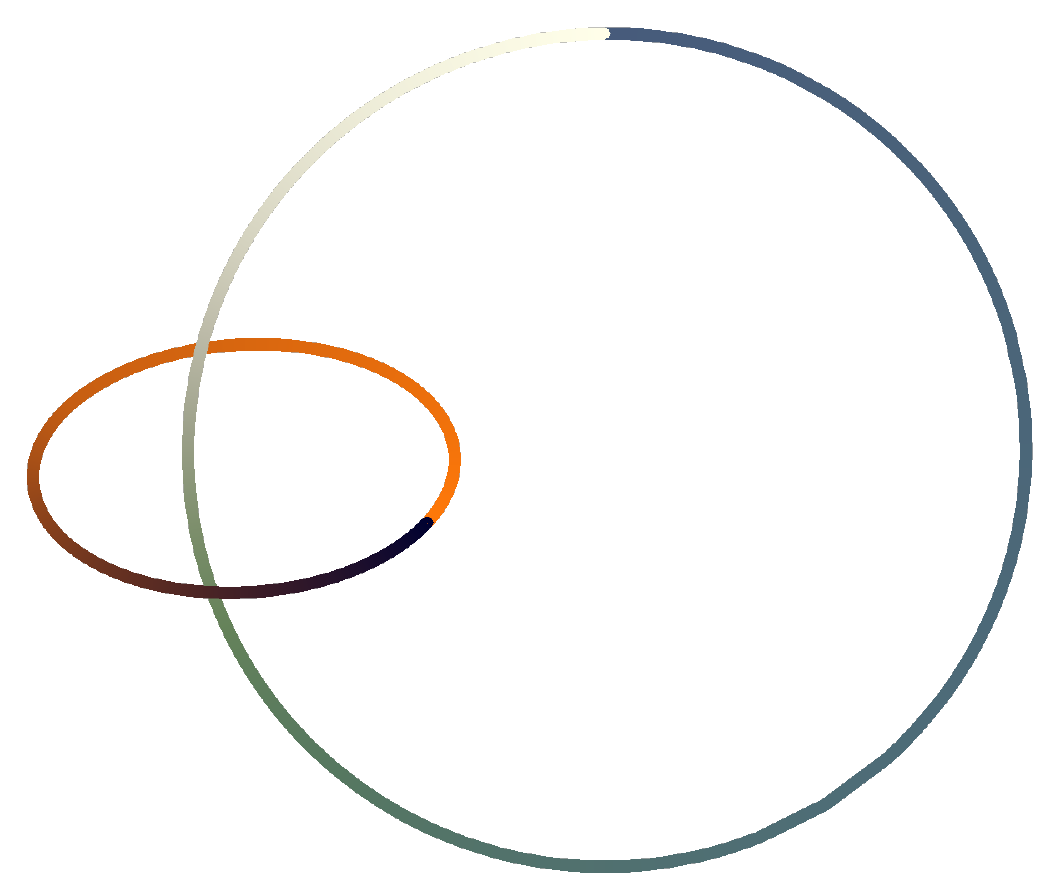}\hspace{2cm}\includegraphics[scale=0.5]{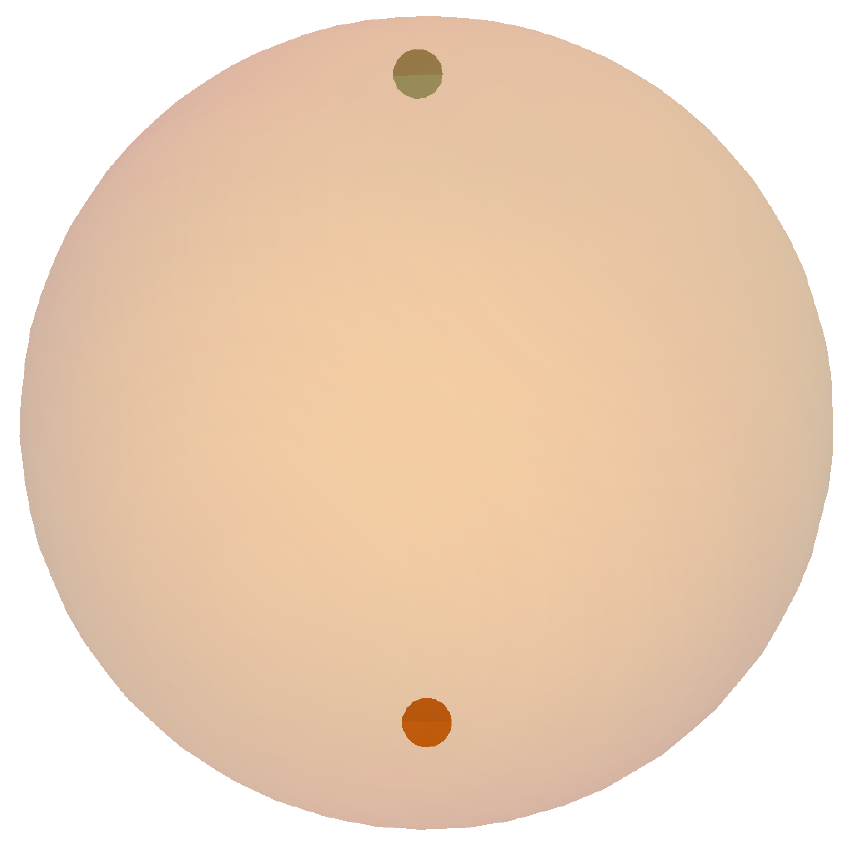}
	\caption{On the left, we plot a pair of Wilson loops in the negative Hopf link configuration on $S^3$ (after stereographic projection). On the right, the two unknots mapped to two points on $S^2\subset S^4$. We interpolate in coloring from light to dark as we go around the  loop in both figures.  The colors are coordinated between the two plots. \label{fig:Hopf}}
	
\end{figure}

Let us think of $S^3$ as the Hopf fibration over a base $S^2$, and take our  next Wilson loop example to be the union of two Hopf fibers located at different base points. Because the Hopf fibers are great circles, their images under $\cS$ are isolated points in the internal $S^2$.\footnote{In this case, these points are just the base points of the fibers, if we identify the $S^2$ base of the Hopf fibration with the internal~$S^2$.}  Let us parametrize the base $S^2$ by $\eta \in [0,\pi/2]$ and $\xi\in [0,2\pi)$, we rescale and identify $t\in [0,4\pi)$ with the fiber coordinate. For every $(\eta,\xi)$, the Hopf fiber
\es{Hopf}{
X = \begin{pmatrix} \sin\eta \cos\le(\xi+t\ov2\ri)
 & \sin\eta \sin\le(\xi+t\ov2\ri) & \cos\eta \cos\le(-\xi+t\ov2\ri) & 
  \cos\eta \sin\le(-\xi+t\ov2\ri) \end{pmatrix} 
}
has an image on $S^2$ given by:
\es{Hopf2}{
\Theta= \begin{pmatrix} \sin2\eta \cos \xi
  & \sin2\eta \sin \xi  & \cos2\eta \end{pmatrix} \,.
 }
See Figure~\ref{fig:Hopf} for an illustration. 
 
As explained above, the dual M2-brane corresponding to each fiber is a $\text{disk}\times S^1$ in $AdS_7$ and a point in $S^2$. While the projection of the M2-branes to $AdS_7$ intersect at $\rho=0$, the M2-branes are disjoint in the full spacetime, as they sit at different points in $S^2$, as can be seen from \eqref{Hopf}. Then the holographic answer trivially factorizes for the Hopf link:
\es{Hopflink}{
\la W^\text{ren}({\rm Hopf})\ra=\la W^\text{ren}(\text{great circle})\ra^2=\lam
}
corresponding to $p=2$,
which matches the Chern-Simons prediction \eqref{Prediction1} because
\es{HopflinkQFT}{
1+ {\rm sl}_\text{FS}({\rm Hopf}) +  \zeta({\rm Hopf})=1+2{\rm lk}(\cK_1,\cK_2)+3=2\,,
}
where we used that in the FS frame the self-linking numbers vanish for each fiber, the linking number of the two fibers is ${\rm lk}(\cK_1,\cK_2)=-1$, and $\zeta({\rm Hopf})=3$.\footnote{The HOMFLY polynomial for the negative Hopf link is
\ie
\cH({\rm Hopf})=z^{-1} \lambda  ({\lambda^{1/2}-\lambda^{-1/ 2} })-\lambda^{1/2} z \,.
\fe
}

In general, whenever the bulk M2-branes corresponding to all disjoint knots in a link are also disjoint in the bulk, holography predicts factorization for the expectation value of the Wilson loop in the strong coupling limit.  Then, if the Chern-Simons prediction \eqref{Prediction1} is true, it has to be the case that for $\cL=\cup_i\cK_i$,
\es{Factorization}{
1+{\rm sl}(\cL) +  \zeta(\cL)= \sum_i (1+ {\rm sl}(\cK_i) +  \zeta(\cK_i)) \,,
}
which is an equality that the Chern-Simons prediction \eqref{Prediction1} obeyed in the case of two Hopf links.  As a trivial generalization, this equality holds for multiple unknots linked pairwise as Hopf links. Below we also verify this equality for torus links. The simplest example where we found that \eqref{Factorization}  does not hold is the Whitehead link (in one orientation).\footnote{
	See Figure~\ref{fig:knotlist} (h) for the Whitehead link. It consists of two unknots with zero linking number. 
	The HOMFLY polynomial for the Whitehead link is
\ie
\cH({\rm Whitehead})=
\frac{\lambda +\lambda  z^4-(\lambda -1)^2 z^2-1}{\sqrt{\lambda } z}
 \,,
\fe
which gives $\zeta =2$ (the mirror Whitehead link has $\zeta=1$).} But it does hold for other non-torus links such as the Borromean rings.  It would be interesting to study the corresponding bulk M2-branes in this and other non-factorizing cases.

\subsubsection{Torus knots and links}

The  torus knot $T_{m,-n}$ with $\gcd(n,m)=1$ on $S^3$ is parametrized, in embedding space, by\footnote{ By $SO(4)$ invariance on $S^3$, we have the equivalence 
	\ie
	T_{m,n}=T_{n,m}  =T_{-m,-n}  
	\fe
On the other hand, under parity $T_{m,n}$ gets mapped to its mirror $T_{m,-n}$. In general for Wilson loop in representation $R$ in a CS theory with gauge group $G$, the expectation values satisfy \cite{Marino:2005sj}
\ie
\la W^{3d}_G(R,\cK) \ra (q,\lambda) =\la W^{3d}_G(R,\cK) \ra (q^{-1},\lambda^{-1})
\fe
where $\overline \cK$ is the mirror (parity transform) of $\cK$ and $q\equiv e^{2\pi i\over k},\lambda\equiv e^{2\pi i h \over k}$ with $h$ the dual Coxeter number of $G$.
	}
  \ie
  X=  \begin{pmatrix} a  \cos (m t) & a  \sin (m t) & \sqrt{1-a^2}  \cos (n t) 
   & \sqrt{1-a^2}   \sin (n t) \end{pmatrix} \,,
  \label{torusknot}
  \fe 
  where  $t\in [0,2\pi)$ and $0\leq a \leq 1$.    
  The first nontrivial example is the well-known trefoil knot, which has $(m,n)=(2,3)$.   The image in $S^2$ of \eqref{torusknot} is:
 \es{TmnImage}{
 \Theta= \begin{pmatrix} A\cos((m-n)t)
  & A\sin((m-n)t) & 
  {(n-(m+n)a^2) \om} \end{pmatrix} \,,
 }
 where $\om\equiv1/\sqrt{m^2a^2+n^2(1-a^2)}>0$ and $A\equiv {(m+n)a\sqrt{1-a^2} \,\om}$. We show a slightly deformed $T_{2,-5}$ with its image in $S^2$ on Figure~\ref{fig:ExampleKnot}.\footnote{We deformed slightly the torus knot in that figure in order for its $S^2$ image to not be a multiply wrapped circle.
  }
   
  The HOMFLY polynomial of the torus knot $T_{m,-n}$  for $n,m> 0$ is\footnote{Note that $ \cH(T_{m,-n})$ is invariant under $m\leftrightarrow n$ although not obvious. See \cite{Giasemidis:2014bfa} for a rewriting that makes this symmetry manifest.}  \cite{Marino:2005sj}
  \ie
  \cH(T_{m,-n})
  =&  { q-1\over \lambda-1}  {(\lambda q^{-1})^{(m-1)(n-1)/2} \over q^m-1} 
  \sum_{\substack{p+i+1=m\\ p,i\geq 0}}
  (-1)^i q^{ni +{1\over 4}(p(p+1)-i(i+1))}
  {\prod_{j=-p}^i (\lambda-q^j)\over [i]![p]!} \,,
  \fe
  where $[i]\equiv q^{i/2}-q^{-i/2}$ and $[i]!=[i][i-1]\cdots[1]$.
 In particular, in the M-theory limit $\lambda \to \infty$, we obtain $\cH(T_{m,-n}) \approx \lambda^{\zeta(T_{m,-n})/2}$ with
   \es{ZetaT}{
  \zeta(T_{m,-n})={mn-\abs{m-n}-1 } \,.
}
Further using the self-linking number in Frenet-Serret frame, ${\rm sl}(T_{m,-n}) =- mn$, we find that the Chern-Simons prediction is
 \es{FinalTorus}{
  \la W^\text{ren}(T_{m, -n})\ra
  &=\lambda^{-{1\over 2}   \abs{m-n}  } \quad \Longleftrightarrow \quad p_\text{FS} =-\abs{m-n} \,.
 }

   Let us compare this to the holographic result. To do that we need the values of the following geometric quantities:
 \es{Quants}{
 L({\cal K})&={1\ov \om}\,, \qquad T_\text{FS}({\cal K})=-{mn \om}\,,
 }
with $\omega$ defined right after \eqref{TmnImage}.   
We also need to evaluate the WZ term, which as discussed in detail below \eqref{WZRelationC} depends on the extension of $\Theta(t)$ into a (topological) disk.

Instead of determining the extension explicitly from the equations \eqref{FirstOrder} satisfied by the M2-brane, we consider the simplest candidate extensions of $\Theta(t)$ in \eqref{TmnImage} that wrap either the southern and northern hemispheres. We will refer to them as $\hat\Theta^{(1,2)}$ to differentiate from the actual M2-brane profile $\Theta(\sigma_1,\sigma_2)$.
These extensions lead to the following integrals for the WZ term
 \es{WZalts}{
 S_\text{WZ}^{(1)}&=\int_a^0 da' \int_0^{2\pi}dt \ \ep_{ijk} \hat\Theta_i^{(1)} \p_t \hat\Theta_j^{(1)} \p_{a'} \hat\Theta_k^{(1)}\,,\\
  S_\text{WZ}^{(2)}&=\int_a^1 da' \int_0^{2\pi}dt \ \ep_{ijk} \hat\Theta_i^{(2)} \p_t\hat \Theta_j^{(2)} \p_{a'}\hat \Theta_k^{(2)}\,,
 }
which evaluate to 
 \es{Quants2}{
 {S^{(1,2)}_\text{WZ}\ov 4\pi}&=\frac12\le({1\ov \om}-{mn \om}\pm (m-n)\ri)\,.
 }
Thus for $m>n$, combining \eqref{Quants2} and \eqref{Quants} with \eqref{LTKrel} and \eqref{WZRelationC}, the simple extension $\hat\Theta^{(1)}$  that wraps the southern hemisphere gives the answer
\es{pdet}{
p_\text{FS}=-(m-n) \,, 
} 
which matches with the field theory prediction  \eqref{FinalTorus}. For $m<n$, the simple extension $\hat\Theta^{(2)}$  that wraps the northern hemisphere 
\es{pdet2}{
p_\text{FS}=(m-n)  
} 
does the job.

It would be interesting to verify from the shape of the M2-brane in the internal $S^2$ whether the simple extensions above are correct.
More precisely, since $p_\text{FS}$ is topological, we only need to see whether $\Theta$ can be continuously deformed into $\hat\Theta^{(1,2)}$.

  A generalization of the torus knots are torus links which we will label by $T_{m,-n}$ with $\gcd (m,n)=r>1$. A torus link consists of $r$ linked torus knots of the type $T_{m/r,-n/r}$ such that the total linking number between each pair of knots is
 \es{LinkNumT}{
  \sum_{i\neq j}{\rm lk}(T^{(i)}_{m/r,-n/r},T^{(j)}_{m/r,-n/r})=-  mn+ {mn\over r} \,.
}
  The simplest example $T_{2,-2}$ is nothing but the negative Hopf link consisting of two linked unknots with linking number $-1$. (The negative Hopf link was analyzed in Section~\ref{HOPF}.)
 The HOMFLY polynomial of a general $T_{m,-n}$ torus link was computed in \cite{Labastida:1993xg,2006math,Stevan:2010jh}. While the HOMFLY polynomial of the torus link is more complicated than for the torus knot, their maximum degrees in $\lambda$ determining the M-theory limit are given by the same expression as \eqref{ZetaT}.
 It is easy to check using this explicit expression and \eqref{LinkNumT} that the expectation value of the Wilson loop forming this link obeys the large $N$ factorization formula \eqref{Factorization},
   \ie
  1+{\rm sl}(T_{m,-n})+\zeta(T_{m,-n})=\sum_{i=1}^r \le(1+ {\rm sl}(T^{(i)}_{m/r,-n/r})+\zeta(T^{(i)}_{m/r,-n/r})\ri)\,.
  \fe  
Another way of saying this is that $p(T_{m,-n})=\sum_{i=1}^r p(T^{(i)}_{m/r,-n/r})$. Note that this equality holds for any framing. It would be interesting to understand in detail the corresponding bulk M2-branes, which we expect to not intersect just like in the Hopf link example that we discussed in Section~\ref{HOPF}.

\subsection{Comments on non-BPS Wilson loops and the Graham-Witten anomaly}
\label{GWANOMALY}

Since the regularized M2-brane action diverges quadratically with the short distance cutoff $\ep\equiv e^{-\rho_c}$, based on the Graham-Witten anomaly \cite{Graham:1999pm} (see also \cite{Berenstein:1998ij}), we could have expected to encounter a logarithmic divergence. The regularized action \eqref{SM2B}, however, does not exhibit such a divergence.

To understand this issue better, we have analyzed non-BPS Wilson loops by solving in an asymptotic expansion the second order equations for minimal area M2-branes that do not follow the appropriate trajectory in the internal $S^4$ to make them BPS\@.  For the divergent terms, we found 
\es{nonBPS}{
S_\text{M2}={\pi R_6 \tau_\text{M2}L^3\ov 4}\int d\sig^1 \ \abs{\dot{X}} \le[{1\ov  \ep^2}+\le(\kappa^2-\Theta'^2\ri)\log \ep+\dots\ri]\,,
}
where $\kappa$ is the curvature of the curve within $S^3$. For a quick review of its definition, and for its expression in terms of $X_i$ see Appendix~\ref{COUNTER}\@. 
From \eqref{Relations1} we see that the logarithmic divergence is absent for the BPS loops we focused on in this paper. The $\kappa^2$ term originates from the Graham-Witten anomaly \cite{Graham:1999pm} for surface operators in the 6d theory.\footnote{On the worldsheet that the bulk M2-brane ends on on $S^5\times S^1$ the Graham-Witten anomaly is the Willmore energy, $S_\text{M2}\supset {\tau_\text{M2}L^3\ov 8}\int d^2\xi \ \sqrt{\ga}\le( K^A_{\al\beta}K^{A,\al\beta} -\frac12 K^{A} K^{A}\ri) \log \ep$.} In fact, it was argued by \cite{Berenstein:1998ij} that because of this logarithmic divergence, the expectation values of surface operators are not well-defined. We see that the BPS loops avoid this conclusion by having a compensating term coming from the scalars.

\section{3d Chern-Simons from  localization}
\label{LOCALIZATION}

In this section, we aim to use supersymmetric localization to derive the 3d CS sector of 5d MSYM on $S^5$ that we conjectured in Section~\ref{CONJECTURE}.  We will proceed by first giving an off-shell formulation of 5d MSYM with gauge group $G$ on $S^5$ that realizes one of the two supercharges $\cQ$ used to define our $1/8$-BPS Wilson loops. The novelty of our choice of $\cQ$ is that it squares to a Killing vector that fixes a great $S^3$ inside $S^5$.\footnote{Recall that for all previous localization computations on $S^5$, the localizing supercharge squares to a Killing vector with no fixed points but rather fixed circles (Hopf fibers over $\mathbb{CP}^2$).}
We  analyze supersymmetric (BPS) equations for the SYM fields with respect to $\cQ$. Without using an explicit reality condition for the fields, we provide evidence that the BPS locus is a certain real slice in the space of complex $G$-connections   on the great $S^3$ weighted by a Chern-Simons action with level \eqref{Gotk}. Moreover the ${1\over 8}$-BPS Wilson loops \eqref{wl} of 5d MSYM descend to familiar Wilson loops in the 3d CS theory as we have conjectured in Section~\ref{CONJECTURE}.  In the end, we will comment on the choice of reality conditions and related issues.

\subsection{5d MSYM with off-shell $\cQ$}

In this section, we find it convenient to write the 5d MSYM theory in terms of a dimensional reduction of 10d fields, whereby we group the 5d gauge field $A_\mu$ and the 5 scalars $\Phi_I$ into an object $A_M$, $M = 1, \ldots, 10$, and we likewise group the four   four-component spinors $\Psi_A$ into a sixteen-component spinor $\Psi$.  We take $A_M$ and $\Psi$ to depend only on the 5d coordinates.  For more details, we refer the reader to Appendix~\ref{app:10dsym}  for the translation between this notation and the 5d notation used in previous sections.  See also Appendix~\ref{app:gamma} for 10d gamma matrices and relevant identities.

To perform SUSY localization, we need an off-shell realization of $\cQ$ (or equivalently $\D_\ve$) for the 5d MSYM action.  The general 5d MSYM action with some off-shell supersymmetry  can be obtained after introducing seven auxiliary fields $K_m$ with $m=1,2,\dots,7$,
\ie
\cL_{\rm YM}=&{1\over 2g_{\rm YM}^2}\tr
\Bigg[
{1\over 2} F_{MN} F^{MN} +\Psi \CC^M D_M\Psi-{1\over 2R}\Psi \Lambda\Psi+{3\over R^2} \Phi^i\Phi_i+{4\over R^2}\Phi^a\Phi_a
\\
&-{2\over 3R}\epsilon^{abc}[\Phi_a,\Phi_b]\Phi_c+K^m K_m
\Bigg] \,,
\label{5dosaction}
\fe
where $a=8,9,10$ and $i=6,7$. More explicitly, \eqref{5dosaction} is invariant under the off-shell supersymmetry transformation $\D_\ve$ given by
\ie
&\D_\ve A_M= i\varepsilon \Gamma_M \Psi  \,,
\\
&\D_\ve\Psi 
=-{i\over 2}F_{MN}\Gamma^{MN}\varepsilon
-{2i\over 5}\Gamma_{\m i}\Phi^i\nabla^\m \ve
-{4i\over 5}\Gamma_{\m a}\Phi^a\nabla^\m \ve
+K^m \n_m \,,
\\
&\D_\ve K^m=-\n^m\Gamma^M D_M\Psi 
+{1\over 2R} \n^m \Lambda\Psi \,,
\label{SUSYos}
\fe 
provided that we choose the pure spinor variables $\n_m$ to satisfy
\ie
\n_m \CC^M \ve=0 \,, \qquad \n_m \CC^M \n_n=\D_{mn} \ve \CC^M \ve \,, \qquad 
\n^m_\A \n^m_\B+\ve_\A \ve_\B={1\over 2} \ve \CC_M \ve {\tilde\CC}^M_{\A\B} \,.
\label{pseq}
\fe
These equations determine $\n_m$ in terms of $\ve$ up to an $\mf{so}(7)$ transformation under which the $\nu_m$ and the $K_m$ transform as a seven-dimensional vector.

From now on, we take $\ve$ to correspond to our supercharge $\cQ$ also introduced in \eqref{lsc}, for which a convenient set of pure spinors is given by\footnote{Here we use hatted indices in $\Gamma_{\hat i}$ and $\tilde\Gamma_{\hat i}$ to denote frame indices so that these gamma matrices are constants.}
\ie
\n_1=&{x_1 \tilde{\CC}_{\hat 1}+x_2 \tilde{\CC}_{\hat 2}\over \sqrt{x_1^2+x_2^2}}{x_1 \CC_{\hat 6} -x_2 \CC_{ \hat7}\over \sqrt{x_1^2+x_2^2}} \ve \,,
\\
\n_i=&{x_1 \tilde{\CC}_{\hat 1}+x_2 \tilde{\CC}_{\hat 2}\over \sqrt{x_1^2+x_2^2}} \CC_{\hat i} \ve \,, \quad i=2,3,4,
\\
\n_j=&{x_1 \tilde{\CC}_{\hat 1}+x_2 \tilde{\CC}_{\hat 2}\over \sqrt{x_1^2+x_2^2}}  \CC_{\widehat{j+3}}\ve \,, \quad j=5,6,7.
\fe

\subsection{BPS configurations}
\begin{figure}
	\centering
	\includegraphics[scale=2]{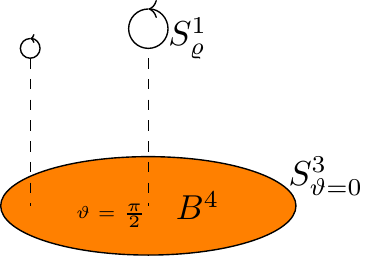}
	\caption{$S^5$ viewed as a fibration of $S^1_\varrho$ over $B^4$. The $S^3$ of interest sits at the boundary of $B^4$ at $\vartheta=0$.}
	\label{fig:s5s3}
\end{figure}

The BPS configurations with respect to $\D_\ve$ are solutions of the equations
\ie
\Psi=\D_\ve\Psi=0 \,.
\fe
Using \eqref{KSEq}, we can write $\D_\ve \Psi$ as the following 16 complex equations:
\ie
&\D_\ve\Psi 
=-{i\over 2}F_{MN}\Gamma^{MN}\varepsilon
+{i\over R}(\tilde\Gamma_{ i}\Phi^i+2\tilde\Gamma_{ a}\Phi^a) \Lambda\ve
+K^m \n_m=0 \,, 
\label{bpseq}
\fe
which we need to solve. To better explain the action of $\cQ^2$, let us write the $S^5$ line element as
 \es{Line}{
ds_5^2=\sin^2 \vartheta d\varrho^2+d\vartheta^2+\cos^2\vartheta  d\Omega_3^2
 }
 (see Appendix~\ref{S5APPENDIX} for details).   The coordinates used in \eqref{Line} make manifest the (singular) fibration of a circle $S^1_\varrho$ over a 4-ball $B^4$ (see Figure~\ref{fig:s5s3}).   The fiber $S^1_\varrho$ achieves maximal size at the center $\vartheta={\pi\over 2}$ of $B_4$,  while the fixed $S^3$ is located at $\vartheta=0$, where the $S^1_\varrho$ shrinks to zero size.  The rotation generator $M_{12}$ in $\D^2_\ve$  acts simplify by translating in the fiber direction.

Instead of working with the explicit components of \eqref{bpseq}, a convenient trick is to first look at the square of the supersymmetry transformation $\D_\ve$ 
\ie
&\D_\ve^2 A_\m=-iv^\n F_{\n\m} + i[D_\m,v^I\Phi_I] \,,
\\
&\D_\ve^2 \Phi_a=-iv^\n D_\n \Phi_a+[\Phi_a,v^I\Phi_I]- \omega_{ab}\Phi^b \,,
\\
&\D_\ve^2 \Phi_i=-iv^\n D_\n \Phi_i+[\Phi_i,v^I\Phi_I]- \omega_{ij}\Phi^j \,,
\\
&\D_\ve^2 \Psi= -iv^N D_N \Psi-{i\over 4} \nabla_\m v_\n \CC^{\m\n} \Psi-{1\over 4} \omega_{IJ} \CC^{IJ}\Psi \,,
\\
&\D_\ve^2 K^m=-iv^M D_M K^m-M^{mn}K_n\,. 
\label{dsq2}
\fe
where 
\ie
&v^\m \pa_\m =-{i\over R}\ve \CC^\m \ve \pa_\m=-{1\over R}\pa_\varrho \,, 
\\
&v^I \Phi_I=-{i\over R}\ve \CC^I \ve  \Phi_I= i \sin\vartheta (\sin \varrho \Phi_7 +\cos \varrho \Phi_6)\,,
\\
&\omega_{ab}={2\over R}\ve \tilde \CC_{ab}\Lambda \ve =0 \,, 
\\
&\omega_{ij}={1\over R}\ve \tilde \CC_{ij}\Lambda \ve =-{i\over R}(\D_{6i}\D_{7j}-\D_{7i}\D_{6j}) \,, 
\\
& M^{mn}\equiv \n^{[m}\CC^\m\nabla_\m \n^{n]}-{1\over 2R} \n^{[m} \Lambda \n^{n]}=0 \,.
\label{dsqpm}
\fe 
These variations in \eqref{dsq} must vanish as a consequence of the BPS equations. We thus obtain the following constraints on the bosonic fields
\ie
& {1\over R} F_{\varrho \m}=  -[D_\m,v^I \Phi_I]   \,, 
\\
& {1\over R}[D_\varrho,\Phi_a]=i[\Phi_a, v^I \Phi_I]  \,, 
\\
& {1\over R}[D_\varrho,\Phi_i]=i[\Phi_i, v^I \Phi_I]-\omega_{ij}\Phi^j  \,, 
\\
& {1\over R}[D_\varrho,K^m]=i [K^m, v^I \Phi_I] \,.
\label{covconst}
\fe
We can define twisted fields
\ie
\hat \Phi_6=& \cos \varrho\Phi_6+\sin\varrho  \Phi_7 \,, \qquad
\hat \Phi_7=\sin \varrho\Phi_6-\cos\varrho  \Phi_7
\fe
and twisted connection
\ie
\cD_\varrho=D_\varrho+iR \sin \vartheta \hat \Phi_6 \,, \qquad \cD_{\vartheta, \zeta,\xi,\Phi}= D_{\vartheta, \zeta,\xi,\Phi} \,.
\fe
Then the equations \eqref{covconst} simply state 
\ie
\cF_{\varrho \m}\equiv i[\cD_\varrho, D_\m]=0 \,,\qquad  [\cD_\varrho, \Phi_a]=0 \,, \qquad [\cD_{\varrho},\hat \Phi_i]=0 \,, \qquad [\cD_\varrho, K^m]=0 \,,
\label{covcssm}
\fe
which implies that all the bosonic fields are covariantly constant along $\varrho$.

The next step is to analyze equation \eqref{bpseq} restricted to $B^4$, for which the reality condition on the fields are crucial and we will comment on this in Section~\ref{sec:reality}.

\subsection{The 3d Chern-Simons action}
Assuming that the smooth BPS configurations are determined by the fields on the boundary $S^3$ of $B^4$, we are ready to derive the action governing the dynamics on the BPS locus.   Using covariance along $\varrho$, the bosonic part of the 5d SYM action becomes
\ie
S={\pi \over  g_{\rm YM}^2} \int_{B^4} d^4 x\sqrt{g_{B^4}} \tr
\Bigg[
{1\over 2} F_{MN} F^{MN}+{3\over R^2} \Phi^i\Phi_i+{4\over R^2}\Phi^a\Phi_a-{2\over 3R}\epsilon^{abc}[\Phi_a,\Phi_b]\Phi_c
+
K^m K_m
\Bigg] \,.
\label{actionB4}
\fe
In Appendix~\ref{App:CSaction}, we show that this induced action on $B^4$ is a total derivative and integrates to an action on the $S^3$ boundary of $B^4$ given by
\ie
S
=&
 { \pi \over  g_{\rm YM}^2} \int_{S^3}  d^3 x\, e^{3\Omega} \Tr \Bigg[
\epsilon^{abc} \cS_c^{\rm i}   (\Phi_a F_{{\rm i}    b} )
+{2\over 3}\epsilon^{ abc }(\Phi_a \Phi_b\Phi_c )
- i e^{-3\Omega}\epsilon^{{\rm i} {\rm j}  {\rm k}   } \cS^a_{\rm i}      (\Phi_a F_{{\rm j k}   } )
\\
&
-{1\over 2} e^{-3\Omega}\epsilon^{\rm ijk}(A_{\rm i}    F_{\rm jk}+ {2\over 3} iA_{\rm i}    A_{\rm j}    A_{\rm k}    )
+2\Phi^a \Phi_a
\Bigg]\,,
\label{actionS3}
\fe
where 
$e^{2\Omega}=1/\left(1+  x_{\rm i}^2/4\right)^2$
 is the usual stereographic conformal factor. Here we have set $R=1$ for convenience  and we will restore $R$ at the end using dimensional analysis.   Note in particular the appearance in \eqref{actionS3} of the $\cS$ matrix defined in \eqref{smat}. 

It is natural to expect that the $S^3$ action \eqref{actionS3} can be rewritten in terms of the twisted connection $\hat A_\mu = A_\mu + i S_{\mu a} \Phi^a$ introduced in \eqref{wl}, which is manifestly $\D_\ve$-invariant. Indeed by using
\ie
e^{3\Omega} \epsilon^{abc} \cS_c^{\rm j}   \cS^{\rm k}_a  \pa_{\rm j}    \cS_b^{\rm i}   \Phi_{\rm i}   \Phi_{\rm k}   
=-\epsilon^{\rm ijk} \cS_{\rm i}   ^b \pa_{\rm j}    \cS^{\rm l}_b \Phi_{\rm k}    \Phi_{\rm l}   
=-2e^{3\Omega} \Phi_{\rm i}   \Phi^{\rm i}   
\fe
and
\ie
{1\over 3}\sqrt{g}\epsilon^{abc}\Phi_a [\Phi_b,\Phi_c] =-{2\over 3}\hat \Phi^3 \,,
\fe
with $\hat \Phi_{\rm i} =\cS_{{\rm i} a}\Phi^a$, the action can be further simplified to
\ie
S=&-{ \pi \over  g_{\rm YM}^2} \int_{S^3} d^3 x   \Tr \Bigg[
-\hat\Phi (d\hat\Phi-2i A\hat\Phi )
- {2\over 3}\hat \Phi^3
+ 2i\hat \Phi F
+
AdA-{2\over 3}iA^3
\Bigg] 
\\
=
&-{ \pi \over  g_{\rm YM}^2} \int_{S^3} d^3 x   \Tr \Bigg[
\hat A d\hat A
-{2\over 3}i\hat A^3
\Bigg] \,.
\label{CSaction}
\fe
As promised, this expression is precisely the 3d Chern-Simons action on $S^3$ at (renormalized) level
\ie
k=i{4\pi^2 R\over g_{\rm YM}^2} =i {R\over R_6}  \,,
\label{CSlvl}
\fe
where we have restored the radius $R$ of $S^5$. 

We emphasize that the action we obtain here should be interpreted as a real CS action with gauge group $G$ and an imaginary level rather than a complex CS action. In the localization computation here, this reality property should arise as a consequence of the reality properties of the bosonic fields $A_M$ that we have not yet specified (see the end of next subsection).  As explained in \cite{Witten:2010cx}, the imaginary level does not need to be quantized and the CS path integral is well-defined on particular contours determined by the gradient flow.\footnote{It would be interesting to make explicit the relation between these contours of \cite{Witten:2010cx} and the contour $\Ga$ inherited from the SUSY localization of the 5d MSYM.
} 
By consistency with the results of \cite{Kim:2012ava}, the path integral contour $\Ga$ from the localization of 5d MSYM on $S^5$  must be such that the partition function 
\ie
Z_{S^5}^\text{pert}=  \int_{\Gamma} D \hat A \,\exp \left[ -{{ i k\over 4\pi}\int \Tr\left( \hat Ad \hat A-{2\over 3}i \hat  A^3\right)}\right] 
\label{5dCSPI}
\fe
correctly reproduces the matrix model in \eqref{ZS5}, which is simply an analytic continuation of the usual CS matrix model. Moreover insertions of the ${1\over 8}$-BPS Wilson loops \eqref{18BPS} on $S^5$ simply correspond to ordinary Wilson loops
\ie
W^{3d}_G(\cK,R)=
{1\over \dim R}\tr_R \cP e^{i
\oint_\cK  \hat A 
}
\fe
inserted in the CS path integral \eqref{5dCSPI}.

We point out that \eqref{5dCSPI} differs from the usual (analytically continued) CS path integral in that the CS coupling $k$ here does not receive further one-loop renormalization. This is a consequence of the difference between the path integral measure $D\hat A$ here and that of the analytically continued CS \cite{Witten:2011zz}.

\subsection{Comments on reality conditions}
\label{sec:reality}

As usual, the idea of supersymmetric localization is to introduce a $\cQ$-exact deformation of the original action,
\ie
S\to S+t \int d^5 x\, \sqrt{g}\, \Tr \D  V\,, \qquad \text{where  } V=(\D_\ve\Psi)^\dag \Psi \,,
\fe
so that for $t\to \infty$, the path integral localizes to solutions of the BPS equation $\D_\ve\Psi=0$. For this to be well-defined, one needs to ensure that the contour of the path integral is such that the localizing term $\D_\ve V$ is $\D_\ve$-closed and has a positive real part.

Here our choice of localizing supercharge $\cQ$ gives \eqref{bpseq}, which can be rewritten as
\ie
\D_\ve\Psi 
=&-{i\over 2}F_{\m\n}\Gamma^{\m\n}\varepsilon
-{1\over 2}[\Phi_I,\Phi_J]\Gamma^{IJ}\varepsilon
-i
D_{\m}\Phi_I\Gamma^{\m I}\varepsilon
+{i\over R}(\tilde\Gamma_{ i}\Phi^i+2\tilde\Gamma_{ a}\Phi^a) \Lambda\ve
+K^m \n_m  \,.
\label{dpsi}
\fe
It is reasonable to assume that the reality conditions of the bosonic fields in MSYM are 
\ie
(A_\m)^\dag= A_\m \,, \qquad  (\Phi_I)^\dag=\Phi_I \,, \qquad (K_m)^\dag= K_m \,,
\fe
which were chosen such that the undeformed action \eqref{5dosaction} has a positive-definite real part.  Using  the conjugation properties of the Killing spinor and auxiliary pure spinors,
\ie
\ve^*=\cC \ve \,, \qquad \nu_{2,3,4}^*=\cC \nu_{2,3,4} \,, \qquad \nu^*_{1,5,6,7}=-\cC \nu_{1,5,6,7} \,,
\fe
where $\cC=C\otimes \hat C$ is the $\mf{so}(5)\times \mf{so}(5)_R$ charge conjugation matrix,\footnote{We emphasize $\cC$ here should not be confused with the charge conjugation matrix for ${\rm Spin}(9,1)$ spacetime symmetry of the 10d SYM in the Lorentzian signature. In terms of the ${\rm Spin}(10)$ chiral spinor representation used here, $\cC$ is represented by a rank-5 gamma matrix.}
we obtain
\ie
-({\D_\ve\Psi })^\dag \cC
=&{i\over 2}F_{\m\n}\ve\tilde\Gamma^{\m\n}
+{1\over 2}[\Phi_I,\Phi_J]\ve\tilde\Gamma^{IJ}
-i
D_\m \Phi_I \ve\tilde\Gamma^{\m I}
+{i\over R}\ve \Lambda(\tilde\Gamma_{ i}\Phi^i+2\tilde\Gamma_{ a}\Phi^a)  
\\
&
-\sum_{m=2,3,4} K_m (\n^m)^T
+\sum_{m=1,5,6,7} K_m  (\n^m)^T \,.
\label{dpsibar}
\fe
Now  the localizing term  
\ie
\D_\ve V=\D_\ve (({\D_\ve\Psi })^\dag  \Psi )= {( \D_\ve\Psi })^\dag \D_\ve  \Psi +{\rm fermionic}
\fe
is by construction positive-definite. The invariance $\int \D^2_\ve V=0$ is also immediate since   the charge conjugation matrix $\cC$, the Killing spinor $\ve$, and the auxiliary spinors $\nu_m$ are all invariant under the bosonic symmetry generators that appear in $\D^2_\ve$:
\ie
\cL_{\partial_\varrho} \ve+{1\over 4}\omega_{ij}\Gamma^{ij}\ve=0 \,, \qquad \cL_{\partial_\varrho} \nu_m+{1\over 4}\omega_{ij}\Gamma^{ij}\nu_m=0 \,.
\fe
One may potentially worry about  the violation of  $\mf{so}(7)$ symmetry for the auxilary fields $K_m$ by the localizing term due to the opposite signs in front of $K_{2,3,4}$ and $K_{1,5,6,7}$ in \eqref{dpsibar}.\footnote{For readers more familiar with the 5d $\cN=1$ localization literature,  $K_{2,3,4}$  here correspond to the $D_I$ auxiliary fields in the 5d $\cN=1$ vector multiplet, and $K_{1,5,6,7}$ are related to the $F_A,\bar F_A$ auxiliary fields in the hypermultiplet \cite{Kim:2012ava}.} In our case, since $\D^2_\ve$ does not induce any $\mf{so}(7)$ rotation, this does not spoil the $\D_\ve$ invariance of the localizing term.

The full set of BPS equations is then given by $\D_\ve \Psi= (\D_\ve \Psi)^\dagger=0$. 
We solve  these 32 equations as follows: 18 of them comes from contractions with $\ve\Gamma^M$ (which are dependent due to the identity \eqref{triality}) and impose that the various fields are covariantly constant; 7 of the equations can be used to determine $K_m$; the remaining 7 equations involve purely the SYM fields and determine their profile in the bulk of $B^4$ in terms of their boundary values on $S^3$.\footnote{Here we focus on the smooth solutions to the BPS equations. There are also singular solutions to the BPS equations $\D_\ve \Psi= (\D_\ve \Psi)^\dagger=0$ with finite classical actions. They are 5d instantons wrapping a great $S^1$ that links the $S^3$ and their contributions to the $S^5$ function are captured by \eqref{Zinst}. 
} 

Demanding  the fields to be smooth everywhere, we find that the BPS equations require the boundary fields  to satisfy
\ie
& \left .F-\hat\Phi\wedge\hat\Phi \right |_{S^3} 
=\left .d_A \hat\Phi \right |_{S^3} \,.
\fe
Consequently the twisted connection is flat
\ie
\hat F=0 \,,
\fe
where the Hodge star is defined with respect to the spherical metric on $S^3$.\footnote{The same constraints arises in the localization computation of Pestun that identifies a 2d Yang-Mills subsector of 4d $\cN=4$ SYM on $S^4$.} 

In other words, with the naive reality condition on the MSYM fields, the BPS locus of is dramatically constrained. In particular for $G=U(1)$, this implies that the CS action we found would vanish on the BPS locus, and our Wilson loops will only have trivial expectation values, which contradicts with what we have found by perturbation theory in Section~\ref{sec:pert}.  This indicates that a refined analysis is needed for the BPS locus, with possibly different reality conditions and/or possible complex saddles taken into account.\footnote{A similar problem happened for the localization computation in \cite{Pan:2017zie} for 4d $\cN=2$ SYM on $S^4$. There the naive reality condition also only gives  trivial solutions to the BPS equations (when fields are assumed to be smooth) and it lead to a contradiction with the expected dependence of the sphere partition function on $g_{\rm YM}$.} We hope to come back to this subtle issue in the future.

\section{Discussion}
\label{CONCLUSION}
To summarize, we have identified a protected sector of ${1\over 8}$-BPS Wilson loop operators in the 5d MSYM theory on $S^5$. They are defined along arbitrary loops that are contained in a great $S^3$ within $S^5$.   Motivated by the results in \cite{Kim:2012ava,Kim:2012qf} for the similarity between the partition functions and circular Wilson loop expectation values in the 5d MSYM theory and the 3d Chern-Simons theory on $S^3$, 
we proposed that the sector of 5d MSYM theory consisting of our more general 5d Wilson loop operators is described by a 3d Chern-Simons theory with the same gauge group as in the 5d MSYM, but a level analytically continued to an imaginary value determined by the 5d gauge coupling.  In particular, the expectation values of our Wilson loops in 5d are topological and compute knot invariants.  Since the 5d MSYM is related by compactification to the 6d $(2,0)$ theory on $S^1 \times S^5$, our Wilson loops lift to a sector of ${1\over 8}$-BPS surface operators in the $(2,0)$ theory, whose correlation functions should also be topological according to our proposal.

We verified this proposal in the weak coupling expansion by explicit Feynman diagram computations in the 5d MSYM on $S^5$. In the strong coupling regime, we considered the large $N$ limit and invoked the holographic description in M-theory. The surface operators are described by probe M2-branes in $AdS_7\times S^4$. The ${1\over 8}$-BPS condition for the surface operator maps to the requirement that the M2-branes wrap calibrated cycles, with the precise calibration form determined by supersymmetry.  Although in general we did not determine the shapes of these calibrated M2-branes, we could nevertheless evaluate the values of the on-shell action.  The results match the prediction from the CS theory description.  Finally, we presented the first steps toward a derivation of the 3d CS sector from supersymmetric localization of the 5d MSYM\@.  Assuming an implicit reality condition, we saw how the 3d CS theory arises as a cohomological sector of the 5d theory, but we did not present a complete proof.   To complete the proof from localization, we would need a better understanding of the admissible reality conditions of the fields of 5d MSYM and of the contributions from any possible complex saddles to the localized partition function. 

The holographic check of our proposal only focuses on the 6d uplift of 5d Wilson loops in the fundamental representation of $SU(N)$ and in that it is valid only at leading order in $1/N$.    It would be interesting to consider 6d surface operators coming from 5d Wilson loops in other representations of the $SU(N)$ gauge group.  For instance, after the uplift to 6d, the Wilson loops in symmetric or antisymmetric tensor product representations of $SU(N)$ correspond to wrapped M5-branes in $AdS_7 \times S^4$ \cite{Camino:2001at, Chen:2007ir, Mori:2014tca}. It would also be interesting to go beyond the leading order in large $N$ by including the effects of the backreaction of the probe M2-branes and quantum corrections.

The $(2,0)$ SCFT is known to have a family of 3d CS sectors which are dual to certain 3d $\cN=2$ gauge theories under 3d-3d dualities \cite{Dimofte:2011py, Dimofte:2011ju, Cordova:2013cea, Lee:2013ida}. Though the CS sectors there were discovered by performing topological twists and the resulting CS theories appear to be quite different from ours, it is not inconceivable that they are related by a deeper structure in the 6d $(2,0)$ theory. 

The 6d $(2,0)$ theory has a protected sector of ${1\over 4}$-BPS operators known as the chiral algebra \cite{Beem:2014kka}. The chiral algebra is defined on a two-plane in the 6d spacetime and was conjectured to coincide with a holomorphic (chiral) W-algebra $\cW_\mf{g}$ associated to an ADE Lie algebra $\mf{g}$. The vacuum character of the chiral algebra can be computed by a particular limit of the $S^1\times S^5$ partition function for the 6d $(2,0)$ theory  and it matches with the known W-algebra character \cite{Beem:2014kka}.  This was refined in \cite{Bullimore:2014upa} by including additional surface defects and codimension-2 defects that preserve the supercharge that defines the chiral algebra. These defects act on the local operators on the chiral algebra plane and the corresponding $S^1\times S^5$ partition function with defect insertions compute characters of certain non-vacuum representations of the chiral algebra  \cite{Bullimore:2014upa}.
 It would be interesting to study how our more general surface defects can further refine the chiral algebra sector. 
 
 Finally, there are additional observables in the 5d MSYM theory on $S^5$ that are mutually supersymmetric with respect to the ${1\over 8}$-BPS Wilson loops on the $S^3$ submanifold. In particular in Appendix~\ref{app:tauwl}, we have introduced a family of ${1\over 4}$-BPS Wilson loops that extend along the $S^1$ fibers over the base $B^4$. Insertions of such Wilson loops in the 5d MSYM path integral would modify the localization computation by introducing point like sources on $B^4$.  This suggests that there exists a generalization of the topological 3d CS sector that we have proposed here for the 5d MSYM theory, given by a certain 4d effective theory with both point and loop operators. It would be very interesting to investigate this construction further.

\subsection*{Acknowledgments}

We thank Hee-Cheol Kim, Joseph Minahan, Satoshi Nawata, Nikita Nekrasov, Wolfger Peelaers, Leonardo Rastelli, and Luigi Tizzano for  useful correspondence and discussions. The research of MM was supported by the Simons Center for Geometry and Physics.  SSP was supported in part by the US NSF under Grant No.~PHY-1820651, by the Simons Foundation Grant No.~488653, and by a Alfred P.~Sloan Research Fellowship.  The work of YW was supported in part by the US NSF under Grant No.~PHY-1620059 and by the Simons Foundation Grant No.~488653.

\appendix

\section{Spinor conventions} 
\label{GAMMACONVENTIONS}
\label{app:gamma}

\subsection{5d spacetime and internal gamma matrices}
Here we record our 5d spinor conventions. In this subsection all indices are taken to be flat.

We denote the 5d spacetime gamma matrices by $\C_\m$ and the internal $\mf{so}(5)_R$ gamma matrices by $\hat\C_I$. They satisfy the usual Clifford algebra
\ie
\{\ga_\m,\ga_\n\}=2\D_{\m\n}1_{4} ,\quad 	\{\hat\ga_I,\hat\ga_J\}=2\D_{IJ}1_{4} \,.
\fe
More explicitly, we choose these gamma matrices to be
\ie
\C_1=\sigma_1\otimes \sigma_1 \,, \quad
\C_2=\sigma_2\otimes \sigma_1 \,, \quad
\C_3=-\sigma_3\otimes \sigma_1 \,,\quad
\C_4=-1_2\otimes \sigma_2 \,, \quad
\C_5=1_2\otimes \sigma_3\,.
\fe
and  
\ie
\hat	\C_1=\sigma_3\otimes \sigma_1 \,, \quad
\hat	\C_2=1_2\otimes \sigma_2 \,, \quad
\hat  \C_3= -\sigma_1\otimes \sigma_1 \,, \quad
\hat  \C_4=-\sigma_2\otimes \sigma_1 \,,  \quad
\hat	\C_5=1_2\otimes \sigma_3 \,.
\fe
Note that these gamma matrices are all hermitian. 
We define the charge conjugation matrix $C$ for the spacetime $\mf{so}(5)$ spinor
\ie
C=i \sigma_2\otimes \sigma_3 \,, 
\fe
which satisfies
\ie
C^2=-1 \,, \quad C^T=-C \,, \quad 
(C\C_\m)^T=-C\C_\m \,, \quad 
(C\C_{\m\n})^T=(C\C_{\m\n})  \,.
\label{SMc}
\fe	
Similarly for $\mf{so}(5)_R$ spinors we have
\ie
\hat C=i \sigma_2\otimes \sigma_3 \,, 
\fe
which satisfy the similar set of conditions as above with internal gamma matrices. 			
The default position of the spinor indices are
\ie
(\C_\m)^\A{}_\B,\quad (\hat\C^I)_A{}^B,\quad C_{\A\B},\quad \hat C^{AB} \,.
\fe

\subsection{10d gamma matrices}
The Euclidean 10d gamma matrices for the chiral spinor of $\mf{so}(10)$  satisfy
\ie
\{\CC^M,\tilde \CC^N \}=2\D^{MN}1_{16}
\fe
and can be chosen to be symmetric $16\times 16$ matrices as
\ie
\CC_\m=i(C\C_\m)\otimes \hat C \,, \quad \CC_I=  -C\otimes (\hat C\hat \C_I) \,, \quad 
\tilde \CC_\m= -i(\C_\m C^{-1})\otimes \hat C^{-1}\,, \quad \tilde \CC_I= - C^{-1}\otimes (\hat \C_I \hat C^{-1}) \,.
\fe
In particular
\ie
\Lambda=-i\Gamma_8 \tilde \Gamma_9\Gamma_{10} =i C\otimes \hat C \hat \C^{12} \,.
\fe
The higher rank Gamma matrices are defined as usual by
\ie
&\CC^{MN}\equiv \tilde \CC^{[M}\CC^{N]}\,,\qquad \tilde\CC^{MN}\equiv \CC^{[M} \tilde\CC^{N]}\,,\qquad 
\CC^{MNP}\equiv  \CC^{[M}\tilde\CC^{N}\CC^{P]}\,,\qquad \tilde\CC^{MNP}\equiv \tilde \CC^{[M}\CC^{N}\tilde\CC^{P]}\,,
\\
&\CC^{MNPQ}\equiv \tilde \CC^{[M}\CC^{N}\tilde\CC^{P}\CC^{Q]}\,,\qquad \tilde\CC^{MNPQ}\equiv \CC^{[M}\tilde \CC^{N}\CC^{P}\tilde \CC^{Q]}\,, 
\\
&
\CC^{MNPQR}\equiv  \CC^{[M}\tilde\CC^{N}\CC^{P}\tilde\CC^{Q}\CC^{R]}\,,\qquad \tilde\CC^{MNPQR}\equiv \tilde \CC^{[M}\CC^{N}\tilde\CC^{P}\CC^{Q}\tilde\CC^{R]}\,.
\fe
In particular $\CC^{MNP},\,\tilde\CC^{MNP} $ are anti-symmetric and $\CC^{MNPQR},\,\tilde\CC^{MNPQR}$ are symmetric. We also have
\ie
(\CC^{MN})^t=-\tilde \CC^{MN}\,,\qquad (\CC^{MNPQ})^t=\tilde \CC^{MNPQ}\,.
\fe

Below we list some useful Gamma matrix identities,
\ie
& \CC^M_{(\A \B}\CC_{M\C)\D}=0
\label{triality}
\fe
and
\ie
&\CC_{PQ}\CC^{MN}=-2\D^M_{[P}\D^N_{Q]}-4\D^{[M}_{[P}\CC_{Q]}{}^{N]}+\CC_{PQ}{}^{MN}\,,
\\
&\CC_M\CC_{NP}=2\D_{M[N}\CC_{P]}+\CC_{MNP}\,,
\\
&\CC_{ABC}\CC_{MN}=\CC_{ABCMN}-3(\D_{N[A}\CC_{BC]M}-\D_{M[A}\CC_{BC]N})+6\D_{M[A}\CC_B \D_{C]N}\,.
\label{10dgammaid}
\fe

\subsection{Explicit $\mf{su}(4|2)$ Killing spinors}

A  basis for the Killing spinors solving \eqref{kseq}, in the stereographic coordinates, takes the form
\ie
\ve  (Q^\A_A)=&
e^{\Omega/2}\left(
1
-{i\over 2r} e^{-\Omega}x^\m \C_\m
\right)\zeta^{(\A)} \otimes \xi_{(A)} \,, 
\\
\ve  (S^\A_A)=& e^{\Omega/2}\left(
1
+{i\over 2r} e^{-\Omega}x^\m \C_\m
\right)C \zeta_{(\A)} \otimes \hat C  \xi^{(A)} \,, 
\fe
where $\zeta$ and $\xi$ are constant spinors of the spacetime $\mf{so}(5)$ rotation symmetry and of $\mf{so}(5)_R$, respectively. More explicitly
\ie
\zeta^{(1)}=\zeta_{(1)}=\begin{pmatrix} 1\\0 \\0 \\0\end{pmatrix} \,, \quad
\zeta^{(2)}=\zeta_{(2)}=\begin{pmatrix} 0 \\1\\0 \\0\end{pmatrix} \,, \quad
\zeta^{(3)}=\zeta_{(3)}=\begin{pmatrix} 0 \\0 \\1  \\0\end{pmatrix} \,, \quad
\zeta^{(4)}=\zeta_{(4)}=\begin{pmatrix} 0 \\0 \\0 \\1 \end{pmatrix} \,,
\fe
and
\ie
\xi^{(1)}=\xi_{(1)}=\begin{pmatrix} 1 \\0 \end{pmatrix} \otimes \begin{pmatrix} 1 \\0 \end{pmatrix} \,, \qquad
\xi^{(2)}=\xi_{(2)}=\begin{pmatrix} 0 \\1 \end{pmatrix} \otimes \begin{pmatrix} 0 \\1 \end{pmatrix} \,.
\fe

\section{Supersymmetry algebras}
\label{ALGEBRAS}

\subsection{6d $(2,0)$ superconformal algebra $\mf{osp}(8^*|4)$}
The 6d $(2,0)$ superconformal algebra $\mf{osp}(8^*|4)$ is generated by
the Poincare supercharges $Q_{\A A}$ and the superconformal charges $S_{B}{}^{\B}$, where $A,B$ are indices for the fundamental representation of $\mf{sp}(4)_R$ symmetry and $\A,\B$ are indices for the chiral and anti-chiral spinor representations of the Lorentz algebra $\mf{so}(5,1)$ (or equivalently the fundamental and anti-fundamental representations of $\mf{sl}(4)$).\footnote{In \cite{Kim:2012ava}, these supercharges are denoted by $Q^{\pm\pm}_{\pm\pm\pm}$ and $S^{\pm\pm}_{\pm\pm\pm}$ subjected to the chiral and anti-chiral constraint respectively for their spacetime spinor representations. Here the $\pm$ are usual 2d spinor indices for the $\mf{so}(2)$ subgroups. For reader's convenience, the translation these notations is 
	\ie
	&{\rm lower}~{\A}: (1,2,3,4)=(++-,+-+,---,-++) \,, \\
	&{\rm upper}~{\B}: (1,2,3,4)=(--+,-+-,+++,+--) \,, \\
	&{\rm lower}~ {A}: {\bf (1,2,3,4)}=(++,-+,--,+-) \,, \\
	&{\rm upper}~ {B}: {\bf (1,2,3,4)}=(--,+-,++,-+)\,.
	\label{indextrans}
	\fe}
The commutation relations are given by
\ie
\{Q_{\A A},Q_{\B B}\}=&\Omega_{AB} P_{\A\B} \,, 
\\
\{S_A{}^\A,S_B{}^\B\}=&\Omega_{AB} K^{\A\B} \,,
\\
\{Q_{\A A},S_B{}^{\B}\}=&\Omega_{AB} (M_\A{}^\B+{1\over 2}\D_\A^\B  H)+\D_\A^\B R_{AB} \,,
\\
[P_{\A\B},K^{\C\D}]=&4\D_{[\B}^{[\C} M_{\A]}{}^{\D]} +\D^{[\C}_{[\A} \D_{\B]}^{\D]} H \,,
\\
[P_{\A\B},M_\C{}^\D]=&2\D_{[\A}^\D P_{\B]\C}+{1\over 2} \D_\C^\D P_{\A\B} \,,
\\
[K^{\A\B},M_\C{}^\D]=&-2\D^{[\A}_\C K^{\B]\D}-{1\over 2} \D_\C^\D K^{\A\B} \,,
\\
[M_\A{}^\B,M_\C{}^\D]=&-2\D^\D_{[\A} M_{\C]}{}^\D \,,
\\
[R_{AB},R_{CD}]=&2\Omega_{A(C}R_{D)B}+2\Omega_{B(C}R_{ D) A}  \,,
\\
[Q_{\A A}, R_{BC}] =& 2 \Omega_{A(B} Q_{\A C) } \,,
\\
[S_{A}{}^{\A}, R_{BC}] =& 2 \Omega_{A(B} S_{C)}{}^{\A} \,,
\\
[H, Q_{\A A}] =& \frac 12 Q_{\A A}  \,,
\\
[H, S_A{}^\A] =& -\frac 12 S_A{}^\A \,,
\fe
where $\Omega_{AB}$ is the skew-symmetric invariant tensor of $\mf{sp}(4)_R$ with $\Omega_{\bf 13}=\Omega_{\bf 42}=1$.\footnote{Here we adopt the natural convention that identifies the 6d charge conjugation matrix $\Omega_{AB}$ with the 5d charge conjugation matrix $\hat C_{AB}$.} At is useful to represent these commutation rules using the oscillator representation \cite{Gunaydin:1984wc} in terms of four pairs of fermionic oscillators $c_\A,\tilde c^\B$, and four   bosonic oscillators $a_A$ which satisfy
\ie
\{c_\A,\tilde c^\B\}=\D_\A^\B \,, \qquad
[a_A,a_B]=\Omega_{AB} \,,
\fe
and then
\ie
&Q_{\A A}=c_\A a_A,\quad S_A{}^\A=\tilde c^\A a_A \,,
\\
&P_{\A\B}=c_\A c_\B,\quad K^{\A\B}=\tilde c^\A \tilde c^\B \,,
 \quad M_\A{}^\B=c_\A \tilde c^\B-{1\over 4} \D^\B_\A c_\C\tilde c^\C \,,
  \quad H={1\over 2}c_\A\tilde c^\A \,,
\\
&R_{AB}=a_{(A} a_{B)} \,.
\fe

\subsection{5d supersymmetry algebra $\mf{su}(4|2)$}
\label{SU42DETAILS}

The superalgebra $\mf{osp}(8^*|4)$ contains a maximal subalgebra
\ie
\mf{osp}(8^*|4) \supset \mf{su}(4|2) \oplus \mf{u}(1) \,.
\fe
Each embedding of the $\mf{su}(4|2)$ algebra into $\mf{osp}(8^*|4)$ can be specified by a choice of the $\mf{u}(1)$ generator that commutes with $\mf{su}(4|2)$.  As mentioned in the main text, for the choice 
\ie
H-{R_{\bf 13}+R_{\bf 24}\over 2} \,, 
\fe
the generators of  $\mf{su}(4|2) $  are:
\ie
\{M_\A{}^\B; \ H-(R_{\bf 13}+R_{\bf 24}); \ R_{\bf 13}-R_{\bf 24}, R_{\bf 12}, R_{\bf 34}; \ Q_{\A {\bf 1} }, Q_{\A \bf{2}}; \ S^\A{}_{\bf{3}}, S^\A{}_{\bf{4}} \} \,,
\fe
or after raising the indices with $\Omega^{AB}$ with  $\Omega^{\bf 13}=\Omega^{\bf 42}=1$,
\ie
\{M_\A{}^\B; \ H-(R_{\bf 1}{}^{\bf 1}+R_{\bf 4}{}^{\bf 4}); \ R_{\bf 1}{}^{\bf 1 }-R_{\bf 4}{}^{\bf4}, R_{\bf 1}{}^{\bf 4}, R_{\bf 4}{}^{\bf 1} ; \ Q_{\A {\bf 1} }, Q_{\A \bf{4}}; \ S^{\A\bf{1}}, S^{\A\bf{4}} \} \,.
\fe	
In the same order as written above, they are the $\mf{so}(6)\times \mf{u}(1)_R \times \mf{su}(2)_R$ generators as well as the supersymmetry generators respectively. We can now think about the $\bf {1,4}$ indices as the $\mf{su}(2)_R$ doublet indices. 

In the main text, for notational convenience we sometimes denote the $\mf{so}(6)$ rotation and $\mf{su}(2)_R\times \mf{u}(1)_R$ generators differently by $M_{ij}$, $R_{ab}$, and $R_{12}$ respectively, making manifest the vector indices.  The relation between the two notations is as follows.   For the $\mf{so}(6)$ rotation generators, we have 
\ie
M_{ij}=& i(\tau_{ij}){}_\A{}^\B M_\B{}^\A \,,
\fe
where $\tau_{ij}=\tau_{[i} \tilde \tau_{j]}$ and $\tau_{i},\tilde \tau_{j}$ with $i,j=1,2,\dots,6$ are $\mf{so}(6)$ gamma matrices in the chiral and anti-chiral bases respectively. In particular
\ie
M_{12}=& M_1{}^1 +M_2{}^2-M_3{}^3-M_4{}^4  \,,
\\
M_{34}=& M_1{}^1 -M_2{}^2-M_3{}^3+M_4{}^4  \,,
\\
M_{56}=& -M_1{}^1 +M_2{}^2-M_3{}^3+M_4{}^4 \,.
\fe
Similarly for the $\mf{su}(2)_R\times \mf{u}(1)_R$ generators, we have
\ie
R_{12}=    R_{\bf 1}{}^{\bf 1}+R_{\bf 4}{}^{\bf 4} \,,\quad 
R_{34}= R_{\bf 1}{}^{\bf 1 }-R_{\bf 4}{}^{\bf 4} \,, \quad 
R_{35}=  iR_{\bf 1}{}^{\bf 4 }- iR_{\bf  4}{}^{\bf 1 }  \,, \quad
R_{45}=   R_{\bf 1}{}^{\bf 4 }+ R_{\bf 4}{}^{\bf 1 }  \,. 
\fe

 \section{Supersymmetric Wilson loops along the fiber}
 \label{app:tauwl}
 In addition to the Wilson loops discussed in the main text that are defined on $S^3$ transverse to Killing vector field $v^\m$ in $\D_\ve^2$,  there are also supersymmetric Wilson loops along the direction of  $v^\m$.\footnote{It would be interesting to systematically classify the supersymmetric loop operators in 5d MSYM following \cite{Dymarsky:2009si}.} In other words, the Wilson loops lie along the $S^1$ fibers over the base $B^4$. In particular, they are invariant under a transverse $\mf{so}(4)$ rotation on $B^4$.
 
 These Wilson loops take the form
 \ie
 W_R(\cK)={1\over \dim R} \tr_R  \cP\exp
 \oint  i\left( A_\m + {\dot x_\m \over \dot x \cdot v} v^I(x)\Phi^I 
 \right)dx^\m \,,
 \label{tauwl}
 \fe
 with $v^I\equiv -\bar\ve \hat \C^I \ve $. Here $\D_\ve W_R(\cK)=0$ follows from the identity
 \ie
 i v^\m \bar \ve  \C_\m - v^I  \bar \ve \hat \C_I=0.
 \fe

 Following the similar analysis as in Section~\ref{18BPS}, one can check that \eqref{tauwl} actually preserves the following 4 supercharges 
 \ie
 \cQ_L^{1}= Q_{1\bf 4}-S^{2 \bf 1} \,, \qquad
 \cQ_L^{2}=  Q_{1 \bf 1}+S^{2\bf 4} \,,  \qquad
 \overline \cQ_L^{2}=   Q_{2 \bf 1}-S^{1\bf 4} \,, \qquad
 \overline \cQ_L^{1}=  Q_{2\bf 4}+S^{1 \bf 1}  \,, 
 \fe
 and they are thus ${1\over 4}$-BPS. The supercharges $\cQ_L^\A$ and $\overline\cQ_L^\B$ are doublets that transform in the same way under $\mf{so}(3)_R$ and the $\mf{so}(3)_l$ subgroup of $\mf{so}(4)$ transverse rotations, while they are singlets under the $\mf{so}(3)_r$ subgroup of $\mf{so}(4)$.
 They satisfy the following anti-commutation relations
 \ie
 &\{\cQ_L^\A, \cQ_L^\B\}= \{\overline\cQ_L^\A, \overline\cQ_L^\B\}=0 \,, 
 \\
 &\{\cQ_L^\A, \overline\cQ_L^\B\}=2\epsilon^{\A\B}(M_{12}+R_{12}) \,.
 \fe
 Therefore the full symmetry preserved by the sector formed by all Wilson loops \eqref{tauwl} is
 \ie
 \left[ {\mf{su}(1|1) \oplus  \mf{su}(1|1) \over \mf{u}(1)} \rtimes (\mf{so}(3)_R \times \mf{so}(3)_l) \right] \oplus \mf{so}(3)_r   \,, 
 \label{taualg}
 \fe
 where the $\mf{u}(1)$ factor in the quotient is generated by $M_{12}+R_{12}$. 
 
 A general Wilson loop of the type \eqref{tauwl} will be ${1\over 4}$-BPS but is only invariant under a subalgebra of the bosonic part of \eqref{taualg}. The special Wilson loop along the great $S^1$ fiber at the center of $B^4$ ($\theta={\pi \over 2}$) preserves the full symmetry \eqref{taualg}.

\section{Analysis of local counter terms}
\label{COUNTER}

\subsection{Curve geometry on $S^3$ }\label{app:curvegeom}
For a curve $\ga(s)$ parameterized by the proper length $s$ relative to a reference point, the Frenet-Serret equations on curved space take the form
\es{FS0}{
{d\ov ds}\,\ga(s)&=t\,,\\
{D\ov Ds}\,t&=\kappa n\,,\\
{D\ov Ds}\,n&=-\kappa t+\tau b\,,\\
{D\ov Ds}\,b&=-\tau n\,,
}
where $(t,n,b)$ are the tangent, normal, and binormal unit vectors, $\kappa$ is the curvature and $\tau$ is the torsion of the curve, and $\frac{D}{Ds}$ denotes the covariant derivative along the curve $\frac{D}{Ds} \equiv t^\alpha \nabla_\alpha$.  These equations uniquely determine $t$;  $n$ is also uniquely determined by \eqref{FS0} and by the requirement that $\kappa > 0$;  and $b$ is uniquely determined by \eqref{FS0} together with a certain handedness condition for the frame:  for a right-handed frame, we have
 \es{Handedness2}{
\epsilon_{\al\beta\ga} t^\al n^\beta b^\ga = 1\,,
}
where $\al,\beta,\dots$ are tangent space indices of $S^3$ and $\epsilon_{\al \beta \ga}$ is an anti-symmetric tensor normalized such that $\epsilon_{123} = \sqrt{g}$.  Consequently, while we always have $\kappa > 0$, the torsion $\tau$ can have either sign.  

Standard discussions of submanifolds involve the extrinsic curvature, which for a curve we can denote by $K_\text{curve}^{A}$, where $A$ runs over the two transverse directions.  The curvature $\kappa$ is expressed in terms of the extrinsic curvature as  $\kappa= \sqrt{K_\text{curve}^{A} K_\text{curve}^{A}}$, and hence it is framing independent.  The torsion $\tau$ is a frame-dependent quantity. It is easy to verify that the definition of torsion given in \eqref{TorsionFT} for a general normal vector field agrees with the $\tau$ defined here, provided we use the $n$ of the Frenet-Serret frame as the normal vector field.

In embedding coordinates, the Frenet-Serret equations take the form:
\es{FS}{
X'&=t\,,\\
t'+X&=\kappa n\,,\\
n'&=-\kappa t+\tau b\,,\\
b'&=-\tau n\,,
}
where we relaxed the unit speed condition and defined $f'\equiv {\dot{f}\ov  \abs{\dot{X}}}$ for the reparameterization-invariant derivative (for example, $X''={1\ov  \abs{\dot{X}}} {d\ov d t} {\dot{X}\ov  \abs{\dot{X}}}$) as in the main text.  If we parameterize the $S^3$ using the stereographic projection from the North pole, and if the handedness condition \eqref{Handedness2} holds in this parameterization, then the handedness condition in the embedding space is
\es{Handedness}{
 \epsilon^{ijkl} t_i n_j b_k X_l = 1\,,
}
with $i, j, k, l$ being tangent indices in $\R^4$ and $\epsilon^{1234} = 1$.

In terms of the embedding coordinates, one can derive the explicit expressions for the curvature and torsion:
\es{Expr}{
\kappa^2&=(X'')^2 -{1}\,,\\
\tau&=-{1\ov\kappa^2}\epsilon_{ijkl} {X_i} X_j' X_k'' X_l'''
}
by starting with the RHS of each equation in \eqref{Expr} and using \eqref{FS} repeatedly.

\subsection{On finite local counter terms}\label{app:loccount}

In the main text we worked with a minimal holographic renormalization scheme. In this appendix, we analyze what local counter terms are allowed based on dimensional analysis. Of course, counter terms should respect both SUSY and Weyl invariance (or restore them, if the regularization scheme breaks them), however here we do not make use of these  additional constraints. 

We want to use the knowledge that the theory is secretly 6-dimensional, and the counter terms should respect the full 6-dimensional coordinate invariance. 
We also want to write down a counter term action that only depends on the intrinsic geometry of the surface operator that becomes the Wilson loop upon dimensional reduction. 
First, we need to understand dimensional analysis, hence we temporarily restore the radii of the $S^5$ and $S^1$, $R$ and $R_6$, and consequently we have $\sum_i X_i^2 = R^2$. We want to write down a dimensionless local action. The energy dimensions of the different quantities that we are working with are the following: 
\es{DimAnal}{
[X_i]&=-1\,,[\tau]=-1\,,\, [\Theta_a]=0\,,[t]=-1\,,\,[R,R_6]=-1\,,\\
[\nabla_\al]&=1\,,\, [A_\mu]=1\,,\,[\gamma_{\al\beta}]=0\,, \, [R_{ABCD}]=2\,,\, [K^A_{\al\beta}]=1\,,
}
where $A_\mu$ is the background R-symmetry gauge field, $R_{ABCD}$ is the Riemann tensor where $A,B,\dots=1,\dots,4$ label transverse directions, and $K^A_{\al\beta}$ is the extrinsic curvature  where $\al,\beta,\dots=1,2$ label tangential directions. 
We would like to construct the most general gauge and diffeomorphism invariant local action. The ingredients that we can use are the Riemann tensor of the ambient space, $S^5\times S^1$, the extrinsic curvature, and the intrinsic Ricci scalar ${\cal R}$ of the string world sheet. A choice of independent terms is
\es{2daction}{
S^\text{(finite)}_\text{counter}=&\tau_\text{M2} L^3\le[\int d^2\xi \ \sqrt{\ga}\le(a_1 K^A_{\al\beta}K^{A,\al\beta}+a_2 K^{A} K^{A}+a_3 Ricci+a_4 R^{A}_{\,\,\,A}+a_5 R^{AB}_{\,\,\,\,\,\,AB}\ri.\ri.\\
&\le.\le.+a_6 D_\al \Theta^{a} D^\al \Theta^{a} \ri)\ri]\,,
}
where $K^A= K^{A,\al}_{\,\,\,\,\,\,\,\,\,\al}\,$, $D_\al=\nabla_\al-i A_\alpha$, and $R_{AB}$ and $Ricci$ are the Ricci tensor and Ricci scalar of the ambient space. The intrinsic Ricci scalar ${\cal R}$ or the Riemann tensor with tangential indices do not make an appearance, as they are not linearly independent of the rest of the tensors.\footnote{Concretely, Gauss' equation implies that the intrinsic Ricci scalar is ${\cal R}=Ricci-2R^A_A+K^A_{\al\beta}K^{A,\al\beta}- K^{A} K^{A}$, and other contractions of the Riemann tensor with tangential indices are given as
\es{Relations}{
R^{\al}_{\,\,\,\al}&=Ricci-R^{A}_{\,\,\,A}\,,\\
R^{\al A}_{\,\,\,\,\,\,\al A}&=R^{A}_{\,\,\,A}-R^{AB}_{\,\,\,\,\,\,AB}\,,\\
R^{\al\beta}_{\,\,\,\,\,\,\al\beta}&=Ricci-2R^{A}_{\,\,\,A}+R^{AB}_{\,\,\,\,\,\,AB}\,.\\
}}$^,$\footnote{Imposing Weyl invariance alone  cannot fix the form of $S^\text{(finite)}_\text{counter}$. Combining the tensors in \eqref{2daction} into combinations that are invariant under Weyl rescalings, $g_{\mu\nu}\to e^{2\om}\, g_{\mu\nu}$, we still have four independent structures:
\es{Weyl}{
\tau_\text{M2} L^3 \le[\int d^2\sig \ \sqrt{\ga}\le(\hat{a}_1\le( K^A_{\al\beta}K^{A,\al\beta} -\frac12 K^{A} K^{A}\ri)+\hat{a}_2 {\cal R}+\hat{a}_3 W^{AB}_{\,\,\,\,\,\,AB}+\hat{a}_4 D_\al \Theta^a D^\al \Theta^a \ri)\ri]\,,
}
where the combination of the extrinsic curvatures is known as the Willmore energy,  $ {\cal R}$ is the Ricci scalar of the induced metric, and  $W^{AB}_{\,\,\,\,\,\,AB}$ is the Weyl tensor projected onto the transverse space. Once we specialize to $S^5\times S^1$ and the 1/8 BPS loops, only the  Willmore energy contributes. It would be interesting to understand if these terms are supersymmetric.}

Let us evaluate this finite counter term action on the 1/8-BPS surface operators of the 6d theory that wrap the M-theory circle. From the 5d perspective we get a Wilson loop characterized by the data of curve geometry on $S^3$ reviewed above in Appendix~\ref{app:curvegeom}, and in terms of this data the resulting expression is:
\es{counterFinal}{
S^\text{(finite)}_\text{counter}=&2\pi R_6 \, \tau_\text{M2}L^3  \int dt \ \abs{\dot{X}}\le(b_1 \kappa^2+b_2 Ricci+b_3 \le(\Theta'\ri)^2\ri)\,,
}
with $b_1=a_1+a_2,\, b_2=a_3+{3\ov 5}a_4+{3\ov 10}a_5,\,b_3=a_6$, where we used that  $S^5$ is a maximally symmetric space. For the BPS operators we consider, the $\Theta^a$ are determined from the first equation in \eqref{Solns}.  An explicit computation starting with this equation and using the identities \eqref{thooftid} for the 't Hooft symbol gives
\es{Relations1}{
\le(\Theta'\ri)^2&=\kappa^2=(X'')^2 -{1\ov R^2}\,,
}
hence we only have two independent counter terms at constant order. Note that upon setting $R=1$ (which implies $Ricci = 20$ for $S^5\times S^1$), the $Ricci$ term becomes the length term we denoted $L(\cK)$ in the main text.

Had we started from the perspective of a Wilson loop in $S^5$, setting the overall factor to $2\pi R_6={g_\text{YM}^2\ov 2\pi}$ (as opposed to an arbitrary function of $g_\text{YM}$) would require a nontrivial justification. If such a justification could be provided, then from the perspective of the curve $\int dt \ \abs{\dot{X}}\tau$ and $S_\text{WZ}[\Theta]$ may look admissible counter terms.\footnote{Adding them to the action would lead to an ambiguous final answer for the Wilson loop expectation value.} Neither of these terms is completely local however, as was discussed at various points in the paper: $\int dt \ \abs{\dot{X}}\tau$ introduces framing dependence, while the value of $S_\text{WZ}[\Theta]$ depends on the extension $\Theta(t)$ into a (topological disk).\footnote{Also $S_\text{WZ}[\Theta]$ only exists if $\Theta_a$ is restricted to $S^2$ (instead of taking values on $S^4$). Because of the Graham-Witten anomaly of non-BPS loops (discussed in Section~\ref{GWANOMALY}) the finite counter terms only make sense for BPS loops, hence imposing the supersymmetric condition that $\Theta_a$ lies on $S^2$ is reasonable.} Thus, a careful argument purely in 5d could rule them out as allowed local counter terms. The higher dimensional origin of the counter term action straightforwardly sets their coefficients to zero.

\section{5d SYM in 10d notation}
\label{app:10dsym}
  In flat space, the action of the ${\cal N} = 2$ SYM theory in 5d can be obtained by dimensionally reducing the 10d SYM action.   In 10d, the SYM Lagrangian takes the simple form  
  \es{LFlatSpace}{
  	\cL={1\over 2g_\text{YM}^2}\tr
  	\Bigg[
  	{1\over 2} F_{MN} F^{MN} +\Psi^T \CC^M D_M\Psi 
  	\Bigg] \,,
  }
  where $\Psi$ is a 16-component Majorana-Weyl spinor, $D_M=\p_M-i A_M$ is the covariant derivative and $F_{MN} =i[D_M,D_N]= \partial_M A_N - \partial_N A_M-i[A_M,A_N]$ is the field strength of the 10d gauge field, with $M, N = 0, \ldots, 9$ being 10d space-time indices raised and lowered with the mostly plus signature metric $\eta_{MN} = \diag\{-, +, \ldots,+\}$.\footnote{The fields $A_M$ are taken to be hermitian here in contrary with the convention of \cite{Pestun:2007rz, Minahan:2015jta}.} The 5d ${\cal N} = 2$ SYM Lagrangian is the same as \eqref{LFlatSpace}, with the only difference being that the fields do not depend on $5$ of the $10$ directions. We take these directions to be $6, 7, 8, 9, 0$, and denote the components of the gauge field in these directions by $\Phi$, \footnote{In the main text, for notational convenience, we have used $\Phi_i$ to denote the scalars $\Phi_{6,7}$ and $\Phi_a$ to denote the scalars $\Phi_{8,9,10}$ where $\Phi_{10}=i\Phi_0$.}
  \es{PhiNotation}{
  	\Phi_M = A_M \,, \qquad M = 6, 7, 8, 9, 0 \,,
  }
  in order to emphasize that they are scalar fields in 5d.
  
  The action for 5d SYM on $S^5$ can be obtained by covariantizing the expression \eqref{LFlatSpace} and adding curvature corrections.  These curvature corrections are fixed by demanding invariance under the ${\cal N} = 2$ supersymmetry algebra on $S^5$, which is $\mf{su}(4|2)$. The bosonic part of this algebra consists of the $\mf{su}(4) \cong \mf{so}(6)$ rotational symmetry of $S^5$ as well as an $\mf{su}(2)_R \oplus \mf{u}(1)_R$ R-symmetry.  The fermionic generators are parameterized by 16-component Majorana-Weyl spinors $\ve$ of the same chirality as $\Psi$, obeying the Killing equation
  \es{KSEq}{
  	D_\mu \ve = \frac{1}{2R} \tilde \Gamma_\mu \Lambda \ve \,,
  }
  where $\Lambda$ is a constant matrix with real entries that obeys $\Lambda^2 = -1$. We take $\Lambda=\Gamma^{890}$ as in \cite{Minahan:2015jta}. The $S^5$ metric is given by
  \ie
  ds^2=e^{2\Omega} dx^2,\quad e^\Omega\equiv {1\over 1+{x^2\over 4R^2}}.
  \fe
  in the stereographic coordinates $x_\m$. The explicit solutions  of the equation \eqref{KSEq} in the standard frame
  are
  \es{ExplicitKS}{
  	\ve=& e^{\Omega\over 2} \left(1+ \frac{1}{2R} x^\m\tilde \CC_{\hat\m}\Lambda \right) \epsilon_s \,,
  }
  with $\epsilon_s$ an arbitrary Majorana-Weyl constant spinor, and thus Eq.~\eqref{KSEq} indeed has a $16$-parameter family of solutions, as appropriate for the number of fermionic generators of $\mf{su}(4|2)$.  The constant matrix $\Lambda$ breaks the $\mf{so}(5)$ R-symmetry of the flat space theory \eqref{LFlatSpace} down to the $\mf{su}(2)_R \oplus \mf{u}(1)_R$ R-symmetry mentioned above, inducing a split of the five scalar fields $\Phi_M$, $M = 6, 7, 8, 9, 0$, into two groups.  With the choice $\Lambda = \Gamma^{890}$, the two groups are $\Phi_i$, with $i= 6,7$ (which are singlets of $\mf{su}(2)_R$ and have charges $\pm 1$ under $\mf{u}(1)_R$), and $\Phi_a$, with $ a= 8, 9, 0$ (which form a triplet of $\mf{su}(2)_R$ and have charge $0$ under $\mf{u}(1)_R$).  See Table~\ref{RSYM}.
  
  The scalars $\Phi_a$ and $\Phi_i$ then appear asymmetrically in the curvature corrections to the Lagrangian and to the SUSY transformation rules.  For the Lagrangian, we have \cite{Hosomichi:2012ek,Minahan:2015jta}\footnote{The path integral for 5d MSYM is defined by \ie
  	Z_{\rm YM}=\int DA D\Psi e^{-\int d^5 x \sqrt{g} \cL}.
  	\fe
  }
  \es{5dS5Action}{
  	\cL_{\rm YM}={1\over 2g_\text{YM}^2}\tr
  	\Bigg[
  	{1\over 2} F_{MN} F^{MN} +\Psi^T \CC^M D_M\Psi-{1\over 2R}\Psi^T \Lambda\Psi+{3\over R^2} \Phi^i\Phi_i+{4\over R^2}\Phi^a\Phi_a
	+{2i\over 3R}\epsilon^{abc}[\Phi_a,\Phi_b]\Phi_c
  	\Bigg] \,,
  }
  with $\epsilon^{890} = - \epsilon_{890} = -1$.   The action is invariant under the SUSY transformations
  \es{SUSYTransfCurved}{
  	&\D_\ve A_M=i \varepsilon^T \Gamma_M \Psi  \,,
  	\\
  	&\D_\ve\Psi 
  	=-{i\over 2}F_{MN}\Gamma^{MN}\varepsilon
  	-{2i\over 5}\Gamma_{\m i}\Phi^i D^\m \ve
  	-{4i\over 5}\Gamma_{\m a}\Phi^a D^\m \ve \,.
  }
  It can be checked that the anti-commutator of two SUSY transformations obeys the relations given by the $\mf{su}(4|2)$ algebra provided (as usual) that the fermion equations of motion are obeyed.
  
  We note that the kinetic term for $\Phi_0$ has the opposite sign in \eqref{5dS5Action}. To define a convergent path integral, we take its Wick rotation $\Phi_0=i\Phi_{10}$ and also make the replacement $\Gamma^0=-i\Gamma^{10},\tilde\Gamma^0=-i\tilde\Gamma^{10}$. From now on, we will work with the Wick rotated action for the 5d MSYM which can be put in the conventional form using the decomposition of the 10d Gamma matrices into the 5d ones in Appendix~\ref{app:gamma}. In particular in the Euclidean theory
  $\Lambda=i C\otimes \hat C \hat \C^{12}$.

\section{Differential geometry on $S^5$}
\label{S5APPENDIX}

We write $S^5$ in the stereographic coordinates as
\ie
ds^2={dx^2\over \left( 1+{ x^2\over 4 R^2}\right)^2} \,,
\fe
where $x$ represents the coordinate of a point in $\R^5$, and $dx^2$ is the line element on flat $\R^5$.  Alternatively, we can make the $S^1_\varrho$ fibration over $D^4$ obvious by writing
 \es{S5MetricFibration}{
ds^2=R^2(d\vartheta^2+ \sin^2\vartheta d\varrho^2+\cos\vartheta^2 d\Omega_3^2) \,,
 }
with $\vartheta\in [0,{\pi\over 2}]$.  The embedding coordinates can be related to the stereographic coordinates via \eqref{stereoS5} and to the coordianates in \eqref{S5MetricFibration} through
\ie
X_1=& R \sin\vartheta \sin\varrho \,, 
\\
X_2=& R \sin\vartheta \cos\varrho \,, 
\\
X_3=& R \cos \vartheta \sin\xi\sin\zeta \sin\phi \,, 
\\
X_4=& R\cos \vartheta \sin\xi\sin\zeta \cos\phi \,, 
\\
X_5=& R \cos \vartheta \sin\xi\cos\zeta \,, 
\\
X_6= &R\cos \vartheta  \cos\xi \,, 
\fe
where $\vartheta\in [0,{\pi\over 2}]$ and $\xi,\zeta\in[0,\pi]$ and $\varrho,\phi\in [0,2\pi]$.

\section{Details of the localization computation}
\label{App:CSaction}
Here we provide some details for the manipulations of the SYM action \eqref{actionB4} on the BPS locus on $B^4$. 
For notational simplicity we will set $R=1$ and only restore units in the end.

We define for convenience 
\ie
\CC_\varrho\equiv -i {\CC_\m v^\m\over v^\m v_\m} \,, 
\fe
which satisfies 
\ie
\ve \CC_\varrho \ve=1 \,,
\fe
and at $\varrho=0$, we have
\ie
\CC_\varrho={i\over \sin\vartheta} \CC_{\hat 1} \,.
\fe
From the BPS equation \eqref{bpseq}, we have on the BPS locus
 \es{KKExpression}{
K^m K_m
=
&{}-({1\over 2}F_{MN}\Gamma^{MN}\varepsilon
-(\tilde\Gamma_{ i}\Phi^i+2\tilde\Gamma_{ a}\Phi^a) \Lambda\ve)^T
\CC_\varrho 
({1\over 2}F_{MN}\Gamma^{MN}\varepsilon
-(\tilde\Gamma_{ i}\Phi^i+2\tilde\Gamma_{ a}\Phi^a) \Lambda\ve) 
\\
= &{1\over 4} F_{MN}F_{PQ} \ve \tilde\CC^{MN}\CC_\varrho \CC^{PQ}\ve
+\Phi_i\Phi^i+4\Phi_a \Phi^a
- F_{MN}\ve \tilde \CC^{MN}\CC_\varrho (\tilde\Gamma_{ j}\Phi^j+2\tilde\Gamma_{ b}\Phi^b) \Lambda\ve \,.
 }
Using the fact that on the BPS locus all fields are covariantly constant in $\varrho$, we drop all $\varrho$ derivatives in \eqref{KKExpression} and evaluate  the integrand at $\varrho=0$.\footnote{A careful analysis shows that the $\varrho$ derivatives all cancel on the BPS locus. Explicitly the $\varrho$ derivative terms are
	\begin{equation*}
	 \begin{aligned}
        2 F_{M\varrho } F^M{}_{Q}\ve \CC^Q \ve
	-2F_{M\varrho} \Phi_j\ve \CC^{M}\tilde\CC_j\Lambda\ve
	-4 F_{M\varrho} \Phi_a\ve \CC^{M}\tilde\CC_a\Lambda\ve 
	&= 2i F_{i\varrho } F^i{}_{Q} v^Q
	-2F_{i\varrho} \Phi_j\ve \tilde\CC^{ij}\Lambda\ve
	-4 F_{b\varrho} \Phi_a\ve \tilde\CC^{ab}\Lambda\ve 
	\\
	&= 2 F_{i\varrho } \omega^{ij}\Phi_j
	-2F_{i\varrho} \Phi_j\omega^{ij}=0 \,.
	\end{aligned}
	\end{equation*} 
   } 
Using the explicit form of the Killing spinor $\ve$, we obtain
\ie
F_{MN}F_{PQ} \ve \tilde\CC^{MN}\CC_\varrho \CC^{PQ}\ve
&=-2F_{MN}F^{MN}+ {i\over \sin \vartheta} F_{MN}F_{PQ} \ve \CC^{MNPQ \hat 1}\ve \,, 
\\
F_{MN}\ve \tilde \CC^{MN}\CC_\varrho \tilde\Gamma_{ j}\Phi^j \Lambda\ve
&={i\over \sin \vartheta} F_{MN} \Phi_j\ve\tilde \CC^{\hat 1 MN j}\Lambda\ve
- {2i\over \sin \vartheta} F_{Mj} \Phi^j \ve\tilde \CC^{M\hat 1}\Lambda \ve \,, 
\\
F_{MN}\ve \tilde \CC^{MN}\CC_\varrho \tilde\Gamma_{a}\Phi^a \Lambda\ve
&={i\over \sin \vartheta} F_{MN} \Phi_a\ve \tilde\CC^{\hat 1 MN a}\Lambda\ve
- {2i\over \sin \vartheta} F_{Ma} \Phi^a  \ve \tilde \CC^{M\hat 1}\Lambda \ve 
\fe
on $B^4$.  Here, we again repeatedly used the fact that all fields are covariantly constant in $\varrho$.  Hence
\ie
K^m K_m
=&-{1\over 2}F_{MN}F^{MN}
+\Phi_i \Phi^i+4\Phi_a\Phi^a 
+{i\over 4\sin\vartheta} F_{MN}F_{PQ}\ve \CC^{MNPQ\hat 1}\ve
\\
&
{}-{i\over \sin \vartheta} F_{MN} \Phi_j\ve \tilde\CC^{\hat 1 MN j}\Lambda\ve
+ {2i\over \sin \vartheta} F_{Mj} \Phi^j \ve\tilde \CC^{M\hat 1}\Lambda \ve
\\
&
{}-{2i\over \sin \vartheta} F_{MN} \Phi_a\ve \tilde\CC^{\hat 1 MN a}\Lambda\ve
+ {4i\over \sin \vartheta} F_{Ma} \Phi^a \ve\tilde \CC^{M\hat 1}\Lambda \ve \,,
\fe
and the bosonic action \eqref{actionB4} becomes
\ie
S=&{\pi\over  g_{\rm YM}^2} \int_{B^4} \sqrt{g_{B^4}} d^4 x \, \sin\vartheta \tr
\Bigg[
4\Phi^i\Phi_i+8\Phi^a\Phi_a-{2\over 3}\epsilon^{abc}[\Phi_a,\Phi_b]\Phi_c \\
&{}+{i\over 4\sin\vartheta} F_{MN}F_{PQ}\ve \CC^{MNPQ\hat 1}\ve
-{i\over \sin \vartheta} F_{MN} \Phi_j\ve\tilde \CC^{\hat 1 MN j}\Lambda\ve
+ {2i\over \sin \vartheta} F_{Mj} \Phi^j \ve\tilde \CC^{M\hat 1}\Lambda \ve \\
&{}-{2i\over \sin \vartheta} F_{MN} \Phi_a\ve\tilde \CC^{\hat 1 MN a}\Lambda\ve 
+ {4i\over \sin \vartheta} F_{Ma} \Phi^a \ve\tilde \CC^{M\hat 1}\Lambda \ve
\Bigg] \,.
\label{baS}
\fe
In order to show that the above integrand is a total derivative, we further compute 
\ie
{1\over 4} F_{MN}F_{PQ}\ve \CC^{MNPQ\hat 1}\ve
&= \ve\CC^{\m  I \n  J\hat 1}\ve D_\m  \tr (\Phi_I F_{\n  J} )
+{1\over 3}\ve\CC^{\m  IJK\hat 1}\ve D_\m  \tr (\Phi_I F_{JK} )
\\
&{}+\ve\CC^{\m  A\n \rho\hat 1}\ve D_\m  \tr (\Phi_I F_{\n  \rho} )
+{1\over 2}\ve\CC^{\m \n \rho\sigma\hat 1}\ve D_\m  \tr (A_\n  F_{\rho  \sigma}
+{2i\over 3} A_\n  A_\rho A_\sigma )
\\
&= D_\m  \Bigg(
\ve\CC^{\m  I \n  J\hat 1}\ve   \tr (\Phi_I F_{\n  J} )
+{1\over 3}\ve\CC^{\m  IJK\hat 1}\ve  \tr (\Phi_I F_{JK} )
\\
&{}+\ve\CC^{\m  I\n \rho\hat 1}\ve  \tr (\Phi_I F_{\n  \rho} )
+{1\over 2}\ve\CC^{\m \n \rho\sigma\hat 1}\ve  \tr (A_\n  F_{\rho \sigma}
+{2i\over 3} A_\n  A_\rho A_\sigma )
\Bigg)
\\
&{}+3\ve \Lambda \CC^{a\n  i\hat 1}\ve  \tr (\Phi_a F_{\n  i}-\Phi_i F_{\n  a})
+{4 \over 3}\sin \vartheta \epsilon^{abc}   \tr (\Phi_a F_{bc})
\\
&{}+2\ve \Lambda \CC^{a\n  \rho\hat 1}\ve  \tr (\Phi_a F_{\n  \rho}) \,.
\fe
In addition, we have
\ie
&- F_{MN} \Phi_j\ve \tilde\CC^{\hat 1 MN j}\Lambda\ve
-2F_{MN} \Phi_a\ve \tilde\CC^{\hat 1 MN a}\Lambda\ve
\\
=&- 2F_{\m  a} \Phi_j\ve \tilde\CC^{\hat 1 \m  a j}\Lambda\ve
-2\sin \vartheta \epsilon^{abc} \Phi_a F_{bc}+4 F_{\m  i} \Phi_a \ve \tilde\CC^{\hat 1 \m   a i}\Lambda\ve-2F_{\m \n } \Phi_a\ve \tilde\CC^{\hat 1\m \n  a}\Lambda\ve \,,
\fe
and similarly
\ie
D_\m  \Tr (\Phi^j \Phi_j \ve \tilde \CC^{\m  \hat 1}\Lambda \ve)= 2\Tr (F_{Mj} \Phi^j )\ve\tilde \CC^{M\hat 1}\Lambda \ve-4i\sin \vartheta \Tr (\Phi^j \Phi_j ) \,, 
\fe
and
\ie
-D_\m  \Tr (\Phi_i \Phi_a \ve \tilde \CC^{   \hat 1 \m a i}\Lambda \ve)=  -\Tr (F_{\m a} \Phi_i )\ve\tilde \CC^{   \hat 1 \m a i}\Lambda \ve
-
\Tr (F_{\m i} \Phi_a )\ve\tilde \CC^{   \hat 1 \m a i}\Lambda \ve \,.
\fe
Combining the five equations above, we can simplify the bosonic action \eqref{baS} to
\ie
S=&{i \pi \over g_{\rm YM}^2} \int_{B^4} \sqrt{g_{B^4}}  d^4 x \, D_\m  \Tr \Bigg[
\ve\CC^{\m  I \n  J\hat 1}\ve   (\Phi_I F_{\n  J} )
+{1\over 3}\ve\CC^{\m  IJK\hat 1}\ve  (\Phi_I F_{JK} )
+\ve\CC^{\m  I\n \rho\hat 1}\ve    (\Phi_I F_{\n  \rho} )
\\
&
+{1\over 2}\ve\CC^{\m \n \rho\sigma\hat 1}\ve  (A_\n  F_{\rho \sigma}+{2i\over 3} A_\n  A_\rho A_\sigma )
+\Phi^j \Phi_j \ve \tilde \CC^{\m  \hat 1}\Lambda \ve
+2\Phi^a \Phi_a \ve \tilde \CC^{\m  \hat 1}\Lambda \ve
- \Phi_i \Phi_a \ve \tilde\CC^{\hat 1 \m   a i}\Lambda\ve
\Bigg]  \,, 
\label{S3bos}
\fe 
which is indeed a total derivative on $B^4$ and integrates to the boundary $S^3$ at $\vartheta=0$,
\ie
S=&{i\pi \over g_{\rm YM}^2}\int_{S^3} \sqrt{g_{S^3}} d^3 x\, n_\m \Tr \Bigg[
\ve\CC^{\m a \n b\hat 1}\ve   (\Phi_a F_{\n b} )
+{1\over 3}\ve\CC^{\m abc\hat 1}\ve  (\Phi_a F_{bc} )
+\ve\CC^{\m a\n\rho\hat 1}\ve    (\Phi_a F_{\n \rho} )
\\
&
+{1\over 2}\ve\CC^{\m\n\rho\sigma\hat 1}\ve  (A_\n F_{\rho \sigma}+{2i\over 3} A_\n A_\rho A_\sigma )
+2\Phi^a \Phi_a \ve \tilde \CC^{\m \hat 1}\Lambda \ve 
\Bigg]  \,, 
\fe
where $n^\m$ is the unit normal to the boundary $S^3$.
In the above equation, we have thrown away the last term from \eqref{S3bos} because it vanishes at $\vartheta=0$ and also the third to last term in \eqref{S3bos} because $\Phi_{6,7}$ decouple (they have a quadratic action and can be integrated out). 

The above expression can be further simplified using the following identities at $\vartheta=0$
\ie
n_\m \ve \tilde \CC^{\m \hat 1}\Lambda \ve= -i \,, \quad 
n_\m \ve \tilde \CC^{\m 345\hat 1}  \ve= {i\over \cos^6 {\xi\over 2}} \,, \quad
n_\m \ve\CC^{\m 890\hat 1}  \ve=-1 \,, 
\quad
n_\m=-\D_{\m2}\cos^2{\xi\over 2} \,, 
\fe 
and the metric on $S^3$ which is determined by the conformal factor
\ie
e^{2\Omega}={1\over \left(
	1+{\sum_{i=3}^5 x_i^2\over 4}
	\right)^2}=\cos^4 {\xi\over 2} \,.
\fe
With all these ingredients,  the action on $S^3$ becomes
\ie
S
=&
-{i\pi \over g_{\rm YM}^2} \int_{S^3}d^3 x \sqrt{g_{S^3}} \,  \Tr \Bigg[
\ve\CC^{\hat 2 a {\rm j} b\hat 1}\ve   (\Phi_a F_{{\rm j} b} )
+{1\over 3}\ve\CC^{\hat 2  abc\hat 1}\ve  (\Phi_a F_{bc} )
+\ve\CC^{\hat 2 a{\rm j}{\rm k}\hat 1}\ve    (\Phi_a F_{{\rm j} {\rm k}} )
\\
&
+{1\over 2}\ve\CC^{\hat 2 {\rm j}{\rm k}{\rm l}\hat 1}\ve  (A_{\rm j} F_{{\rm k} {\rm l}}+{2i\over 3} A_{\rm j} A_{\rm k} A_{\rm l} )
+2\Phi^a \Phi_a \ve \tilde \CC^{\hat 2 \hat 1}\Lambda \ve
\Bigg]
\\
=&
{\pi \over g_{\rm YM}^2} \int_{S^3}  d^3 x \,e^{3\Omega} \Tr\Bigg[
\epsilon^{abc} \cS_c^{\rm j}   (\Phi_a F_{{\rm j} b} )
+{2\over 3}\epsilon^{ abc }(\Phi_a \Phi_b\Phi_c )
-i e^{-3\Omega}\ve^{{\rm i}{\rm j}{\rm k}} \cS^a_{\rm i}   (\Phi_a F_{{\rm j} {\rm k}} )
\\
&
-{1\over 2} e^{-3\Omega}\epsilon^{{\rm j}{\rm k}{\rm l}}(A_{\rm j} F_{{\rm k} {\rm l}}+{2i\over 3} A_{\rm j} A_{\rm k} A_{\rm l} )
+2 \Phi^a \Phi_a  
\Bigg]  \,,
\fe
where in the 2nd equality we used the tensor $\cS_{a{\rm i}}$ defined in \eqref{smat} and the following identities,
\ie
\ve \CC^{\hat 1\hat 2 \hat 3 \hat 4 \hat 5}\ve=-i
\,, \quad
\ve \CC^{\hat 1\hat 2 \hat 8 \hat 9 \hat{ 10}}\ve=1
\,, \quad
\ve\CC^{\hat 1\hat 2 {\rm j} a  b}\ve   =-i\epsilon^{abc} \cS^{\rm j}_c \,, 
\quad
e^{3\Omega}\ve\CC^{\hat 1\hat 2 {\rm j} {\rm k}  a}\ve=\epsilon^{{\rm i} {\rm j}{\rm k}} \cS_{\rm i}^a \,.
\fe

\bibliographystyle{ssg}
\bibliography{5d3d}

\end{document}